\documentclass[review]{elsarticle}

\usepackage[colorlinks,citecolor=blue,linktoc=all,linkcolor=cyan]{hyperref}
\usepackage{graphicx}

\usepackage[T1]{fontenc}
\usepackage{dsfont}               
\usepackage{mathrsfs}            
\usepackage{slashed}              
\usepackage{amsmath}
\usepackage{amssymb}
\usepackage{amsbsy}
\usepackage{amsfonts}

\numberwithin{equation}{section}
\numberwithin{table}{section}
\numberwithin{figure}{section}

\journal{Progress in Particle and Nuclear Physics}

\usepackage{titlesec}
\usepackage{sectsty}
\titleformat{\section}{\normalfont\Large\bfseries}{\thesection}{1em}{}
\titleformat{\subsection}{\normalfont\large\bfseries}{\thesubsection}{1em}{}
\titleformat{\subsubsection}{\normalfont\normalsize\bfseries}{\thesubsubsection}{1em}{}

\usepackage{wrapfig}
\usepackage{xspace}
\usepackage{dcolumn}

\graphicspath{{./Figures/}}

\newcommand{\snn} {\sqrt{s_{_{\rm NN}}}}
\newcommand{\elab}{E_{\rm{lab}}}
\newcommand{\ekin}{E_{\rm{kin}}}
\newcommand{\txt}[1]{\rm{#1}}
\newcommand{\ceft}{$\chi$EFT\xspace}

\begin{document}
	
	\begin{frontmatter}
		
		\title{Dense Nuclear Matter Equation of State from Heavy-Ion Collisions}


		\author[INT]{Agnieszka~Sorensen\corref{mycorrespondingauthor}}
		\cortext[mycorrespondingauthor]{Corresponding author}
		\ead{agnieszka.sorensen@gmail.com}
		\author[TubingenU]{Kshitij~Agarwal}
		\author[FRIB,MSUChem]{Kyle~W.~Brown}
		\author[WMU]{Zbigniew~Chaj\k{e}cki}
		\author[FRIB,MSU]{Pawe{\l}~Danielewicz}
		\author[OhioU]{Christian~Drischler}
		\author[LANL]{Stefano~Gandolfi}
		\author[TAMUPhysandAstro,TAMUCyclo]{Jeremy~W.~Holt}
		\author[UA]{Matthias~Kaminski}
		\author[TAMUPhysandAstro,TAMUCyclo]{Che-Ming~Ko}
		\author[FRIB]{Rohit~Kumar}		
% ...
		\address[INT]{Institute for Nuclear Theory, University of Washington, Seattle, WA 98195, USA}
		\address[TubingenU]{Physikalisches Institut, Eberhard Karls Universit\"{a}t T\"{u}bingen, D-72076 T\"{u}bingen, Germany}
		\address[FRIB]{Facility for Rare Isotope Beams, Michigan State University, East Lansing, MI 48824, USA}
		\address[MSUChem]{Department of Chemistry, Michigan State University, East Lansing, MI 48824, USA}
		\address[WMU]{Department of Physics, Western Michigan University, Kalamazoo, MI 49008, USA}
		\address[MSU]{Department of Physics and Astronomy, Michigan State University, East Lansing, Michigan 48824, USA}
		\address[OhioU]{Department of Physics and Astronomy, Ohio University, Athens, OH 45701, USA}
		\address[LANL]{Theoretical Division, Los Alamos National Laboratory, Los Alamos, NM 87545, USA}
		\address[TAMUPhysandAstro]{Department of Physics and Astronomy, Texas A\&M University, College Station, TX 77843, USA}
		\address[TAMUCyclo]{Cyclotron Institute, Texas A\&M University, College Station, TX 77843, USA}
		\address[UA]{Department of Physics and Astronomy, University of Alabama, Tuscaloosa, AL 35487, USA}
% ...

		\begin{abstract}
		The nuclear equation of state (EOS) is at the center of numerous theoretical and experimental efforts in nuclear physics. With advances in microscopic theories for nuclear interactions, the availability of experiments probing nuclear matter under conditions not reached before, endeavors to develop sophisticated and reliable transport simulations to interpret these experiments, and the advent of multi-messenger astronomy, the next decade will bring new opportunities for determining the nuclear matter EOS, elucidating its dependence on density, temperature, and isospin asymmetry. Among controlled terrestrial experiments, collisions of heavy nuclei at intermediate beam energies (from a few tens of MeV/nucleon to about 25 GeV/nucleon in the fixed-target frame) probe the widest ranges of baryon density and temperature, enabling studies of nuclear matter from a few tenths to about 5 times the nuclear saturation density and for temperatures from a few to well above a hundred MeV, respectively. Collisions of neutron-rich isotopes further bring the opportunity to probe effects due to the isospin asymmetry. However, capitalizing on the enormous scientific effort aimed at uncovering the dense nuclear matter EOS, both at RHIC and at FRIB as well as at other international facilities, depends on the continued development of state-of-the-art hadronic transport simulations. This white paper highlights the essential role that heavy-ion collision experiments and hadronic transport simulations play in understanding strong interactions in dense nuclear matter, with an emphasis on how these efforts can be used together with microscopic approaches and neutron star studies to uncover the nuclear EOS.
	 	\end{abstract}		
	\end{frontmatter}
	

    \newpage
	\tableofcontents
	


\newpage
\section{Introduction} 
\label{sec:introduction}

The equation of state (EOS) is a fundamental property of nuclear matter, describing its emergent macroscopic behavior originating from the underlying strong interactions. For ordinary nuclear matter, the EOS controls the structure of nuclei through the binding energy \cite{Day:1967zza}, the incompressibility \cite{Blaizot:1980tw,Moller:2012pxr}, or the neutron-skin thickness in neutron-rich nuclei \cite{Horowitz:2001ya,Reed:2021nqk}. 
The EOS also determines the properties of nuclear matter at extreme densities and/or temperatures, corresponding to conditions produced in experiments colliding heavy nuclei~\cite{HOTQCDWhitePaper,Meehan:2016qon,HADES:2022gdr,FRIB400,Senger:2021dot} or observed in neutron stars \cite{Ozel:2016oaf} and neutron star mergers \cite{Bauswein:2012ya}. 
Far beyond describing the properties of matter composed of only protons and neutrons, the EOS can also reflect the appearance of new degrees of freedom, e.g., strange particles in the cores of neutron stars \cite{Balberg:1998ug,Chatterjee:2015pua,Oertel:2016xsn,Roark:2018boj,Gerstung:2020ktv} or quarks and gluons in ultrarelativistic heavy-ion collisions \cite{Aoki:2006we,Aoki:2006br,HotQCD:2018pds,Borsanyi:2020fev}, or the emergence of new states of matter, e.g., chirally-restored matter \cite{Brown:1995qt,Berges:1998rc,Alford:1998mk}, meson condensates \cite{Kaplan:1987sc,Son:2000xc,Pethick:2015jma}, or quarkyonic matter~\cite{McLerran:2007qj,McLerran:2018hbz,Jeong:2019lhv}. 

\begin{figure}[!b]
    \centering
    \includegraphics[width=0.99\linewidth]{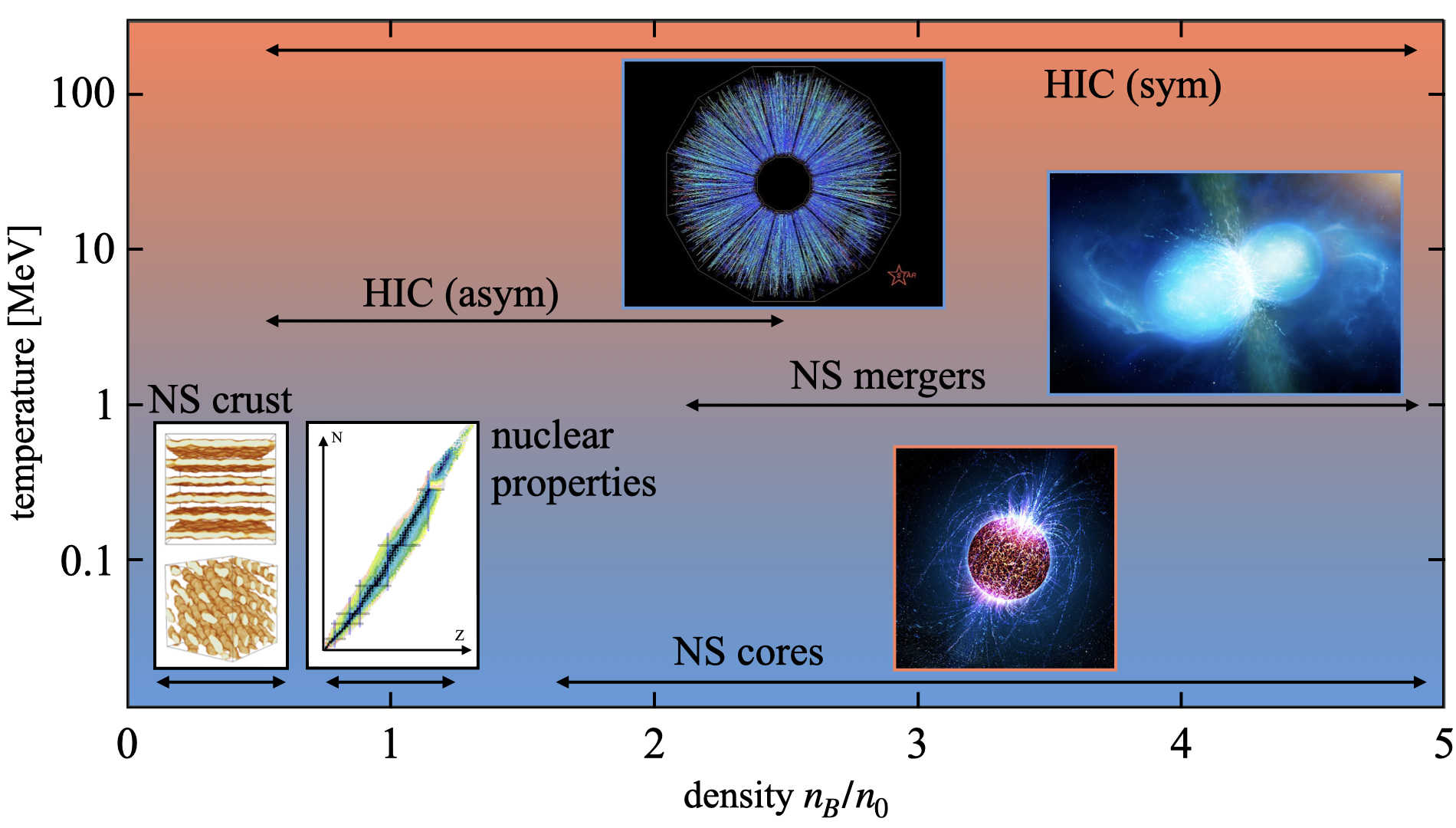}
    \caption{Schematic depiction of the ranges of density and temperature probed in experiments and astronomical observations sensitive to the EOS of nuclear matter (\textit{counterclockwise from bottom left}): neutron star (NS) crust physics, including nuclear pasta structures; properties of nuclei; structure of neutron stars; dynamics of neutron star mergers; and outcomes of heavy-ion collisions (HIC) which can probe both symmetric and asymmetric matter. Figures adapted from \cite{Caplan:2016uvu,Nuclear_chart_NNDC,Neutron_star_isolated,Neutron_star_merger,RHIC_gold_collision}.
    }
    \label{fig:observable_ranges}
\end{figure}

In heavy-ion collision experiments, the EOS is studied by detecting particles emerging from the collision zone and measuring observables sensitive to the properties of nuclear matter. 
Interpretation of these observables, including quantitative constraints on the EOS, requires comparisons of experimentally measured observables to results obtained in dynamic simulations. 
This White Paper highlights the essential role of hadronic transport simulations of heavy-ion collisions in advancing our understanding of the EOS. 
It also elucidates the many connections between inferences of the EOS from heavy-ion collision data and other efforts aiming to describe and understand the properties of nuclear matter.

\subsection{Constraining the nuclear matter EOS using heavy-ion collisions}

\begin{wrapfigure}{l}{0.42\textwidth}
  \centering
    \vspace{-6mm}
    \includegraphics[width=0.42\textwidth]{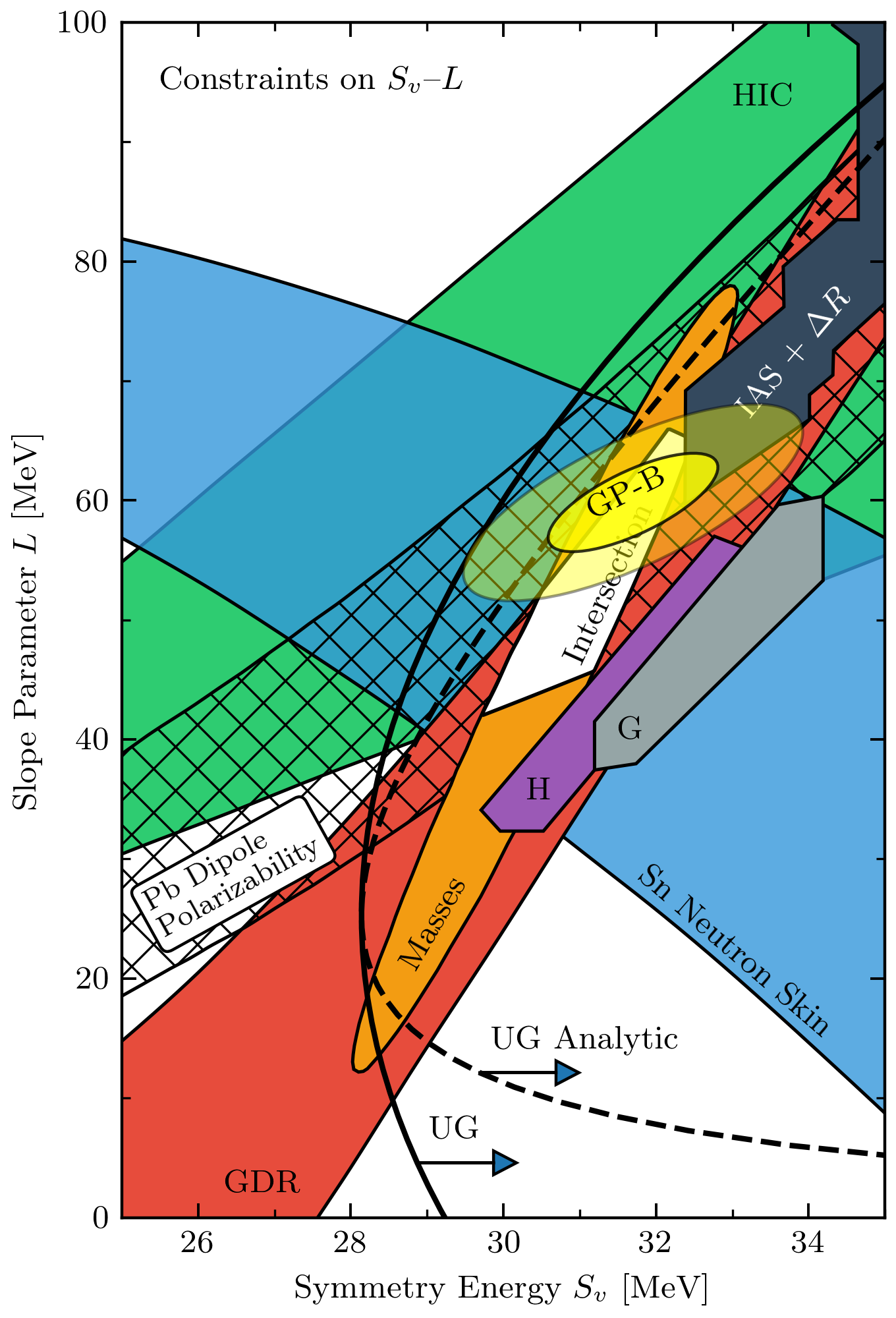}
    \vspace{-6.75mm}
  \caption{Constraints on the zeroth ($S_v$) and first ($L$) coefficient of the symmetry energy expansion. Experimental constraints are derived from heavy-ion collisions (HIC)~\cite{Tsang:2008fd}, neutron-skin thicknesses of Sn isotopes \cite{Chen:2010qx}, giant dipole resonances (GDR) \cite{Trippa:2008gr}, the dipole polarizability of $^{208}$Pb~\cite{Tamii:2011pv,Roca-Maza:2013mla}, nuclear masses \cite{Kortelainen:2010hv}, and isovector skins (IAS+$\Delta R$) \cite{Danielewicz:2016bgb}. Also shown are constraints from $\chi$EFT (GP-B) \cite{Drischler:2020hwi}, microscopic neutron-matter calculations \mbox{(H, G)}~\cite{Hebeler:2010jx,Gandolfi:2011xu}, and from the unitary gas limit (UG)~\cite{Tews:2016jhi}. Figure from~\cite{Drischler:2020hwi}.
  }
  \vspace{-4mm}
  \label{combined_S_and_L_constraints}
\end{wrapfigure}
The last decade has brought tremendous progress in extracting the EOS as a function of baryon density~$n_B$, temperature $T$, and the isospin asymmetry~$\delta$ (or, equivalently, the proton fraction) from a variety of experimental and astronomical data as well as theoretical calculations. 
Many-body theory, based on sophisticated approaches with input from
nucleon scattering or nuclear structure data, can now state the EOS below and near the saturation density $n_0$ with meaningful uncertainties~\cite{Hebeler:2015hla,Lynn:2019rdt,Drischler:2021kqh,Drischler:2021kxf} (see Section~\ref{sec:microscopic_calculations_of_the_EOS}, ``Microscopic calculations of the EOS''). 
New classes of experiments have extracted the thickness of neutron skins in nuclei~\cite{Abrahamyan:2012gp,PREX:2021umo,Reed:2021nqk,CREX:2022kgg,Reinhard:2022inh,Zhang:2022bni}, shedding light on the isospin-dependence of the EOS (or, equivalently, the symmetry energy) near or below~$n_0$.
High-energy heavy-ion collisions~\cite{BRAHMS:2004adc,PHENIX:2004vcz,PHOBOS:2004zne,STAR:2005gfr} have constrained the EOS of the quark-gluon plasma at high temperatures and small baryon densities~\cite{Pratt:2015zsa}, while ongoing experimental efforts worldwide focus on the EOS of nearly-symmetric dense baryonic matter, probed in collisions at intermediate energies.
Meanwhile, collisions at lower energies have led to experimental constraints on the symmetry energy at sub- and suprasaturation densities~\cite{Zhang:2007hmv,Zhang:2014sva,SRIT:2021gcy,SpRIT:2020blg,Lynch:2021xkq}. 
Most remarkably, a revolution in the quality and breadth of astronomical observations, highlighted by the first simultaneous detection of gravitational waves and electromagnetic signals from a neutron-star merger~\cite{LIGOScientific:2017vwq}, ushered in a new era of multi-messenger astronomy (see Section~\ref{sec:neutron_star_theory}, ``Neutron star theory''). 
Together with the newly available experimental capabilities at the Facility for Rare Isotope Beams (FRIB), there are unprecedented opportunities to probe the isospin-dependence of the EOS through astronomical and terrestrial measurements.

Among the experimental efforts discussed above, heavy-ion collisions probe the widest range of baryon densities and, moreover, represent the only means to address the EOS away from~$n_0$ in controlled terrestrial experiments, see Fig.~\ref{fig:observable_ranges}. 
Indeed, heavy-ion reactions at beam kinetic energies from a few tens of MeV/nucleon to about 25 GeV/nucleon in the fixed-target frame probe the EOS of hadronic matter at baryon densities from a few tenths to about 5 times $n_0$. 
Controlling the properties of matter produced in these experiments is possible by varying the beam energy, collision geometry, and isotopic composition of the target and projectile. Insights and constraints obtained from transport model analyses of these experiments are relevant both for our understanding of nuclear matter as found on Earth and for our understanding of neutron stars from crust to core.

Within ongoing efforts, the STAR experiment's Beam Energy Scan (BES) fixed-target (FXT) program \cite{Meehan:2016qon} at the Relativistic Heavy Ion Collider (RHIC) at the Brookhaven National Laboratory (BNL), which collided gold nuclei at
intermediate beam energies and which completed data taking 
in 2022, leads the U.S. efforts to constrain the EOS of 
\begin{wrapfigure}{r}{0.38\textwidth}
  \centering
  \vspace{-2mm}
    \includegraphics[width=0.38\textwidth]{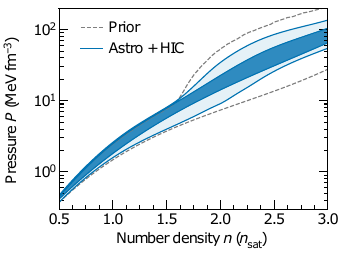}
    \vspace{-7,mm}
  \caption{Pressure in neutron star matter as a function of density from a Bayesian analysis combining nuclear theory and data from multi-messenger neutron-star observations and heavy-ion collisions \cite{Huth:2021bsp}: the dark blue and light blue region corresponds to the 68\% and 95\% credible interval, respectively, while the gray dashed line shows the 95\% bound obtained in $\chi$EFT calculations and used as a prior. Figure from~\cite{Huth:2021bsp}.
  }
  \vspace{-4mm}
  \label{Huth_et_al_panel_d}
\end{wrapfigure}
nearly-symmetric nuclear matter at high baryon densities up to around $5n_0$, corresponding to densities present in the 
deep interiors of neutron stars. 
Among comparable efforts in Europe, the HADES experiment \cite{HADES:2009aat} at GSI, Germany, probes matter at densities up to $2.5n_0$. 
Preliminary results from these contemporary efforts, as well as measurements from other heavy-ion collision experiments in the past, have led to competitive constraints on the EOS of symmetric nuclear matter~\cite{Danielewicz:2002pu,Fuchs:2003pc,Lynch:2009vc,LeFevre:2015paj,Oliinychenko:2022uvy}, with future measurements expected to shed more light on its high-density behavior. 
Detailed constraints on the isospin-dependence of the EOS can be obtained by varying the isospin content of the target and projectile nuclei. 
Here, the ability to use radioactive isotopes, as in, e.g., intermediate-energy heavy-ion collision experiments at RIKEN and FRIB, is crucial to resolve the subtle effects arising from changes in the isospin asymmetry of colliding systems~\cite{FRIB400}.

Above all, obtaining constraints on the EOS from heavy-ion measurements would not have been possible if not for advances in theory, and in particular for the collaborative effort to test the robustness and quantify the uncertainties of hadronic transport simulations (see Section \ref{sec:model_simulations_of_HICs}, ``Transport model simulations of heavy-ion collisions''). 
At the same time, much remains to be learned, as tight constraints on both the symmetric and asymmetric EOS at higher densities have so far remained elusive. This is predominantly due to model uncertainties~\cite{TMEP:2022xjg}, which themselves are rooted in the inherent complexity of nucleus-nucleus collisions and the challenging task of describing all processes contributing to the final state observables.

\subsection{Connections to fundamental questions in nuclear physics}
\label{sec:connections_to_fundamental_physics}

The wealth of data from efforts conducted in recent years has brought forward fascinating questions challenging our understanding of strong interactions.

Following the successful BES-I campaign at RHIC, questions remain about the structure of the QCD phase diagram at large baryon densities, where the sign problem prevents predictions from lattice QCD calculations \cite{Karsch:2001cy} and extrapolations of lattice QCD results become unreliable~\cite{Borsanyi:2020fev}. 
Surprisingly, the expected disappearance of quark-gluon plasma signatures has not been unequivocally observed in BES-I, with some observables suggesting that the QCD first-order phase transition may be located in the statistically demanding low-energy region probed by BES-II \cite{BES_II_WhitePaper}, including the region probed by the currently analyzed BES-II FXT data~\cite{STAR:2021yiu,STAR:2021fge}. 
If this is the case, then constraining the EOS at lower densities and describing the approach to the transition from the hadronic side, which would manifest as a softening of the EOS, will be crucial for a robust interpretation of BES-II measurements. 
Due to the largely out-of-equilibrium evolution of collision systems probing that region of the QCD phase diagram, hadronic transport simulations will play a dominant role in describing the dynamics of the collisions, and therefore in constraining the EOS of nearly-symmetric dense nuclear matter.

Understanding the physics of neutron-rich matter across a range of densities is necessary not only to explain the properties of rare neutron-rich isotopes and the structure of neutron stars, but also to constrain microscopic interactions in isospin-asymmetric nuclear matter. 
At low densities, this challenge is addressed by experimental and theoretical analyses 
\begin{wrapfigure}{l}{0.45\textwidth}
  \centering
    \vspace{0.mm}
    \includegraphics[width=0.45\textwidth]{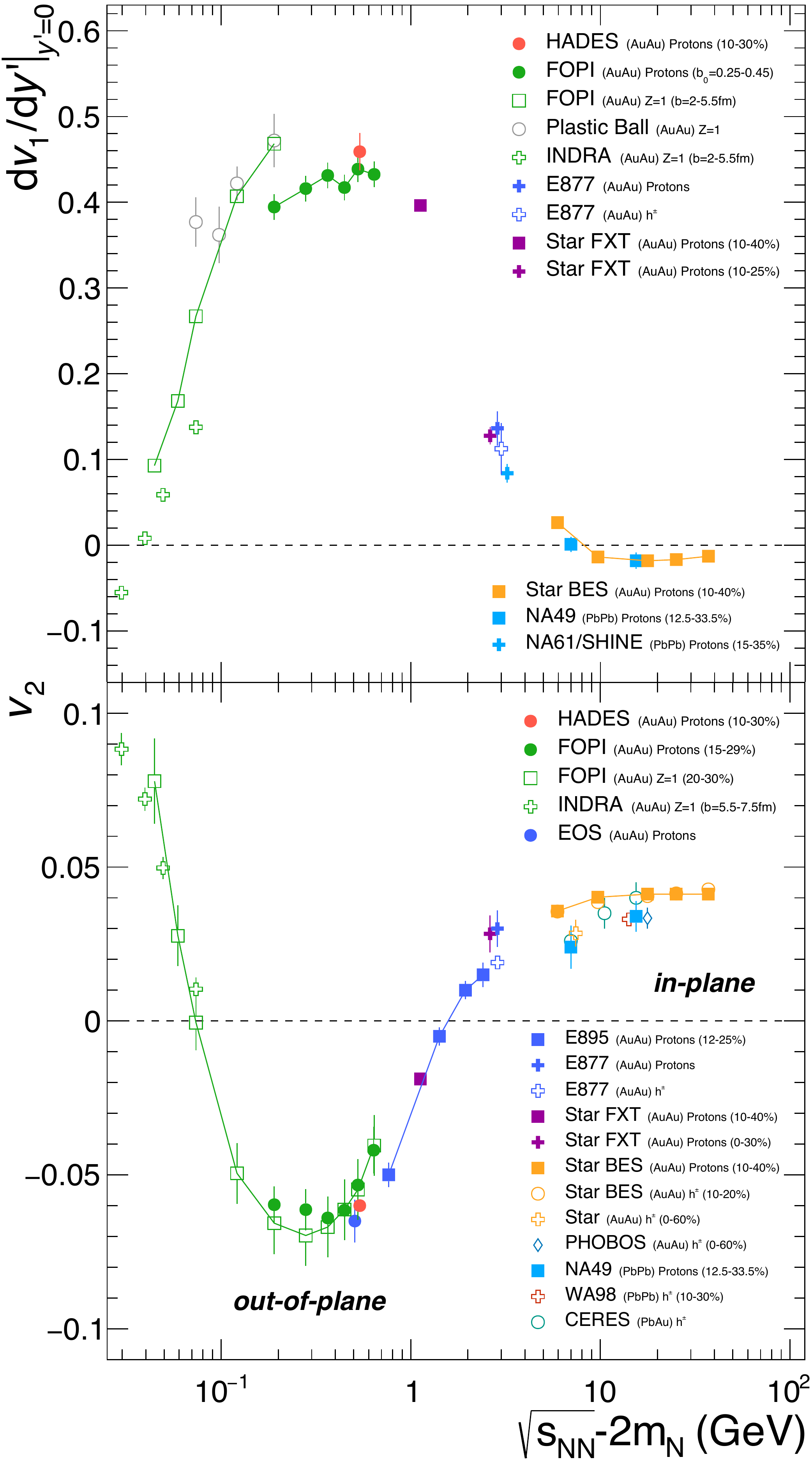}
    \vspace{-7mm}
  \caption{Compilation of the world data on the slope of the directed flow at mid-rapidity ($dv_{1}/dy|_{y^{\prime} = 0}$, \textit{top}) and the elliptic flow ($v_2$, \textit{bottom}) as functions of the reduced center-of-mass energy $\sqrt{s_{NN}} - 2m_{N}$ for protons, $Z=1$ nuclei, and inclusive charged particles. 
  Figure modified from~\cite{HADES:2022osk}. 
  }
  \vspace{-2mm}
  \label{fig:v1_v2_world_data}
\end{wrapfigure}
of nuclear structure observables~\cite{Hergert:2020bxy,Johnson:2019sps,Tews:2020hgp}. 
An important objective of nuclear many-body theorists is to accurately 
calculate these observables and reliably deduce the EOS using microscopic interactions derived within the framework of chiral effective field theory ($\chi$EFT)~\cite{Epelbaum:2008ga,Machleidt:2011zz,Machleidt:2021ggx}.
Probing the symmetry energy over a range of densities wider than found in nuclei is possible through heavy-ion collisions and neutron star studies. 
Often, knowledge of the isospin-asymmetric EOS is encoded in terms of constraints on the Taylor expansion coefficients of the symmetry energy around~$n_0$.
Numerous analyses yield consistent constraints on the first few expansion coefficients~\cite{Drischler:2020hwi} (see Fig.~\ref{combined_S_and_L_constraints}), although they rely on an assumption that the expansion remains accurate away from~$n_0$.
The recent advent of Bayesian inference techniques allows one to pursue a different approach, within which the isospin-asymmetric EOS is described in terms of the dependence of the pressure on baryon density~\cite{Huth:2021bsp} (see Fig.~\ref{Huth_et_al_panel_d}). 
Moreover, Bayesian analyses can shed more light on densities at which measurements constrain the symmetry energy and quantify the uncertainties of the extracted EOS. 
As a result, combining diverse measurements and using advanced analysis techniques can lead to significantly tighter constraints, especially on the high-density behavior of the symmetry energy (or, equivalently, on the higher-order symmetry energy expansion coefficients), so far poorly known.

Constraints on the EOS of neutron-rich matter at high densities have been dramatically affected by discoveries of heavy neutron stars~\cite{Fonseca:2021wxt,NANOGrav:2019jur}. 
Combined with the properties of all known compact stars, these observations indicate that while the EOS of neutron-rich matter is relatively soft around (1--2)$n_0$, the pressure steeply rises with density for $n_B \gtrsim 2n_0$~\cite{Miller:2021qha,Essick:2020flb}. 
In fact, multiple analyses show that describing the known population of neutron stars is only possible for EOSs in which the speed of sound in neutron-star matter breaks the conformal limit at high densities, that is exceeds $1/\sqrt{3}$ of the speed of light $c$ for $n_B \gtrsim 2n_0$~\cite{Bedaque:2014sqa,Alford:2015dpa,Tews:2018kmu,Fujimoto:2019hxv, Marczenko:2022jhl}. 
This striking behavior remains to be understood. 
In particular, it is currently not known whether the speed of sound exceeds $c/\sqrt{3}$ above certain densities at all isospin fractions of nuclear matter or, alternatively, only in neutron-rich matter. 
Robust constraints on the symmetric matter EOS at $n_B \gtrsim 2n_0$, obtained from heavy-ion collisions at intermediate to high beam energies, would answer this question as well as put constraints on the isospin-dependent part of the EOS through comparisons with the EOS inferred from neutron star studies, thus uncovering the magnitude of isospin-related effects at high baryon density.

.

\subsection{Upcoming opportunities}

The next decade will be an era of high-luminosity heavy-ion collision experiments at high baryon density with modern detector and analysis procedures, as well as detailed studies of the symmetry energy with collisions of proton- and neutron-rich isotopes.

Many of the discoveries of the BES program in ultra-relativistic heavy-ion collisions at RHIC, e.g., the discovery of the triangular flow~\cite{Mishra:2007tw,Sorensen:2010zq} and elliptic flow fluctuations~\cite{PHOBOS:2006dbo}, illustrate that modern analyses of heavy-ion collisions bring new quality to the understanding of the underlying processes. 
Because of this, revisiting the intermediate to high beam 
energies, previously explored at the AGS at BNL as well as at SIS18 at GSI and now explored by the STAR FXT program 
and the HADES experiment, is imperative to enable putting tighter constraints on the EOS of dense nuclear matter. 
Moreover, the future CBM experiment \cite{Senger:2021dot,Durante:2019hzd} at the Facility for Antiproton and Ion Research (FAIR), Germany, will be able to measure interaction rates exceeding those currently used by several orders of magnitude, allowing for exploration of multiple high-statistics observables~\cite{Almaalol:2022xwv}. 
Furthermore, the explored beam energy range is where lower-order flow observables, reflecting the collective motion of the colliding system due to the underlying hadronic EOS, are particularly
\begin{wrapfigure}{r}{0.45\textwidth}
  \centering
  \vspace{-5.5mm}
    \includegraphics[width=0.45\textwidth]{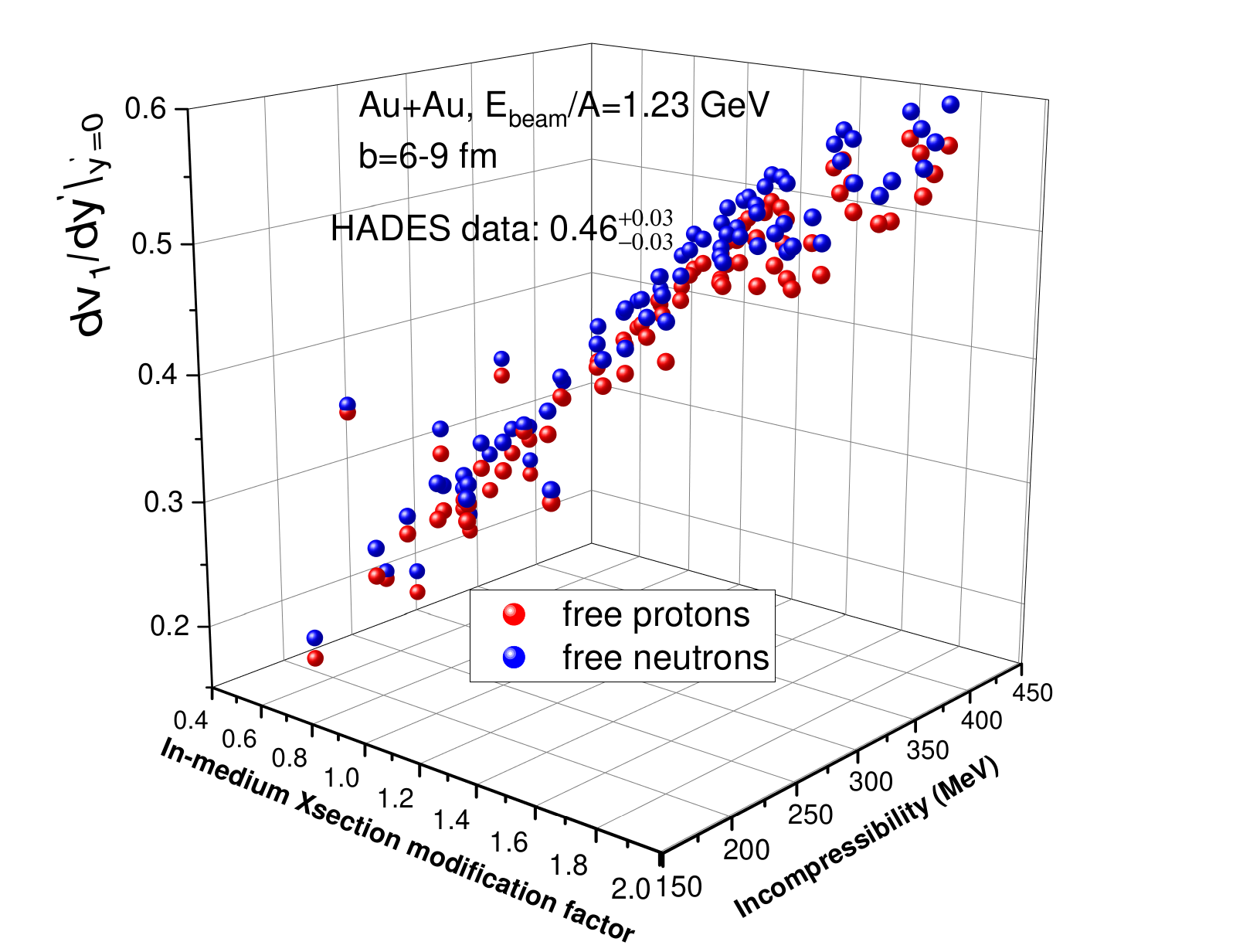}
    \includegraphics[width=0.45\textwidth]{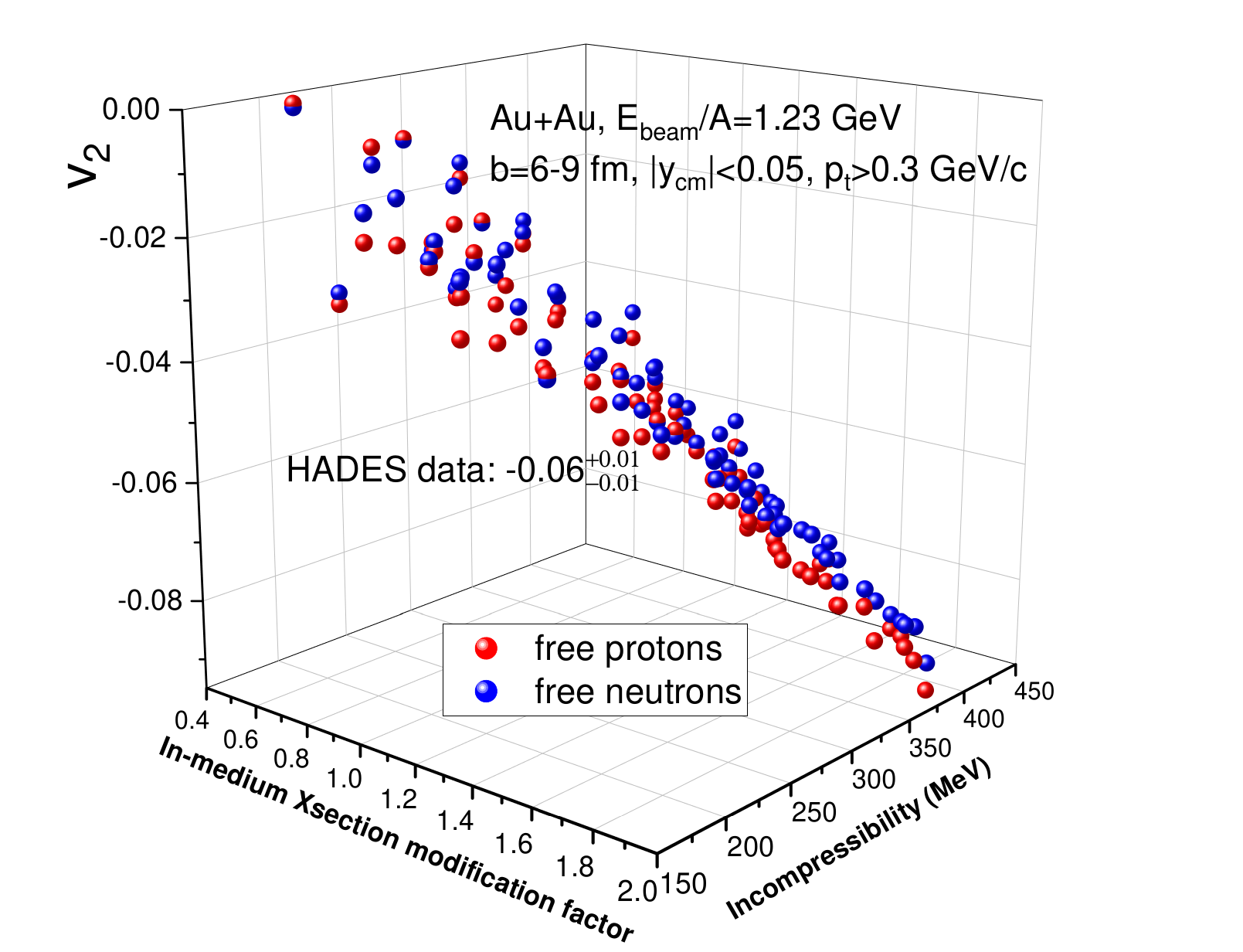}
    \vspace{-6.5mm}
  \caption{Predicted slope of the directed flow at mid-rapidity ($dv_{1}/dy|_{y^{\prime} = 0}$, \textit{top}) and elliptic flow ($v_2$, \textit{bottom}) as functions of the incompressibility and the in-medium nucleon-nucleon scattering cross section modification factor, generated in simulations of Au+Au reactions using the isospin-dependent BUU (\texttt{IBUU}) transport model~\cite{Li:1997px,Li:2008gp}. Figure from Ref.~\cite{Li:2023ydi}. 
  }
  \vspace{-6mm}
  \label{fig:v1_v2_80_np}
\end{wrapfigure}
prominent (see Fig.~\ref{fig:v1_v2_world_data}). 
Therefore, the corresponding precision measurements carry with them the opportunity to bring 
a richer perspective and a better understanding of the physics underlying the complex dynamics of nuclear matter at extreme conditions (see Section~\ref{sec:experiment_snm}, ``Experiments to extract the EOS of symmetric nuclear matter''). 
This advancement can only occur provided a simultaneous development of hadronic transport simulations, as only a detailed understanding of various factors affecting the dynamics of heavy-ion collisions can lead to meaningful descriptions of the experimental data, and, consequently, more robust constraints on the EOS of nearly-symmetric nuclear matter (see Section~\ref{sec:model_simulations_of_HICs}, ``Transport model simulations of heavy-ion collisions''). 
As an example of the sensitivity of observables to various details of the underlying physics, Fig.~\ref{fig:v1_v2_80_np} shows the dependence of the slope of the directed flow (top panel) and of the elliptic flow at midrapidity (bottom panel) on the stiffness of the EOS, parametrized by the incompressibility, and on the in-medium nucleon-nucleon scattering cross-section modification factor.

Unprecedented possibilities are on the horizon for studies of the isospin-dependence of the EOS, which is critical for connecting heavy-ion collision measurements to astrophysical observations.
The difficulties in using nuclei with significant variations in the isospin asymmetry, along with the paucity of neutron measurements at midrapidity, have in the past greatly restricted the capability to put tight constrains on the EOS of asymmetric nuclear matter. 
Fortunately, at this time modern neutron detectors are available for heavy-ion measurements in many facilities, including at accelerators performing collisions at high beam energies such as GSI, while radioactive beam measurements are entering a new era at RIKEN and FRIB.
FRIB will provide proton- and neutron-rich beams of not only the highest-intensity worldwide, but also characterized by the widest currently accessible range of the isospin asymmetry~\cite{FRIB_beams,FRIB400}. 
Establishing a strong heavy-ion program at FRIB will therefore enable previously inaccessible exploration of the symmetry energy (see Section~\ref{sec:experiment_asym}, ``Experiments to extract the symmetry energy'').
Moreover, the proposed FRIB400 beam energy upgrade would not only allow exploration of densities up to around $2n_0$, but it would also provide increased resolution of the isospin-dependence of the EOS~\cite{FRIB400}. In particular, among observables sensitive to the symmetry energy~\cite{Xu:2019hqg,Colonna:2020euy}, both charged pion yields and the absolute magnitude of the elliptic flow (see Fig.\ \ref{fig:v1_v2_world_data}) significantly increase between the current top FRIB beam kinetic energy of 200 MeV/nucleon in the fixed-target frame and the proposed 400 MeV/nucleon~\cite{FRIB400}.

The increase in available computing power and advances in statistical methods make it possible to perform wide-ranging comparisons of heavy-ion collision simulations with experimental data (e.g., using Bayesian analysis), allowing one to vary multiple model assumptions at the same time as well as to put robust uncertainties on the obtained constraints.
Furthermore, given the wealth of the upcoming independent data, e.g., from heavy-ion collision experiments, neutron star observations, and microscopic nuclear theory calculations, global analyses of complementary efforts have likewise a strong potential for putting tight constraints on the EOS (see Section \ref{sec:combined_constraints}, ``The EOS from combined constraints'').

Beyond the much-needed interpretation of intermediate energy heavy-ion collisions, advances in transport theory can lead to significant contributions to other areas of nuclear physics. Recently, attention has been given to cross-cutting opportunities for employing state-of-the-art hadronic transport codes in studies supporting space exploration and advanced medical treatments (see Section \ref{sec:applications_of_hadron_transport}, ``Applications of hadronic transport''). Transport theories may also be used in tests of extensions of hydrodynamic approaches supporting far-from-equilibrium evolution (see Section~\ref{sec:hydrodynamics}, ``Hydrodynamics''), which are a focus of intense studies due to their importance for modeling heavy-ion collisions at high energies. Finally, constraining the dense nuclear matter EOS through interpretations of heavy-ion collision measurements may have other profound consequences, including helping to answer fundamental questions about the possible existence of dark matter in the cores of neutron stars or providing the impetus for studies of nuclear systems in fractional dimensions (see Section~\ref{sec:exploratory_directions}, ``Exploratory directions'').

\subsection{Scientific needs}

The next-generation experimental measurements of observables sensitive to the nuclear matter EOS are imminent. Capitalizing on the enormous worldwide scientific effort aimed at uncovering the dense nuclear matter EOS through heavy-ion collision experiments is contingent on enhanced theory support. In particular, the development of transport theories based on microscopic hadronic degrees of freedom, which are the only means of interpreting measurements from heavy-ion collisions at intermediate beam energies, must be strengthened and expanded. Support for individual scientists, and in particular creating opportunities for early career researchers, is imperative to maintain the health of and diversify the U.S.\ hadronic transport community. Collaborative research programs are needed to enable both a systematic advancement of hadronic transport models as well as explorations of new directions in microscopic descriptions of heavy-ion collisions.
With this support, innovative theory research will enable the exploration of the dense nuclear matter EOS and help fully realize the potential of the U.S.\ nuclear physics facilities.

\newpage
\section{The equation of state from $0$ to $5n_0$}
\label{sec:theory}

Efforts to determine the equation of state (EOS) of nuclear matter are at the forefront of nuclear physics. An EOS contains fundamental information about the properties of a many-body system (see, e.g., Section~\ref{sec:connections_to_fundamental_physics}), and is, in essence, any nontrivial relation between the thermodynamic properties of a given type of matter. In nuclear physics, the form of the EOS that is most often pursued is the relation between energy per baryon or pressure and baryon density~$n_B$, isospin asymmetry~$\delta$, and temperature~$T$. For symmetric matter, the isospin excess vanishes ($\delta=0$), and for asymmetric matter the energy per baryon or pressure are commonly partitioned into a part corresponding to symmetric matter and the remainder, which contains all information about the isospin-dependence of the EOS. Due to the charge invariance of strong interactions, the latter part is (to a good accuracy) quadratic in $\delta$ at densities around the nuclear saturation density $n_0$, relevant to nuclear experiments and astrophysical observations. The quadratic coefficients in the expansion around $\delta = 0$ are independent of $\delta$, and are often referred to as the symmetry energy (denoted as $S(n_B)$ at $T=0$) or symmetry pressure, respectively. These, together with the EOS of symmetric matter, are then sufficient to describe the EOS of nuclear matter at any isospin asymmetry. 

While many approaches to constraining the nuclear matter EOS are pursued, here we describe three research areas which have the capability to constrain the EOS over wide ranges of density: inferences of the EOS from comparisons of experimental measurements to model simulations of heavy-ion collisions (Section \ref{sec:model_simulations_of_HICs}), microscopic calculations of the EOS using chiral effective field theory (Section \ref{sec:microscopic_calculations_of_the_EOS}), and EOS inferences from neutron star studies (Section \ref{sec:neutron_star_theory}). To fully utilize the opportunity behind these complementary research directions, efforts must be made, both within these areas as well as across the different approaches, to describe the underlying physics consistently and to combine different data sets in a well-controlled way. This is expanded upon in Section~\ref{sec:combined_constraints}.

Given the wealth of data expected in the near future from heavy-ion collision experiments, nuclear structure studies, and multi-messenger astronomical observations, a concerted theoretical effort aimed at a consistent interpretation of precise measurements across varying thermodynamic conditions is needed. Some efforts of this nature are already underway. The Nuclear Physics from Multi-Messenger Mergers (NP3M) NSF Focused Research Hub \cite{NP3M_website} develops theoretical models and numerical simulations of dense and hot matter to connect multi-messenger observations of neutron stars to the underlying merger dynamics. The goals of the Network for Neutrinos, Nuclear Astrophysics, and Symmetries (N3AS) NSF Physics Frontier Center \cite{N3AS_website} include developing simulations of supernovae, mergers, kilonovae, and cooling neutron stars incorporating the most advanced treatments of the underlying neutrino and nuclear matter microphysics, enabling robust connections between observations and fundamental neutrino, dense matter, and dark matter properties. The Modular Unified Solver of the Equation of State (MUSES) NSF CSSI Framework \cite{MUSES_website} is developing a cyberinfrastructure with novel tools to answer critical interdisciplinary questions in nuclear physics, and will provide, e.g., modules for generating ensembles of EOSs relevant for heavy-ion collisions and neutron stars or neutron star mergers. Finally, the Beam Energy Scan Theory (BEST) DOE Topical Collaboration in Nuclear Theory \cite{BEST_website}, operating in the years 2016--2020, supported the development of hadronic transport models for the description of the final state of heavy-ion collision experiments performed within the Beam Energy Scan (BES) program at the Relativistic Heavy Ion Collider (RHIC) \cite{An:2021wof}.

At this time, results from the fixed-target (FXT) campaign of the BES are imminent, while future experiments at intermediate beam energies are being planned worldwide. In this regime, hadronic transport simulations are currently the only way to interpret these measurements and use them to understand the properties of QCD interactions at high baryon densities and finite temperatures. Consequently, there is an urgent need for a collaborative theoretical research program aimed at a further development of hadronic transport models as well as at explorations of new directions in microscopic descriptions of heavy-ion collisions. Possible research directions that could be addressed within such an effort are described in Section \ref{sec:transport_opportunities}.

\subsection{Transport model simulations of heavy-ion collisions}
\label{sec:model_simulations_of_HICs}

\begin{wrapfigure}{l}{0.45\textwidth}
  \centering
  \vspace{-5.75mm}
    \includegraphics[width=0.44\textwidth]{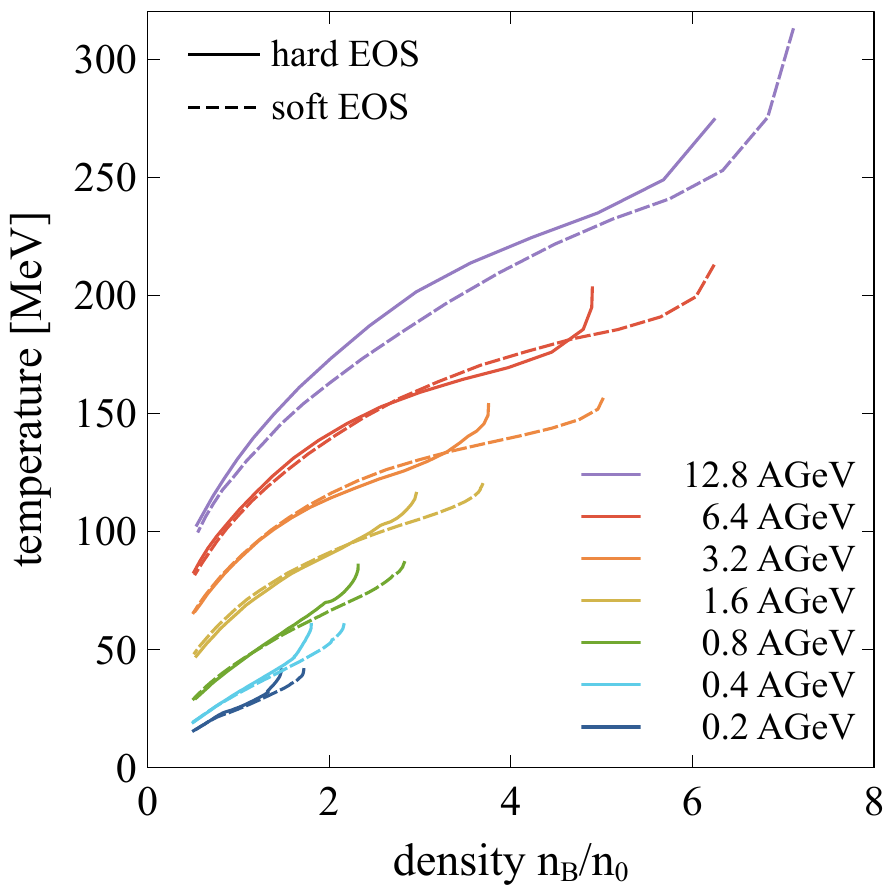}
    \vspace{-2.5mm}
  \caption{Phase diagram trajectories of the central region (cubic box of volume $27\ \rm{fm}^3$) in Au+Au collisions at zero impact parameter, obtained from \texttt{UrQMD} simulations with a soft or a hard (characterized by $K_0 = 200$ or $K_0 = 380\ \txt{MeV}$, respectively) EOS~\cite{Bass:1998ca,Bleicher:1999xi,Steinheimer:2022gqb}. The trajectories only follow the evolution at times when temperature is fairly well-defined, from the moment of the highest compression to densities around~$0.5n_0$.
  }
  \vspace{-4mm}
  \label{fig:ranges_UrQMD}
\end{wrapfigure}

Heavy-ion collisions at low to intermediate beam energies provide the means to probe nuclear matter at different baryon densities (from subsaturation to several times the saturation density), temperatures (from a few MeV to well above one hundred), and neutron to proton ratios (from nearly symmetric nuclear matter, where $N_n/N_p \approx 1$ and $\delta \approx 0$, to very neutron-rich matter, where $N_n/N_p \approx 2$ and $\delta \approx 0.25$). An illustrative calculation of the beam-energy-dependence of heavy-ion collision trajectories in the $T$-$n_B$ phase diagram, obtained from simulations using two schematic EOSs, can be seen in Fig.~\ref{fig:ranges_UrQMD} (note that the trajectories are only evaluated at times when temperature extraction is fairly well-defined). These wide ranges of system properties accessed in heavy-ion collisions position them as a perfect tool to extract the nuclear matter EOS, test predictions and extrapolations from regions of the QCD phase diagram accessed by other approaches, and provide a valuable input to nuclear theory and nuclear astrophysics calculations. For example, the density- and momentum-dependence of the nuclear potential in both symmetric and asymmetric matter, and thus of the corresponding EOS, can shed light on modeling effective nuclear interactions in the medium \cite{Gandolfi:2011xu,Hebeler:2013nza,Hagen:2013yba,1501.05675} or constrain approaches using the density functional theory~\cite{Erler:2012qd,Goriely:2013xba,Chen:2014sca}.

\begin{figure}[!b]
    \centering
    \includegraphics[width=0.99\linewidth]{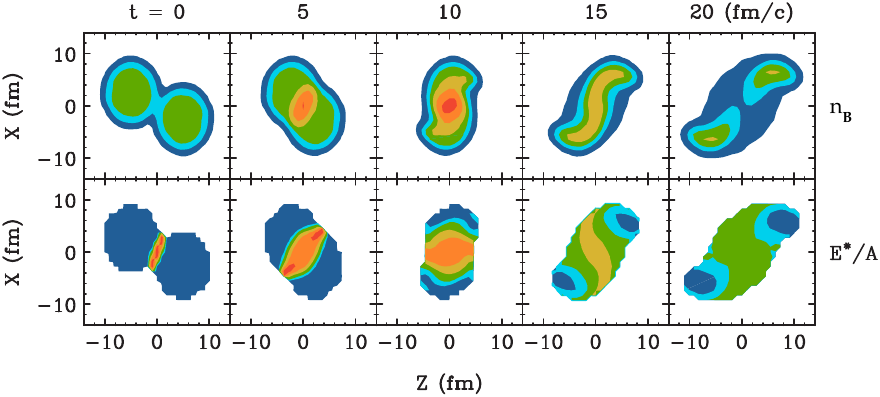}
	\vspace{-2mm}	
    \caption{Contour plots of the system-frame baryon density $n_B$ (\textit{top row}) and local excitation energy $E^{*}/A$
    (\textit{bottom row}) as a function of time (\textit{columns from left to right}), obtained from a transport simulation \cite{Shi:2001fm} of the $^{124}$Sn+$^{124}$Sn reaction at beam kinetic energy $\ekin = 800\ A\rm{MeV}$ ($\snn = 2.24\ \rm{GeV}$) and impact parameter $b$ = 5 fm. The contour lines for the density use increments of $0.4n_0$, starting from $0.1 n_0$, while the contour lines for the local excitation energy correspond to the values of $E^{*}/A = \{5, 20, 40, 80, 120\}\ \rm{MeV}$; the energy contour plots have been suppressed for baryon densities $n_B < 0.1 n_0$. Figure modified from \cite{Shi:2001fm}.
    }
    \label{fig:Collision}
\end{figure}

However, systems created in heavy-ion collisions are short-lived, and their dynamics is out of equilibrium over significant fractions of the total collision time. The evolution of a colliding system,
which strongly depends on both the energy 
and centrality of the collision, progresses through initial compression, growth of the compression zone, development of flows, and overall decompression with a gradual local equilibration throughout the process, see Fig.~\ref{fig:Collision}. The inherent complexity of the evolution means that the corresponding transport equations cannot be solved directly due to their high non-linearity, and therefore any inferences from heavy-ion collision experiments require comparisons to results of simulations. These are obtained using transport models which are able to describe the non-equilibrium evolution of nuclear matter over substantial ranges of density, as well as naturally include baryon, strangeness, and charge diffusion, and describe effects due to the interplay between the evolving collision zone and the spectators, which are crucial for a correct description of, e.g., flow observables. Beyond modeling the dynamics of the collisions, the dependence of the evolution on single-particle interactions provides a connection allowing one to use transport models for inferring equilibrium properties such as the EOS~\cite{Danielewicz:2002pu,Oliinychenko:2022uvy}, transport coefficients~\cite{Barker:2016hqv}, as well as the in-medium properties and cross-sections of hadrons \cite{GulminelliTrautmannYennello,Xu:2019hqg,ColonnaCollisionDynamics}.

\subsubsection{Transport theory}

At its core, transport theory aims to describe the time evolution a dissipative system composed of a large number of particles (here, a system of two heavy nuclei colliding at an energy per nucleon which is typically larger than the Fermi energy) in terms of the one-body phase-space distribution function in a semi-classical approximation. 
The theoretical foundations of transport theory include the BBGKY hierarchy of coupled equations for reduced density matrices \cite{boercker_degenerate_1979} as well as the equations for nonequilibrium single-particle Green's functions, known as the Kadanoff--Baym equations \cite{KadanoffBaym,Danielewicz:1982kk,Botermans:1990qi}. In turn, the nonequilibrium Green's function theory was built on the equilibrium theory developed by Martin and Schwinger \cite{Martin:1959jp}, retaining some of its features, hence in this context the approach is often also called the Martin--Schwinger (or Schwinger--Keldysh) formalism (see also Section~\ref{sec:hydrodynamics}).

To arrive at transport equations, one employs (among others) a Wigner transformation and coarse-graining as well as a gradient expansion. The Wigner transformation and coarse-graining lead to positive-definite phase-space distributions~\cite{Hillery:1983ms} that can be efficiently sampled with Monte-Carlo techniques, while the gradient expansion yields, for each particle species, the force acting on a particle and the particle's velocity as gradients of its total energy with respect to the spatial position and momentum, respectively. Knowledge of the kinematics of all particles, together with the elementary collision rates, drives the evolution in the phase space. Finally, to arrive at a set of Vlasov--Boltzmann-like equations, one employs the quasi-particle approximation, neglecting details of the spectral functions and treating all particles as on-shell (we note here that while there are some transport codes with off-shell particle treatment, e.g., \cite{Effenberger:1999uv,Cassing:1999mh,Moreau:2019vhw}, this approach is still an outstanding challenge in the transport theory, as will be discussed further below). Alternative approaches to arriving at a transport theory for heavy-ion collisions include using the relativistic Landau quasiparticle theory~\cite{Baym:1975va} or, in approaches starting from a molecular picture, representing the global wavefunction as a product (sometimes antisymmetrized) of single-particle Gaussian wavepackets \cite{Ono:1992uy}.

\begin{wrapfigure}{l}{0.47\textwidth}
  \centering
  \vspace{-1.5mm}
    \includegraphics[width=0.47\textwidth]{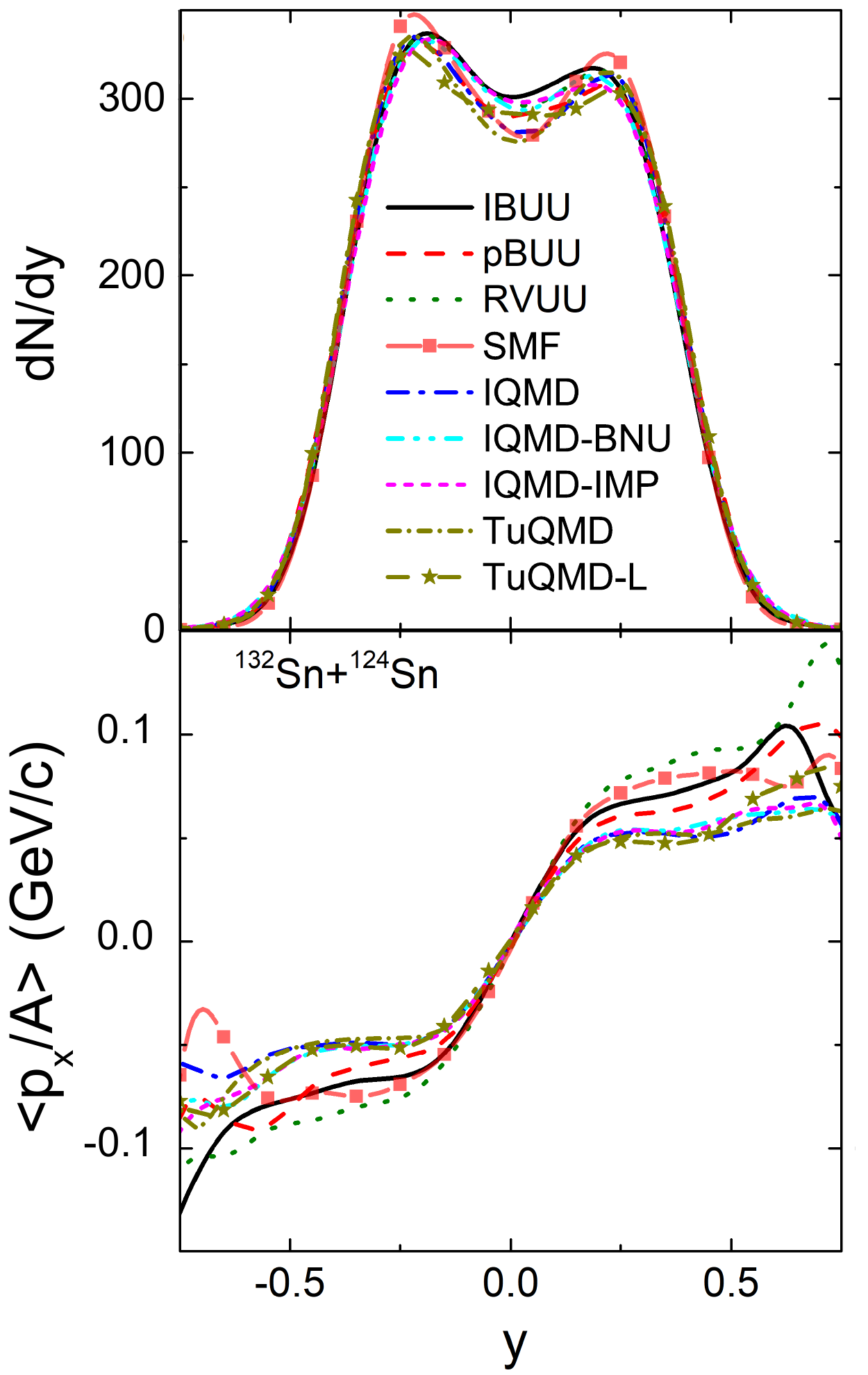}
    \vspace{-9mm}
  \caption{Comparison of rapidity distributions (\textit{top}) and transverse flow of nucleons (\textit{bottom}) as functions of the scaled rapidity, obtained with different transport codes (identified in the legend) within the TMEP initiative. The results shown were obtained for $^{132}$Sn+$^{124}$Sn collisions at beam kinetic energy $\ekin = 270\ A\rm{MeV}$ ($\snn = 2.01\ \rm{GeV}$) and impact parameter $b=4\,\text{fm}$, using controlled input models for the EOS and the cross sections as well as identically initialized nuclei. Figure from~\cite{Xu:2023iyy}.
  }
  \vspace{-4mm}
  \label{fig:FlowComparison}
\end{wrapfigure}
The particle species considered in transport theory depend on the collision energy and may range from nucleons, through pions and the delta resonances, to higher resonances, kaons, and hyperons. Some transport formulations further incorporate light clusters (e.g., deuterons, tritons, and $^{3}$He nuclei) as independent degrees of freedom, with recent extensions also including alpha particles~\cite{Ono:2019jxm} which appear abundantly in experiments and are of particular importance for collisions at fixed-target beam energies on the order of hundreds of MeV/nucleon. In some of these approaches, clusters are produced through multi-particle reactions, as discussed further below. For the lowest energy collisions, nonrelativistic formulations of the transport theory may be employed, but the majority of the available codes are relativistic, with many addressing collisions at energies from tens of MeV/nucleon to at least a few GeV/nucleon (see \cite{TMEP:2022xjg,Reichert:2021ljd,ColonnaCollisionDynamics} for reviews).

Transport approaches can be generally divided into those concentrating on a single-particle characterization of the colliding system and those attempting to describe many-particle correlations. Both types of approaches are highly complex and nonlinear, and the relevant equations are solved by simulations.
The single-particle approaches typically solve a set of Boltzmann--Vlasov-type equations~\cite{TMEP:2022xjg,Friman:2011zz} (also known as the Boltzmann--Uehling--Uhlenbeck, or BUU equations) in which the evolution of the system is governed by a mean-field evolution of the phase space distribution (Vlasov equations) and a collision term which drives the dissipation (the Boltzmann collision term). While, in principle, the Boltzmann--Vlasov equation is deterministic, numerical solutions contain numerically-induced fluctuations due to the fact that the evolution is obtained using the method of test particles, in which the continuous distribution function is represented by a large, but finite, number of test particles sampling the phase space. To include fluctuations of a physical origin, one can add a fluctuation term to the two-particle collision term, thus arriving at the Boltzmann--Langevin formulation \cite{ColonnaCollisionDynamics,Lin:2018wjj}.

In contrast, quantum molecular dynamics (QMD) approaches include classical many-body correlations in the ansatz of the many-body wave function \cite{TMEP:2022xjg,Li:2018bus}, which is postulated as a product of single-particle wave packets of a fixed width, with the width regulating the amount of fluctuations and correlations in QMD \cite{TMEP:2021ljz} (this width is usually fixed to reproduce realistic properties of nuclei). In Anti-Symmetrized Molecular Dynamics (\texttt{AMD}) \cite{Ono:1992uy}, the product wave function is anti-symmetrized and the formulation includes Pauli correlations in the propagation as well as in, to a certain extent, the collision term.

The fact that hadronic transport approaches are built on firm theoretical foundations has been crucial for the continued development of simulation frameworks. Reaching back to the roots of the nuclear transport theory has made it possible to resolve ambiguities which would be otherwise hard to tackle by purely phenomenological means, including descriptions of cluster production \cite{Staudenmaier:2021lrg}, low relative-velocity correlations (Hanbury--Brown-Twiss correlations) \cite{Danielewicz:1992pei}, and off-shell transport \cite{Effenberger:1999uv,Cassing:1999wx,Cassing:2000ch,Friman:2011zz,Buss:2011mx}. The strong theoretical foundation of transport theory has also been effective in ensuring covariance of the theory and preserving conservation laws in case of interactions that stray beyond outcomes of field-theoretic models, in particular interactions employing energy density functionals \cite{Baym:1975va,Danielewicz:1999zn,Sorensen:2020ygf,Oliinychenko:2022uvy} which are often needed for realistic descriptions of bulk properties of nuclear matter. The many-body theoretical origin of transport equations provides a connection between the mean fields, entering the drift terms, and the collision integrals, both of which originate from consistent self-energies in the many-body theory, and also provides expressions for spectral functions describing the widths of all particles  \cite{Botermans:1990qi,Danielewicz:1982kk,Danielewicz:1982ca}. These theoretical foundations both serve as a basis for currently used frameworks as well as offer a means for future improvements and expansions.

An important effort to validate conclusions reached from comparing transport model results to data has been recently intensified by the formation of the Transport Model Evaluation Project (TMEP) \cite{TMEP:2022xjg}. Within this endeavor, predictions from different models are compared in controlled settings (e.g., ensuring the same physical input such as the EOS, initial densities, and cross sections), oftentimes with comparisons to known results that can be achieved analytically or by other methods. Similar controlled comparisons of complex simulations have been done in other fields of physics: from atomic traps, through ultra-relativistic heavy-ion collisions, to core-collapse supernova calculations \cite{Xu:2004mz,Lepers:2010be,OConnor:2018sti,Just:2018djz}, and they are known to be very fruitful for their respective fields. The TMEP analyses not only enable identifying models that produce outlier predictions, but also determine details of implementation or physical assumptions behind the diverging results. 
An example of such a comparison of codes for simulations of heavy-ion collisions at lower energies, with controlled input, can be seen in Fig.\ \ref{fig:FlowComparison}, showing results for rapidity distributions (left) and the transverse flow (right) \cite{Xu:2023iyy}. In general, the codes agree with each other reasonably well, however, differences between the codes are visible and, moreover, can be traced to specific model choices in the simulations. For example, the generally lower values of the transverse flow in the case of QMD codes are a result of an approximation used in the evaluation of a non-linear term in the mean-fields, which becomes relevant when density fluctuations become large, as often occurs in QMD \cite{Yang:2021gwa,Colonna:2021xuh}. 
Beyond identifying this and similar problems, the Project has yielded recommendations for optimal algorithms used in transport codes, e.g., for ensuring obeying the Pauli principle in elementary two-body collisions \cite{Zhang:2017esm} or for integration of equations of motion with mean-fields \cite{Colonna:2021xuh}. Moreover, the Project has identified a set of tests for transport codes that ensure their credibility when addressing different heavy-ion collision observables. Stringent tests of hadronic transport codes are especially important for studies aimed at constraining the nuclear symmetry energy, which, compared to other model parameters, has a comparatively weak effect on heavy-ion observables and which therefore demands maximal precision from transport simulations. Below, we will also discuss the role that such comparisons can play in determining the uncertainty of transport model investigations.

\subsubsection{Selected constraints on the EOS obtained from heavy-ion collisions}
\label{sec:selected_constraints}

\begin{figure*}[!t]
\includegraphics[width=0.49\linewidth]{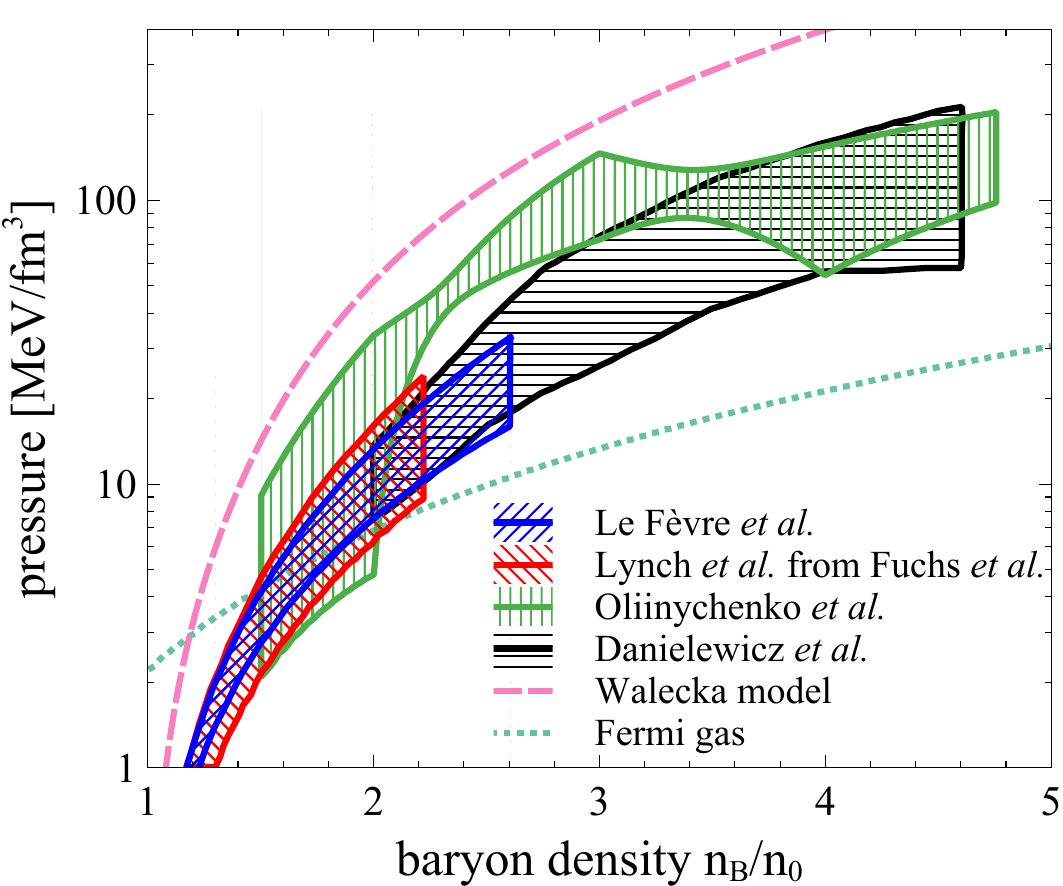}
\includegraphics[width=0.49\linewidth]{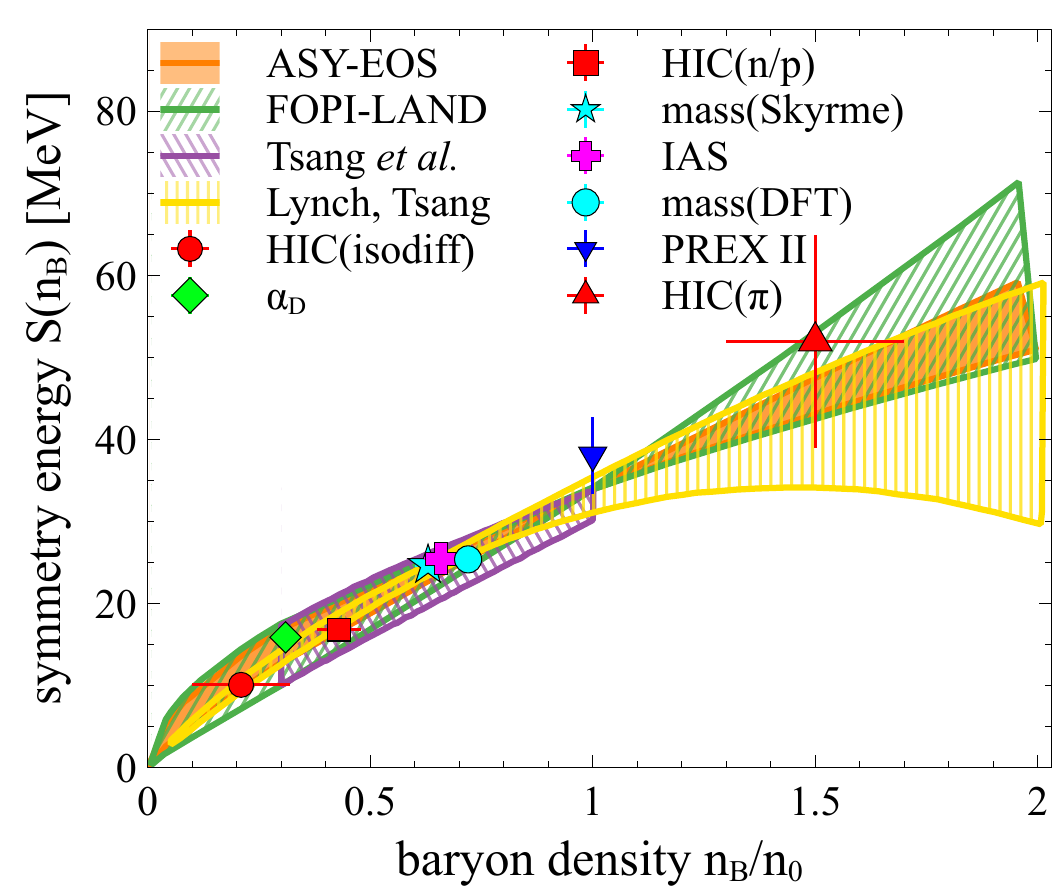}
\caption{\textit{Left:} Selected constraints on the symmetric EOS obtained from comparisons of experimental data to hadronic transport simulations in \cite{Danielewicz:2002pu} (region with black horizontal stripes), \cite{Fuchs:2003pc, Lynch:2009vc} (region with red forward stripes), \cite{LeFevre:2015paj} (region with blue backward stripes), and \cite{Oliinychenko:2022uvy} (region with green vertical stripes); see text for more details. Also shown are results of analytical calculations for the free Fermi gas (green dotted line) and in the linear Walecka model (pink dashed line).
\textit{Right:} Selected constraints on the symmetry energy obtained from comparisons of hadronic transport simulations to experimental data in \cite{Tsang:2008fd} (region with purple forward stripes), \cite{Russotto:2011hq} (region with green backward stripes), \cite{Russotto:2016ucm} (the solid orange region), and~\cite{Lynch:2021xkq} (the red circle, square, and triangle symbols). Also shown are symmetry energy constraints obtained in~\cite{Lynch:2021xkq} based on a novel interpretation of analyses of dipole polarizability $\alpha_D$ \cite{Zhang:2015ava} (green diamond), of nuclear masses in DFTs~\cite{Kortelainen:2010hv,Kortelainen:2011ft} (cyan dot symbol) and in Skyrme models \cite{Brown:2013mga} (cyan star symbol), of Isobaric Analog States (IAS) energies \cite{Danielewicz:2013upa} (magenta plus symbol), and of PREX-II experiment \cite{PREX:2021umo} (blue inverted triangle symbol), as well as the $68\%$ confidence region consistent with the best fit of experimental data points presented in~\cite{Lynch:2021xkq} (region with yellow vertical stripes).
} 
\label{fig:joint_constraints_EOS}
\end{figure*}

A selection of important constraints on the EOS obtained from heavy-ion collisions can be found in Fig.\ \ref{fig:joint_constraints_EOS} for both symmetric matter (pressure as a function of density, left panel) and asymmetric matter (symmetry energy as a function of density, right panel). Additional constraints and measurements are also discussed in Sections \ref{snm-low}, \ref{snm-high}, \ref{selected_constraints_experiment}, and~\ref{sec:Constraints}. All quoted beam energies $\ekin$ are the single-beam kinetic energies per nucleon, in units of $A\rm{MeV}$ or $A\rm{GeV}$.

We note here that while many results are reported in terms of constraints on the incompressibility $K_0$, in the context of heavy-ion collision studies of the EOS, $K_0$ should be understood as a parameter which specifies the behavior of the EOS in the range of densities probed by a given study. For example, in the case of experiments probing mostly densities above $2n_0$, constraints on $K_0$ are only indicative of the behavior of the EOS above $2n_0$, and in particular \textit{do not} constrain the behavior of the EOS around $n_0$. This subtle, and often confusing, point is a consequence of simple parametrizations of the EOS used in many transport codes, where the only parameter controlling the behavior of the EOS \textit{both} around $n_0$ and at higher densities is $K_0$. Recently, flexible parametrizations of the EOS have been developed (see, e.g., \cite{Sorensen:2020ygf,Oliinychenko:2022uvy}) and implemented (e.g., in hadronic transport code \texttt{SMASH} \cite{Weil:2016zrk,Sorensen:2021zxd}) which allow one to vary the incompressibility $K_0$ and the high-density behavior of the EOS independently, in turn enabling description of non-trivial features at high density such as, e.g., a phase transition.

The collective behavior of matter created in the collisions, especially the directed and elliptic flow, has been shown to be a very sensitive probe of the EOS \cite{Stoecker:1980vf,Hartnack:1994ce,Danielewicz:2002pu,LeFevre:2015paj,Nara:2021fuu}. In contrast to collisions at the Fermi energies, where all nucleons within nuclei participate in the collisions, and unlike in collisions at ultrarelativistic energies, where the evolution of the colliding nuclei can be understood in terms of participant nucleons, at intermediate energies the interplay between the expanding collision zone and the dynamics of the spectators are key ingredients to understanding experimental results. A seminal constraint on the symmetric nuclear matter EOS \cite{Danielewicz:2002pu} in the density range $(2$--$4.5) n_0$ was obtained by comparing measurements of collective flow from heavy-ion collisions \cite{Gustafsson:1988cr,EOS:1994kku,E877:1997zjw,E895:2000maf} at beam energies $\ekin = 0.15$--$10\ A\rm{GeV}$ (corresponding to nucleon-nucleon center-of-mass energies $\snn = 1.95$--$4.72\ \rm{GeV}$) with results from hadronic transport simulations using EOSs with different values of the incompressibility at saturation density $K_0$. The outcome of this study suggests a symmetric-matter EOS to lie between those labeled with $K_0 =210 \ \text{MeV}$ and $K_0 = 300 \ \text{MeV}$ (see the region with black horizontal stripes in the left panel of Fig.\ \ref{fig:joint_constraints_EOS}). For densities in the range  $(1.0$--$2.5) \, n_{0}$, probed in collisions below $\ekin \lesssim 1.5\ A\rm{GeV}$ ($\snn \lesssim 2.5\ \rm{GeV}$), the EOS may be inferred from meson yields \cite{Fuchs:2000kp, Liu:2020jbg, Cozma:2021tfu}. Indeed, subthreshold production of strange mesons (specifically, $K^+$ and $K^0$), which interact weakly with nuclear matter, depends on the highest densities sampled in the collision, which in turn depend on the stiffness of the EOS \cite{Aichelin:1985rbt}. In \cite{Fuchs:2003pc}, ratios of experimentally measured kaon yields in Au+Au and C+C collisions have been reproduced in hadronic transport simulations with soft mean-field interactions yielding $K_0 = 200\ \rm{MeV}$ and an EOS \cite{Lynch:2009vc} consistent with the constraint from \cite{Danielewicz:2002pu} (see the region with red forward stripes in the left panel of Fig.\ \ref{fig:joint_constraints_EOS}). In~\cite{LeFevre:2015paj}, the elliptic flow data measured at $\ekin = 0.4$--$1.5\ A\rm{GeV}$ ($\snn = 2.07$--$2.52\ \rm{GeV}$) by the FOPI collaboration \cite{FOPI:2011aa} were used together with simulations from Isospin Quantum Molecular Dynamics (\texttt{IQMD}) \cite{Aichelin:1991xy,Hartnack:1997ez} to constrain the incompressibility at $K_0 = 190 \pm 30 \txt{MeV}$, again indicating a rather soft EOS (see the region with blue backward stripes in the left panel of Fig.\ \ref{fig:joint_constraints_EOS}). Recently, new measurements by the STAR collaboration from the fixed target (FXT) program at RHIC have become available, providing an opportunity to expand the set of world data utilized to deduce the baryonic EOS. A Bayesian analysis study~\cite{Oliinychenko:2022uvy}, in which the speed of sound was independently varied in specified intervals of baryon density (thus providing a more flexible EOS at higher densities), suggests a tension between the E895 \cite{E895:1999ldn,E895:2000maf,E895:2001axb,E895:2001zms} and STAR \cite{STAR:2020dav,STAR:2021yiu} data. Using only the STAR measurements, the study \cite{Oliinychenko:2022uvy} further found that EOSs which simultaneously describe the slope of the directed flow and the elliptic flow, in the considered energy range of $\ekin = 2.9$--$8.9\ A\rm{GeV}$ ($\snn = 3.0$--$4.5\ \rm{GeV}$), are relatively stiff at lower densities and relatively soft at higher densities (see the region with green vertical stripes in the left panel of Fig.\ \ref{fig:joint_constraints_EOS}). However, the model used in that work did not include the momentum dependence of the EOS, which has been to shown to result in a spuriously stiff EOS at intermediate densities. As such, the study should be treated as a proof of principle that a tight constraint on the EOS at high densities can be achieved by using a combination of precise data, flexible forms of the EOS used in simulations, state-of-the-art models, and advances in analysis techniques.

The symmetry energy contribution to the EOS can be studied at low collision energies $\ekin \lesssim 1.0\ A\rm{GeV}$ ($\snn \lesssim 2.32\ \txt{GeV}$), where in particular observables such as charged pion yields \cite{Li:2002qx} or neutron and proton flow \cite{Li:2000bj,Li:2014oda} have been proposed as sensitive to the asymmetric contribution to the EOS. Some of the constraints derived from such studies are shown in the right panel of Fig.~\ref{fig:joint_constraints_EOS}, where, in addition to the usual EOS constraint bands, symbols with uncertainty bars represent results from analyses in which the symmetry energy has been determined for the most sensitive density of a given measurement. At incident energies below $\ekin = 100\ A\rm{MeV}$ ($\snn = 1.93\ \rm{GeV}$), low densities are probed after the initial impact and compression of the projectile and target~\cite{Tsang:2008fd,Morfouace:2019jky}. Since the symmetry potentials for neutrons and protons have opposite signs, emission of a particular nucleon type is enhanced or suppressed depending on the asymmetry. A~comparison of the experimental measurements of isospin diffusion and the ratio of neutron and proton spectra in collisions of $^{112}$Sn+$^{124}$Sn at $\ekin = 50\ A\rm{MeV}$ ($\snn = 1.90\ \rm{GeV}$) to results from \texttt{ImQMD} \cite{Zhang:2006vb} simulations produced a constraint on the symmetry energy for densities $(0.3$--$1)n_{0}$~\cite{Tsang:2008fd} (see the region with purple forward stripes in the right panel of Fig.\ \ref{fig:joint_constraints_EOS}). Collisions at higher energies ($\ekin > 200\ A\rm{MeV}$, or $\snn > 1.97\ \rm{GeV}$) probe the EOS at $n> n_0$. In the FOPI-LAND experiment, constraints on the symmetry energy were obtained from studies of the ratio of the elliptic flow of neutrons and hydrogen nuclei in Au+Au collisions at $\ekin = 0.4 \ A\rm{GeV}$ ($\snn = 2.07 \rm{GeV}$)~\cite{Russotto:2011hq}, while the ASY-EOS experiment used neutron to charged fragments ratios measured in Au+Au collisions \cite{Russotto:2016ucm} (see the region with green backward stripes and the solid orange region, respectively, in the right panel of Fig.\ \ref{fig:joint_constraints_EOS}). In~\cite{Lynch:2021xkq}, a comprehensive analysis was performed with the goal of identifying the values of the symmetry energy at densities to which given experiments are most sensitive. Using the isospin diffusion in collision systems with different proton to neutron ratios \cite{Zhang:2007hmv}, neutron to proton energy spectra in Sn+Sn systems \cite{Zhang:2014sva}, and spectral pion ratios measured by the S$\pi$RIT collaboration in Sn+Sn collisions at $\ekin = 270\ A\rm{MeV}$ ($\snn = 2.01\ \rm{GeV}$)~\cite{SRIT:2021gcy,SpRIT:2020blg}, that work \cite{Lynch:2021xkq} put constraints on the values of the symmetry energy at about $0.2n_0$, $0.4n_0$, and $1.5n_0$, respectively (see the red circle, square, and triangle symbols in the right panel of Fig.~\ref{fig:joint_constraints_EOS}). Also shown in the right panel of Fig.~\ref{fig:joint_constraints_EOS} are symmetry energy constraints obtained in~\cite{Lynch:2021xkq} based on a novel interpretation of the analyses of dipole polarizability $\alpha_D$ \cite{Zhang:2015ava} (green diamond symbol), of nuclear masses in DFTs \cite{Kortelainen:2010hv,Kortelainen:2011ft} (cyan dot symbol) and in Skyrme models~\cite{Brown:2013mga} (cyan star symbol), of the Isobaric Analog State (IAS) energies \cite{Danielewicz:2013upa} (magenta plus symbol), and of the PREX-II experiment result \cite{PREX:2021umo} (blue inverted triangle symbol), as well as the $68\%$ confidence region consistent with the best fit of experimental data points presented in~\cite{Lynch:2021xkq} (region with yellow vertical stripes), where identifying the specific densities at which the measurements constrain the symmetry energy allowed for a comparatively tight constraint at low densities.

\subsubsection{Challenges and opportunities}
\label{sec:transport_opportunities}

Selected results presented in Fig.\ \ref{fig:joint_constraints_EOS} showcase significant achievements in determining the EOS
and, simultaneously, the need to develop improved transport models to obtain tighter and more reliable constraints. Answering this need will require support for a sustained collaborative effort within the community to address remaining challenges in modeling collisions, in particular in the intermediate energy range ($\ekin \approx 0.05$--$25\ A\rm{GeV}$, or $\sqrt{s_{NN}}\approx 1.9$--$7.1\ \rm{GeV}$). In the following, we will address selected areas where we see the need for such developments: (1) comprehensive treatment of both mean-field potentials and the collision term in transport codes, (2) use of microscopic information on mean fields and in-medium cross sections, such as discussed in Section~\ref{sec:microscopic_calculations_of_the_EOS}, in transport, (3) better description of the initial state of heavy-ion collisions in hadronic transport codes, (4) deeper understanding of fluctuations in transport approaches, which affect many aspects of simulations, (5) inclusion of correlations beyond the mean field into transport, which is crucial for a realistic description of, e.g., light-cluster production, (6) treatment of short-range-correlations in transport, which are tightly connected to multi-particle collisions as well as off-shell transport, (7) sub-threshold particle production, (8) connections between quantum many-body theory and semiclassical transport theory, (9) investigations focused on extending the limits of applicability of hadronic transport approaches, (10) studies of new observables, e.g., azimuthally resolved spectra, to obtain tighter constraints on the EOS, (11) the question of quantifying the uncertainty of results obtained in transport simulations, and (12) the use of emulators and flexible parametrizations for wide-ranging explorations of all possible EOSs. Fortunately, advances in transport theory as well as the greater availability of high-performance computing make many of these improvements possible. Support for these developments will lead to a firm control and greater understanding of multiple complex aspects of the collision dynamics, allowing comparisons of transport model calculations and heavy-ion experiment measurements to provide an important contribution to the determination of the EOS of dense nuclear matter, which, in particular, cannot be determined by any other method at intermediate densities (1--5)$n_0$.

\vspace{5mm}
{\textbf{Comprehensive treatment of mean-field potentials and the collision term}}

Over the last two decades, driven by specific experimental needs, the refinement of hadronic transport codes has diverged into two complementary branches: Codes which were applied to describing experiments at very low to low energies ($\ekin \lesssim 1.5 \ A\text{GeV}$, or $\snn \lesssim 2.5 \ \text{GeV}$), such as \texttt{IQMD}~\cite{Stoecker:1986ci,Hartnack:1997ez}, \texttt{AMD}~\cite{Ono:1992uy,Ikeno:2016xpr} and \texttt{pBUU}~\cite{Danielewicz:1991dh,Danielewicz:1999zn}, have become progressively better at describing the momentum- and isospin-dependence of the interaction, while codes which were primarily designed for simulations of relativistic and ultra-relativistic heavy-ion collisions ($\ekin \gtrsim 25 \ A\text{GeV}$, or $\snn \gtrsim 7 \ \text{GeV}$), such as \texttt{SMASH}~\cite{Weil:2016zrk} or \texttt{UrQMD}~\cite{Bass:1998ca,Bleicher:1999xi}, were developed to offer a fully relativistic evolution as well as scattering and decay modes taking into account a large number of established particle and resonance species. As heavy-ion collisions are entering an era of precision data on symmetric nuclear matter at high densities (e.g., in experiments at HADES, BES FXT, and future CBM, probing densities up to several times the saturation density) and on asymmetric nuclear matter at normal and supranormal densities (e.g., at FRIB and future FRIB400, probing densities up to twice the saturation density), where features of both diverging branches of hadronic transport codes are important, a vigorous development of transport models is needed to incorporate all relevant physics. 

For example, numerous studies show the importance of including the momentum-dependence of the interactions, which is observed in elastic scattering of hadrons off nuclei. Moreover, momentum-dependence naturally occurs in microscopic effective interactions~\cite{Botermans:1990qi,Sammarruca:2021bpn} where it contributes to the calculated mean fields, whether near or away from saturation density. Incorporating single-particle energies with momentum dependence different than that in free space, which is often quantified with effective masses, is crucial in hadronic transport both for studies of symmetric nuclear matter~\cite{Gale:1987zz, Aichelin:1987ti, Danielewicz:2002pu, Nara:2021fuu} as well as studies of the symmetry energy and its relation to effects such as the neutron-proton effective mass splitting \cite{Das:2002fr,Li:2003zg,Li:2018lpy} (see also Section~\ref{sec:density-dependence_effective_masses} for more discussion on effective masses and the nuclear symmetry energy). 

As another example, inclusion of high-mass resonant degrees of freedom, which can be produced in substantial amounts at collision energies $\ekin \gtrsim 1 \ A\text{GeV}$ ($\snn \gtrsim 2.32 \ \text{GeV}$), can significantly affect the evolution of the system. In particular, since some part of the collision energy can be used for resonance production (so that a non-negligible fraction of the nuclear matter becomes ``resonance matter''), it may affect both the degree of initial compression and the spectrum of mesons emitted and absorbed throughout the evolution, which has consequences for the collective flow \cite{Hombach:1998wr} and thus directly influences extraction of the EOS. Therefore, inclusion of all relevant resonant species and ensuring correct description of meson production and absorption is crucial for a reliable inference of the EOS.

While some of the theoretical and implementation solutions needed to improve the models in the intermediate energy region have already been established, others will require devising new approaches. When possible, best practices need to be carried over across the domains, as has been exemplified in, e.g., the development of the \texttt{SMASH} code, which combines multiple well-tested implementation solutions from \texttt{UrQMD}, \texttt{GiBUU} \cite{Buss:2011mx}, and \texttt{pBUU}.

\vspace{5mm}
{\textbf{Microscopic input to transport}}

One of the most prominent opportunities for improvement in transport models concerns implementations of the EOS informed by state-of-the-art many-body studies. Such efforts are especially timely given that sophisticated microscopic calculations of the properties of nuclear matter are
currently becoming available for large ranges of baryon density, temperature, and isospin fraction (see Section \ref{sec:microscopic_calculations_of_the_EOS} for more details). To incorporate the effects of the resulting EOSs in hadronic transport calculations, the corresponding Lorentz-covariant single-particle potentials as well as the in-medium interactions (both as functions of density, asymmetry, and momentum) are needed. A~particular challenge is to determine the connection between the EOS inferred from a transport calculation and the zero-temperature EOS obtained from microscopic calculations \cite{Xu:2019ouo}, or even the finite-temperature EOSs that are becoming increasingly available  \cite{Wellenhofer:2014hya,Keller:2022crb}. In a heavy-ion collision, the medium progresses through a set of non-equilibrium states that relax toward a local equilibrium, however, the nature of the local equilibrium also evolves during the collision due to the system expansion, so that even if the system approaches a local equilibrium at any given moment of the evolution, that agreement is only temporary. Errors incurred due to differences between non-equilibrium and equilibrium states of high-density matter contribute to the systematic error in inferring the EOS when comparing transport to experimental data (see Fig.~\ref{fig:joint_constraints_EOS} and~\cite{Danielewicz:2002pu}). Here, the availability of state-of-the-art microscopic calculations at finite temperature could reduce systematic errors in connecting the finite- and zero-temperature EOSs.   
Moreover, the use of microscopic input would provide a consistency between the effective in-medium cross sections in the collision term and the mean fields used in the propagation of the phase space distribution. It could also help address the question of the extent to which nonlocalities in the microscopic theory should be reflected in the propagation and the collision term \cite{Danielewicz:1995ay,Morawetz:1999zz} (where, in particular, departures from standard approaches modify the entropy to take a form different than that obtained in the Landau quasiparticle theory \cite{Baym:1975va,pines_theory_1966}). Microscopic calculations could also inform the treatment of strange degrees of freedom in hadronic transport through a better understanding of nucleon-hyperon and hyperon-hyperon interactions, which can be pursued within, among others, lattice QCD approaches (see, e.g., \cite{HALQCD:2019wsz}). 

To accelerate progress at the interface of the transport description of heavy-ion collisions and microscopic nuclear matter theory, direct collaboration of practitioners in the two research areas is required to assess how the needs of transport simulations can be answered by what can be currently calculated in microscopic theories. Conversely, the use of microscopic interactions in transport could validate the many-body theory results in regions of density and temperature which are only accessible by heavy-ion collisions \cite{Zhang:2018ool}.

\vspace{5mm}
{\textbf{Initial state}}

Numerous studies point toward the dependence of outcomes of heavy-ion collision experiments on details of the initial conditions. In ultrarelativistic heavy-ion collisions, understanding these effects have led to the discovery of higher order flow harmonics \cite{Mishra:2007tw,Sorensen:2010zq} and flow fluctuations~\cite{Sorensen:2009cz}. (Interestingly, the importance of the initial state for experimental outcomes also positions heavy-ion collisions at high energies as an unusual, but complementary probe of nuclear structure, see, e.g., a white paper on \textit{Imaging the initial condition of heavy-ion collisions and nuclear structure across the nuclide chart} \cite{Bally:2022vgo}.) Given the high sensitivity of flow observables to both the EOS and the initial state of collisions, the impact of the initial conditions on outcomes of heavy-ion collisions needs to be thoroughly understood in order to narrow the constraints on the EOS of both symmetric and asymmetric matter. Aspects of initial conditions that need to be considered include event-by-event fluctuations of the initial state \cite{Mishra:2007tw,Sorensen:2010zq,Sorensen:2009cz}, relative distributions of neutrons and protons and shell effects \cite{Stone:2017jpa}, and correlations tied to deformation \cite{Cao:2010bc} or short-range correlations~\cite{Danielewicz:1990cd}. Some of these elements will be further discussed below in the context of the dynamics of heavy-ion collisions.

\vspace{5mm}
{\textbf{Fluctuations}}

Fluctuations of the phase space distribution are an important ingredient of transport simulations. In particular, fluctuations of the one-body density are important for including the consequences of the dissipation-fluctuation theorem in the reaction dynamics as well as for describing effects due to the largely unknown, neglected many-body correlations, thus going beyond the mean-field description. The question of how to include them properly and of their consistency with the nucleon-nucleon correlations explicitly implemented in transport theories, however, has not been completely clarified. As discussed above, fluctuations are included in a different manner in the two families of transport approaches. While in the BUU transport fluctuations can be introduced by the Langevin extension of the Boltzmann--Vlasov equation, which adds a fluctuation term to the collision term (and which is still rarely implemented), in the molecular dynamics approach fluctuations are introduced in a classical way by using finite-size particles, the width of which regulates the amount of fluctuations. Fluctuations then affect the outcome of simulations in many ways, including by regulating the formation of intermediate-mass fragments (IMFs) which appear through the growth of fluctuations in regions of spinodal instability. It was also shown in box calculations that fluctuations have a strong influence on the efficiency of Pauli-blocking \cite{Zhang:2017esm} and even on the calculation of the force in the Vlasov term 
for QMD codes in which non-linear parametrizations of the fields are used \cite{Colonna:2021xuh}.

\vspace{5mm}
{\textbf{Correlations}}

Correlations in transport simulations strive to address intermediate-range correlations beyond the mean-field picture, that is beyond fluctuations of the one-body density. Physically, such correlations also contribute to one-body density fluctuations, but at the same time they have other additional impacts, including, e.g., influencing the production of light clusters (LCs), that is light nuclei up to the alpha particle which are copiously produced in heavy-ion collisions. However, the mean-field models used in transport calculations are usually not detailed enough to realistically describe very light nuclei with their particular spin-isospin structure reflecting strong quantum effects. An additional complication results from the fact that in a collision, clusters often appear in the nuclear medium where their properties are drastically changed (e.g., the binding energy of clusters is reduced with increasing density until the Mott point, at which they dissolve). Currently, most codes describe the production of clusters by using a cluster-finding algorithm, based on particle proximity in coordinate and/or momentum space (coalescence) toward the end of the evolution, which in more advanced versions also takes into account criteria related to the binding energy of the produced clusters \cite{LeFevre:2019wuj}. However, these late-stage algorithms do not take into account the dynamic role played by both correlations and LCs in the evolution of the collision. One of the known approaches to this problem has been to consider LCs as separate degrees of freedom, with their own distribution functions and corresponding transport equations, where the collision terms can lead to creation or destruction of clusters (\texttt{pBUU}, \texttt{SMASH}) and which in particular can also take into account the in-medium modifications of clusters. However, this approach becomes increasingly complex as heavier clusters are characterized by more and more production channels, and consequently it is significantly challenging to include, e.g., alpha particles. Another approach is to modify the phase space of the correlated nucleons according to the Wigner function of the cluster, but then to propagate them after the collision again as nucleons, which still requires using a cluster-finding step at the end  (this is done in, e.g., the \texttt{AMD} code \cite{Hillery:1983ms}, within which it has also been demonstrated that the clustering effects may influence pion production~\cite{Ikeno:2016xpr}). In both approaches, the production and destruction of clusters necessarily requires multi-particle collisions to ensure energy-momentum conservation. Finally, at lower incident energies the LC production can also be described in terms of the catalyzing effect of spectator nucleons in few-particle collisions~\cite{Danielewicz:1991dh,Ono:2019jxm}. To explain LC production in high-energy collisions, where LCs are produced in numbers that cannot be obtained through nucleon catalysis due to the relatively few nucleons present in the final stages of these collisions, a similar mechanism of catalysis by pions \cite{Oliinychenko:2018ugs,Staudenmaier:2021lrg,Sun:2021dlz} can be invoked.

\vspace{5mm}
{\textbf{Short-range correlations}}

A particular aspect of describing correlations in transport simulations is the treatment of short-range-correlations (SRCs), which have been measured in nucleon knock-out experiments \cite{Arrington:2011xs,Hen:2014nza,Hen:2016kwk,Arrington:2022sov}. Along with the experiments, microscopic many-body calculations show that SRCs introduce a high-momentum tail (HMT) into the nucleon momentum distribution and, moreover, reduce the kinetic symmetry energy relative to the Fermi gas kinetic energy, which is a consequence of the fact that SRCs are more pronounced in symmetric relative to asymmetric matter \cite{Xu:2012hf,Vidana:2011ap,Carbone:2011wk,Carbone:2013cpa,Rios:2013zqa,Hen:2014yfa,Li:2014vua,Cai:2015xga} (see also Section~\ref{sec:SRC_effects_on_nuclear_EOS}). Phenomenological methods have been used to include SRCs in transport models, e.g., by initializing nuclei with a HMT, but such a procedure does not take into account the dynamic role of SRCs in the initial state, which in the case of the on-shell semiclassical equations of motion results in obtaining non-stationary, excited states of nuclei.
In on-shell transport approaches, three- and many-body collisions, incorporated into transport codes within varying approximations, have been suggested as a way of treating SRCs. In particular, in an investigation~\cite{Bertsch:1995ig} of three-body collisions for pion production processes (e.g., $NNN \rightarrow NN\Delta$), it was found that SRCs between two of the incident nucleons give a noticeable contribution to pion yields. Another approach~\cite{Bonasera:1992kjm}, based on a mean-free-path approximation to the collision integral, observed large effects also on bulk observables. The incorporation of $n$-body collisions in transport equations within a schematic cluster approximation was also studied \cite{Batko:1991xd}, however, the effects were found to be rather small. So far, none of these methods have been widely exploited in the description of heavy-ion reactions. Since HMTs are tied to the tails of the nucleon spectral functions (away from the quasiparticle peaks), a consistent description of SRCs should involve an off-shell transport formulation.

\vspace{5mm}
{\textbf{Threshold effects}}

An important influence of mean-field potentials in heavy-ion transport appears in the form of threshold shifts and the related subthreshold production of particles. Thresholds of particle production are modified in a medium since the mean-field potentials have to be taken into account in the energy-momentum balance of a two-body collision. Specifically, when the mean-field potentials are momentum-dependent and/or as a consequence of other model assumptions for the mean-field potentials of the produced particles, the thresholds are shifted away from their free-space values. This may strongly change the production rates of particles. Moreover, the threshold shifts make it necessary to involve other nucleons, besides the two collision partners in the process, to ensure the energy-momentum conservation. Various schemes to achieve this locally or globally have been in use \cite{Cozma:2014yna,Zhang:2018ool}. Indeed, explaining recent heavy-ion collision subthreshold pion yields, measured by the S$\pi$RIT Collaboration \cite{SpRIT:2020blg}, required invoking many-body elementary effects in the form of mean-field effects on thresholds in two-particle collisions  \cite{SRIT:2021gcy,Cozma:2021tfu}. However, because the physics invoked in describing the threshold effects is similar to that invoked for other multi-particle effects, alternative multi-particle options remain to be investigated, including producing pion degrees of freedom in multi-particle collisions or in the aftermath of an off-shell propagation between binary collisions. (We note here that there is a physics overlap between these mechanisms and the impact of SRCs on pion production \cite{Danielewicz:1990cd,Bertsch:1995ig,Effenberger:1999uv}.) Notably, theoretical explorations find sequences of on-shell binary processes to dominate the production at higher beam energies \cite{Bertsch:1995ig,Cassing:1999mh,Cassing:2000ch}, and no comparable difficulties have been encountered in describing the data \cite{FOPI:2006ifg,Hong:2013yva} by transport models without multi-particle effects. The contrasting struggles of transport models which do not include threshold or other multi-particle effects of this type \cite{SpRIT:2020blg}, together with expected further theoretical explorations and future measurements of the subthreshold production in heavy-ion collisions, offer exciting possibilities for gaining understanding of the more exotic in-medium processes.

\vspace{5mm}
{\textbf{Interface between quantum many-body theory and semiclassical transport theory}}

In deriving the transport theory in the nonequilibrium Green's functions formalism, one considers the evolution of the one-body Green functions up to 1-body level in terms of self-energies.  In the semiclassical limit, these self-energies yield both the potentials and the in-medium cross sections in the collision terms of the transport equations \cite{Botermans:1990qi,Danielewicz:1982kk,Danielewicz:1982ca,Buss:2011mx}. In the usual applications of the transport theory, these driving terms are modeled with phenomenological density functionals and effective in-medium cross sections, respectively. Within a more microscopic approach, the self-energies would be calculated in the $T$-matrix approximation similar to that employed in the Brueckner theory. The Hermitian part of the resulting self-energy, in a given intrinsically consistent approximation, would yield the potential for the drift term in transport. The anti-Hermitian part would yield widths of the spectral functions as well as contribute to the feeding and depletion in the transport collision integrals. By retaining additional terms in the expansion leading to transport, one may arrive at equations where the phase-space evolution is coupled to the evolution of the spectral functions. With some additional approximations, the spectral functions for particles (including those which are stable in free space such as, e.g., nucleons) have been implemented in \texttt{GiBUU} \cite{Buss:2011mx}, \texttt{PHSD} \cite{Linnyk:2015rco,Moreau:2019vhw}, and \texttt{PHQMD} \cite{Aichelin:2019tnk} codes. A study \cite{Lehr:2000ua}, performed within the approach developed in \texttt{GiBUU}, showed that the momentum distribution naturally develops a HMT in off-shell transport.

Further development of these aspects will become more important as data and theory become more precise. Their impact on inferences of the symmetry energy from heavy-ion collision data based on, e.g., charged pion subthreshold production yields, can be particularly consequential, but has yet to be systematically investigated. Fully quantum transport approaches with SRCs (or equivalent content), without any semi-classical expansions as are present in current off-shell transport approaches, remain a long-term goal, and progress in this area has not ventured yet beyond schematic models \cite{Arrizabalaga:2005tf, Lin:2019ngf}. However, increasing computational power combined with emulation techniques may make such efforts more realistic and enable, e.g., a seamless integration of the treatment of shell effects in the initial state and collision dynamics.

\vspace{5mm}
{\textbf{Limits of hadronic transport}}

In heavy-ion collisions at higher beam energies, increasing numbers of nucleon resonances and heavier mesons are produced, which at a certain point start to significantly affect the evolution of the system (this includes the emergence of ``resonance matter'', mentioned above). In some transport codes, e.g., \texttt{UrQMD} or \texttt{SMASH}, a large number of such states is included. Eventually, however, as one approaches energies at which quark-gluon plasma (QGP) is produced, sub-nuclear or partonic degrees of freedom may be needed for a correct description of systems created in heavy-ion collisions. Notably, some transport codes include both hadronic and partonic degrees of freedom simultaneously (e.g., \texttt{PHSD}~\cite{Linnyk:2015rco}).

The separation of scales underlying transport theory, which can be encompassed in the condition that the duration of a collision is short compared to the mean free flight time, is increasingly undermined when the systems probe large densities. The latter also lead to fast equilibration and the associated hydrodynamic behavior of the system. In principle, transport theory can be used to model hydrodynamics, even in cases when the degrees of freedom employed in transport are not the ones truly underlying the system. At some point, however, modeling heavy-ion collisions in terms of hydrodynamics becomes more straightforward, especially given the fact that the influence of the degrees of freedom believed to be appropriate can be easily included both in the EOS and the transport coefficients, which can be pursued within microscopic approaches \cite{Liu:2016ysz,Liu:2021rjf}. At the same time, hydrodynamic modeling of heavy-ion collisions may face certain challenges below $\ekin \approx 25\ A\txt{GeV}$ ($\snn \approx 7\ \txt{GeV}$), where not only is the assumption of near-equilibrium not very well satisfied, but where one also needs to take into account, among others, the hydrodynamic evolution of all conserved charges (e.g., baryon number $B$, strangeness $S$, and electric charge $Q$), possible evolution through unstable regions of the phase diagram, or the influence of the spectators. Extensions of hydrodynamics that may make it possible to consistently apply it in this regime are discussed in Section \ref{sec:hydrodynamics}.

\vspace{5mm}
{\textbf{New observables}}

Upcoming precision data will further bring unprecedented observables that could be previously considered only in theory, such as triple-differential spectra tied to a fixed orientation of the reaction plane \cite{HADES:2020lob,Danielewicz:2021vqq,Berkowitz:2022pgz,HADES:2022osk} not only for protons (see Fig.~\ref{fig:HADES}) and most abundant mesons, but also for deuterons, tritons, light nuclei, and hypernuclei. The potential of such spectra for the determination of the EOS is still to be fully explored, but a preliminary investigation~\cite{Danielewicz:2021vqq} indicates a rich structure with spectra which exhibit a maximum away from the beam direction, characterized by slopes dependent on azimuthal angle and slope discontinuities. Models that might have agreed with each other in describing low-order Fourier coefficients of flow will likely find describing such detailed observables difficult. Challenges remain even at the level of the low-order
coefficients, as many models now reproduce proton flow, but not Lambda or pion flow (see, e.g., Fig.\ \ref{fig:STAR}). Understanding the relations between observables for various particle species will lead to constraints on the physics driving the evolution of heavy-ion collisions in simulations and, through that, to understanding cluster formation, hyperon yields, in-medium interactions with of strange hadrons, and more (see also the white paper on \textit{QCD Phase Structure and Interactions at High Baryon Density: Continuation of BES Physics Program with CBM at FAIR}~\cite{Almaalol:2022xwv}).

\vspace{5mm}
{\textbf{Quantifying uncertainties of transport predictions}}

In the era of multi-messenger physics, where information on the EOS is derived from different areas of physics such as nuclear structure, nuclear reactions, and astrophysics, the ability to assess the uncertainty of a particular result is of crucial importance. This problem is especially relevant for evaluations of constraints on the EOS from transport simulations of heavy-ion reactions, since it has been found that using different transport models to describe the same data can lead to very different conclusions. As found in the TMEP comparisons (see \cite{TMEP:2022xjg} for a review), even with controlled input the results from different models may vary considerably due to different implementation strategies which in themselves are not dictated by the underlying physics. In such a situation, calculating the mean and variance of different model predictions is not a reliable way of determining the uncertainties. An approach currently considered for ensuring a robust quality control in combining inferences from different models is to weigh the models with a Bayesian weight which could be based, e.g., on the performance of a given model in benchmark tests and/or its ability to reproduce all key observables of a given reaction (for example, flow observables, particle multiplicities, and spectra). Bayesian analysis can be also used for model selection through a comparison of results from a list of available models with data, during which one assigns to each model a probability of being correct based on the quality of the fit. However, this approach implicitly assumes that among the considered models there is at least one ``true'' model (also known as the $\mathcal{M}$-closed assumption), which is often not fulfilled. Efforts have been taken to analyze data with an $\mathcal{M}$-open assumption, where the existence of a perfect model is not assumed. For nuclear physics efforts, this is being attempted within the Bayesian Analysis of Nuclear Dynamics (BAND) group \cite{Phillips:2020dmw} by using Bayesian model mixing, where information from different models is combined for inference.

\vspace{5mm}
{\textbf{Emulators and flexible EOS parametrizations}}

Robust explorations of the possible physics underlying various observables often necessitate repeating the calculations many times for different combinations of physics parameters. When high event statistics is needed, the computational task can easily overwhelm the available computational resources. An additional computational strain often arises from assessing Bayesian probability distributions for any conclusions. Increasingly, emulators are going to be used for this task, with some steps having been already made \cite{SpRIT:2020blg, Liyanage:2022byj,Oliinychenko:2022uvy}. Notably, similar issues emerge in the area of applications of hadronic transport \cite{Ciardiello:2020dtr} (see also Section \ref{sec:applications_of_hadron_transport}).

For explorations focused on the EOS, it may be of advantage to fit various possible EOSs with flexible relativistic density functionals as suggested in \cite{Sorensen:2020ygf,Sorensen:2021zxd}. This approach, given the freedom in varying both the functional form of the EOS as well as
the EOS parameters, is particularly amenable to Bayesian analyses (see, e.g., \cite{Oliinychenko:2022uvy} for a Bayesian
analysis with a parametrization of the EOS in terms of the functional dependence of the speed of
sound on density).

\vspace{5mm}

The above list of issues facing the application of transport theory to heavy-ion collisions highlights the fact that this approach to putting tighter constraints on the EOS rests on overcoming certain challenges. In simple terms, one attempts here to use a very dynamic and complex non-equilibrium process to obtain information describing a relatively simple and well-defined system, namely the equilibrated EOS of nuclear matter for different densities, temperatures, and isospin asymmetries. To achieve this in a reliable way, multiple complex issues of many-body physics have to be well controlled. On the other hand, several of the needed improvements are relatively well-understood, and tackling some of the unresolved problems poses an exciting intellectual challenge. As a reward for undertaking this effort, one gains the opportunity to obtain information on the EOS in a region which cannot be accessed through any other means: For densities below saturation, there is strongly constraining information from nuclear structure, with significant contributions coming also from low-energy heavy-ion collisions. Astrophysical observations on neutron stars and neutron star mergers are mainly sensitive to densities above about $3n_0$. The gap between these domains can only be filled with intermediate energy heavy-ion collisions, and transport studies are the essential tool to extract the information on the EOS from experimental data.

\subsection{Microscopic calculations of the EOS} 
\label{sec:microscopic_calculations_of_the_EOS}

Over the past decade, many-body nuclear theory has made significant progress in deriving microscopic constraints on the nuclear EOS at low densities from chiral effective field theory (\ceft)~\cite{Hebeler:2015hla,Lynn:2019rdt,Drischler:2021kqh,Drischler:2021kxf}.
The progress has been driven by improved two-nucleon (NN) and three-nucleon (3N) interactions, rigorous uncertainty quantification, and algorithmic and computational advances in the frameworks used to solve the many-body Schr{\"o}dinger equation with these interactions (see also the recent white paper on \textit{Dense matter theory for heavy-ion collisions and neutron stars}~\cite{Lovato:2022vgq}).

\subsubsection{Status}

Chiral EFT~\cite{Epelbaum:2008ga,Machleidt:2011zz,Hammer:2019poc,Epelbaum:2019kcf,Piarulli:2019cqu} provides a systematic way to construct nuclear interactions consistent with the low-energy symmetries of QCD, using nucleons (N's), pions ($\pi$'s), and (in the case of delta-full \ceft), $\Delta$-resonances ($\Delta$'s) as the relevant effective degrees of freedoms. 
Nuclear interactions in \ceft~are expanded in powers of  momenta or the pion mass over a hard scale at which \ceft breaks down; this breakdown scale is expected to be of the order of the $\rho$-meson mass, $\Lambda_b \approx 600$ MeV.
At each order in the EFT expansion, only a finite number of diagrams enter the description of the interaction according to a chosen power counting scheme, of which the Weinberg power counting has been predominant. 
(Several alternative power counting schemes have been developed, see, e.g., \cite{Kaplan:1996xu,Nogga:2005hy,Long:2011xw,Long:2012ve}; however, these approaches are not well-developed at higher expansion orders and in some cases fail to reproduce properties of nuclear matter at leading- or next-to-leading order.)
At the leading-order (LO) in Weinberg's power counting one includes contribution from the one-$\pi$ exchange between two nucleons as well as momentum-independent contact interactions, which allow one to describe key features of the nuclear interaction already at the lowest order. 
At next-to-leading-order (NLO), two-$\pi$ exchanges are included as well as momentum-dependent contact interactions, and similarly, more involved terms appear at higher orders. 
The various low-energy coupling constants are determined from fits to experimental data, e.g., the $\pi$-N couplings are fit to $\pi$-N scattering, while those describing NN short-range interactions are fit to NN scattering data.
The advantage of \ceft over phenomenological approaches is that multi-nucleon interactions, such as the important 3N interactions, naturally emerge in the EFT expansion and, moreover, are consistent with the NN sector.
The contribution of chiral EFT 3N forces to the neutron-matter EOS naturally includes both the interaction as well as method uncertainties, typically adding a repulsion of about 2--4 MeV at nuclear saturation density~\cite{Hebeler:2009iv,Kruger:2013kua, Tews:2015ufa,Drischler:2016djf}.
Forces involving increasingly more nucleons are correspondingly more suppressed, e.g., the leading contribution to 3N forces (four-nucleon (4N) forces) appears at N$^2$LO (at N$^3$LO) in Weinberg's power counting. 
Furthermore, there are only two new low-energy couplings appearing in the three- and four-body forces to N$^3$LO, which govern the strengths of the intermediate- and short-range contribution to the leading 3N forces, respectively.
Consequently, \ceft 3N and 4N interactions at N$^3$LO are completely determined by constraints on the coupling constants obtained from NN and $\pi$-N scattering, usually resulting in tight constraints on very neutron-rich matter from \ceft.

\begin{figure*}[t]
    \includegraphics[width=0.95\textwidth]{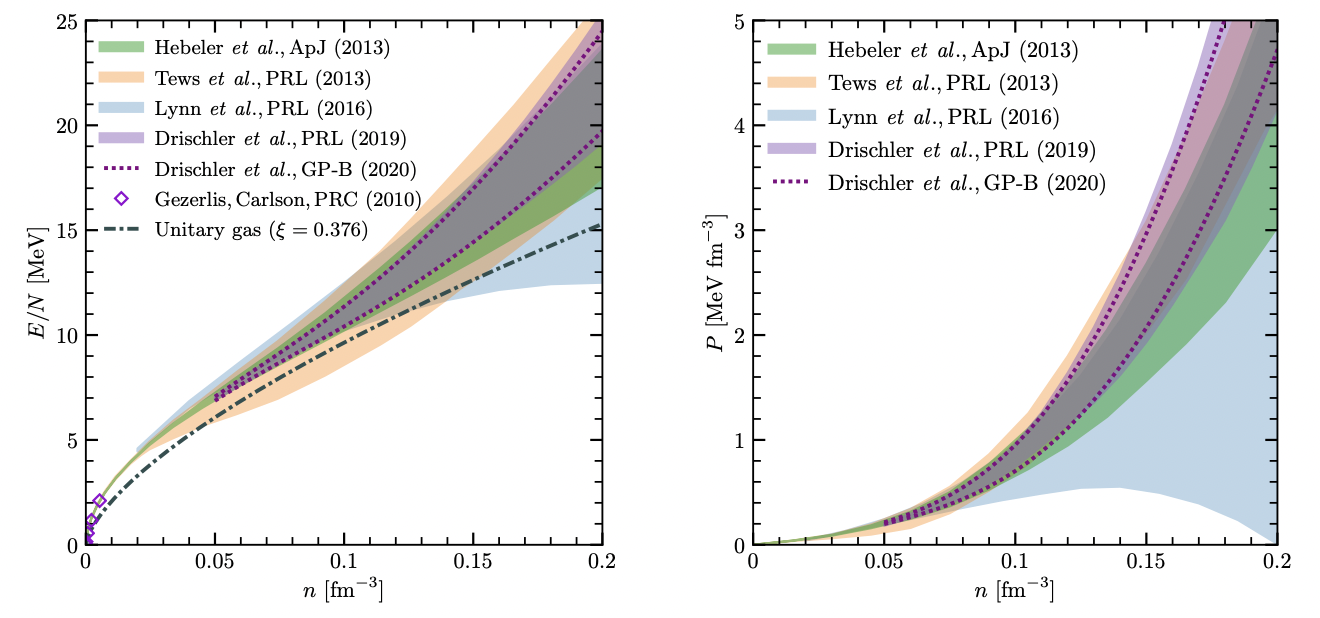}
    \caption{Comparison of the energy per particle $E/N$ (\textit{left}) and the pressure $P$ (\textit{right}) as functions of density for pure neutron matter in different many-body calculations using interactions from \ceft. The left panel also shows low-density QMC results of Ref.~\cite{Gezerlis:2009iw} and the conjectured unitary-gas lower bound on the energy per particle of pure neutron matter from Ref.~\cite{Tews:2016jhi}.
    Figure from Ref.~\cite{Huth:2020ozf}.}
    \label{fig:PNMcomparison}
\end{figure*}

Another key feature of \ceft is that order-by-order calculations in the \ceft expansion have enabled estimation of theoretical uncertainties due to truncating the chiral expansion at a finite order~\cite{Epelbaum:2014efa,Drischler:2020hwi,Drischler:2020yad,Drischler:2021kxf}.
Quantifying and propagating these EFT truncation errors enables meaningful comparisons between competing nuclear theory predictions, see Fig.~\ref{fig:PNMcomparison}, and/or constraints from nuclear experiments and neutron-star observations in the multi-messenger astronomy era~\cite{Essick:2021kjb}.  
Such comparisons are facilitated by Bayesian methods in a statistically rigorous way~\cite{Essick:2020flb,Drischler:2021kxf,Essick:2021kjb} to take full advantage of the great variety of empirical EOS constraints we anticipate in the next decade.

Chiral EFT also provides nuclear Hamiltonians governing the interactions in nuclear systems. 
However, to calculate properties of a many-body system, computational methods able to solve the Schroedinger equation for this system are necessary.
Among various frameworks used to solve the nuclear many-body problem in dense matter, quantum Monte Carlo (QMC) methods and many-body perturbation theory (MBPT) have been the main tools employed to study the physics of neutron-star matter in recent years.
Both methods have recently made tremendous advances in predicting properties of nuclei and calculating the nuclear matter EOS~\cite{Carlson:2014vla,Lynn:2019rdt,Tews:2020wrl,Gandolfi:2020pbj,Hu:2021trw,Tichai:2020dna,Drischler:2021kxf}.

QMC frameworks, such as the auxiliary field Diffusion Monte Carlo (AFDMC) method, are based on imaginary-time propagation of a many-body wave function and enable us to extract ground-state properties of a nuclear many-body system with high statistical precision~\cite{Carlson:2014vla,Lynn:2019rdt}.
Their nonperturbative nature also allows for the treatment of nuclear interactions at high momentum cutoffs, providing important insights into nuclear interactions at relatively short distances that may help to improve the modeling of \ceft interactions. 
QMC calculations of binding energies, radii, and electroweak transitions of nuclei up to $A=16$~\cite{Gezerlis:2013ipa,Gezerlis:2014zia,Lynn:2017fxg,Tews:2021bqc,Piarulli:2014bda,Piarulli:2016vel,Baroni:2018fdn} using \ceft NN and 3N interactions are in very good agreement with experimental data~\cite{Lonardoni:2017hgs,Lonardoni:2019ypg,Piarulli:2017dwd,King:2020wmp}. 
QMC methods were also used to calculate the EOSs of matter up to about twice the nuclear saturation density $n \approx 2n_{0}$~\cite{Lynn:2015jua,Tews:2018kmu,Tews:2018iwm,Piarulli:2019pfq,Lovato:2022apd}. 
The calculated EOSs include estimates of systematic truncation uncertainties, and are commonly used to constrain properties of neutron stars~\cite{Tews:2018kmu,Al-Mamun:2020vzu,Dietrich:2020efo}.

The past decade has also seen a renaissance for many-body perturbation theory (MBPT) calculations in nuclear physics~\cite{Drischler:2021kxf,Tichai:2020dna}. 
Key to this development has been the discovery that nuclear potentials with momentum-space cutoffs in the range $400\,{\rm MeV}\lesssim \Lambda \lesssim 500$\,MeV (not to be confused with the breakdown scale of \ceft, $\Lambda_b$) are sufficiently soft to justify the use of perturbation theory methods~\cite{Hebeler:2009iv} (see \cite{Hoppe:2017lok} for a Weinberg eigenvalue analysis).
Such low-momentum potentials can be obtained from renormalization group methods~\cite{Bogner:2009bt} or by directly constructing chiral effective field theory potentials at a coarse resolution scale. Furthermore, recent advances in automatic diagram generation~\cite{Arthuis:2020tjz} combined with automatic code generation~\cite{Drischler:2017wtt} and high-performance computing have led to a fully automated approach to MBPT calculations in nuclear physics~\cite{Drischler:2021kxf}, in which chiral two- and multi-nucleon forces can be included to high orders in the chiral and MBPT expansions.
MBPT has been demonstrated to be a computationally efficient and versatile tool for studying the nuclear EOS as a function of baryon number density $n_B$, isospin asymmetry $\delta = (n_n-n_p)/(n_n+n_p)$, and temperature $T$~\cite{Wellenhofer:2014hya,Wellenhofer:2017qla,Keller:2020qhx,Keller:2022crb} with implications for neutron star structure~\cite{Drischler:2021kxf} and astrophysical simulations~\cite{Carbone:2019pkr}; here, $n_n$ and $n_p$ correspond to the neutron and proton densities, respectively.
In particular, MBPT allows us to compute the EOS of neutron-star (i.e., $\beta$-equilibrated) matter explicitly, which can help improve isospin asymmetry expansions of the low-density nuclear EOS such as the standard quadratic expansion~\cite{Drischler:2013iza,Drischler:2015eba,Wellenhofer:2015qba,Wellenhofer:2017qla,Wen:2020nqs,Somasundaram:2020chb}.
MBPT also allows us to study nuclear properties other than the nuclear EOS, including the linear response and transport coefficients that could be used to inform more accurate numerical simulations of supernovae and neutron-star mergers.
Furthermore, MBPT for (infinite) nuclear matter has been used to construct a microscopic global optical potential with quantified uncertainties based on \ceft NN and 3N interactions~\cite{Whitehead:2020wwb,Holt:2022piv}.~
Altogether, MBPT calculations of nuclear matter properties can provide important constraints that enable microscopic interpretations of future nuclear reaction experiments~\cite{Hebborn:2022vzm} (e.g., at the Facility for Rare Isotope Beams) and neutron star observations.

To date, theoretical predictions for the nuclear EOS, optical potentials, and in-medium NN scattering cross sections have been computed at finite temperature at various levels of approximation starting from fundamental two- and multi-nucleon forces. 
These quantities are inputs to transport model simulations~\cite{Bertsch:1988ik,Aichelin:1991xy} of heavy-ion collisions used to extract constraints on the properties of hot and dense nuclear matter (see Section \ref{sec:model_simulations_of_HICs} for more details). In transport simulations, the EOS, single-particle potentials, and in-medium NN cross sections are usually obtained from effective phenomenological interactions \cite{Xu:2014cwa,Dutra:2012mb} that are fitted to the properties of finite nuclei and cold nuclear matter, and then extrapolated into the finite-temperature regime. 
Recently, some effort has been devoted to benchmarking \cite{Xu:2019ouo} the temperature dependence of these effective interactions against predictions from \ceft or directly using EFT constraints in fitting effective interactions~\cite{Brown:2013pwa,Zhang:2017hvh,Du:2021rhq}. 
To enable such comparisons, the free energy of homogeneous nuclear matter as a function of temperature, baryon number density, and isospin asymmetry has been calculated using \ceft interactions up to second order in many-body perturbation theory \cite{Wellenhofer:2014hya} and within the Self-Consistent Green’s Function (SCGF) approach \cite{Carbone:2018kji}, which resums particle-particle and hole-hole ladder diagrams to all orders. 
The resulting EOS has been shown to be consistent with the critical endpoint of the symmetric nuclear matter liquid-gas phase transition~\cite{Wellenhofer:2014hya,Carbone:2018kji} as well as the low-density/high-temperature pure neutron matter EOS from the virial expansion~\cite{Wellenhofer:2015qba}. 
Furthermore, single-particle potentials have been computed at finite temperature at the Hartree-Fock level \cite{Rrapaj:2014yba}, from G-matrix effective interactions \cite{Schnell:1998zz}, and in SCGF theory~\cite{Rios:2020oad,Carbone:2019pkr}. 
Of particular importance is the associated nucleon effective mass, which is obtained from a momentum derivative of the single-particle energy. 
The nucleon effective mass is directly related to the density of states and hence governs entropy generation at finite temperature, with consequences for the dynamical evolution of core-collapse supernovae and neutron star mergers.
Finally, in-medium NN scattering cross sections have been computed at finite density and zero \cite{Sammarruca:2013bda} as well as at finite \cite{Schnell:1998zz} temperature using high-precision nuclear forces. 
Overall, improved effective field theory methods can provide much-needed microscopic inputs for transport simulations of heavy-ion collisions and astrophysical simulations, leading to a consistent description of microscopic processes affecting their dynamical evolution.

\subsubsection{Challenges and opportunities}

To fully capitalize on experimental and observational data and extract key information on fundamental questions in nuclear physics, continued progress in nuclear theory is crucial.
The combination of \ceft with modern computational approaches like machine learning, artificial intelligence, emulators, and Bayesian inference have provided EOS results for a wide range of densities, and at various proton-to-neutron asymmetries and temperatures, with quantified uncertainties~\cite{Keller:2022crb,Drischler:2020yad} (see also \cite{Boehnlein:2021eym} for a broader review of machine learning in nuclear physics).
Future progress in the development of fundamental interactions, combined with these tools, will increase the precision of the results and enable us to answer open problems in chiral EFT.
Among these, the most pressing is at which densities and how $\chi$EFT breaks down~\cite{Tews:2018kmu,Drischler:2020yad}.
In particular, for studies of neutron-star mergers it is of great importance to describe dense matter at finite temperatures~\cite{Wellenhofer:2015qba,Carbone:2019pkr,Keller:2020qhx}, however, these might influence the breakdown of the theory in dense matter.
In the next decade, it will be crucial to reliably determine how far one can push the \ceft approach in nucleonic matter.

While microscopic calculations have been very successful in calculating properties of nuclei and homogeneous matter at densities up to 1--2 times the nuclear saturation density, we need improved microscopic descriptions of neutron-rich dense matter beyond that regime, at a few times nuclear saturation density and finite temperatures, with quantified uncertainties. 
This can be achieved by employing models derived within relativistic mean-field or density functional theory that are firmly rooted in microscopic theory at lower densities.
Such models will be very important to connect theoretical calculations within the framework of \ceft to heavy-ion collision experiments at accelerator facilities around the world.
Heavy-ion collision experiments at intermediate beam energies bridge the low- and high-density regimes of the EOS and provide complementary information to that obtained from nuclear structure or neutron-star studies~\cite{Huth:2021bsp} (see Section \ref{sec:model_simulations_of_HICs}).
Robust inferences from the experimental data will require more accurate predictions from transport theory, which strongly depend on, among others, mean-field or density functional models.
It will be imperative to test and constrain such models for the EOS with more rigorous microscopic calculations.
Beyond their use in hadronic transport simulations, these models are also a crucial input for calculations of properties of neutron star crusts (see Section \ref{sec:neutron_star_theory}).

Additional theoretical constraints might be provided by high-density calculations within the framework of perturbative QCD (pQCD)~\cite{Kurkela:2009gj}, which can be applied at very high densities of the order of 40 times the nuclear saturation density, where the strong interactions among quarks become perturbative.  Constraints on the EOS based on pQCD, together with assumptions on causality and stability, have been used to constrain the EOS at lower densities probed in the core of neutron stars~\cite{Kurkela:2014vha,Annala:2017llu,Komoltsev:2021jzg,Gorda:2022jvk}.
However, it has been found that the constraining power of pQCD calculations is strongly dependent on the way in which they are implemented~\cite{Gorda:2022jvk,Somasundaram:2022ztm}.
Future studies have to establish to what extent pQCD constraints are robust at densities of the order of several times nuclear saturation density, and how constraining future higher-order calculations may become. 
In this regard, improved microscopic calculations of the nuclear EOS using the functional renormalization group~\cite{Leonhardt:2019fua,Braun:2022jme} will provide important insights.

\subsection{Neutron star theory}
\label{sec:neutron_star_theory}

\subsubsection{Status}

Measurements of the EOS, masses of neutron-rich isotopes far from the band of stability, and experimental constraints on nucleon effective masses provide essential input into neutron star models, progressing our understanding of the structure and dynamics of these astronomically 
important objects. Several properties of neutron stars, including the mass-radius relation and their tidal deformabilities, can be calculated once the EOS is provided. This, in turn, enables us to constrain the EOS once those properties are observed \cite{Lattimer:2021emm}.

Nuclear EOSs for neutron stars can be constructed from, for example, \textit{ab initio} calculations and density functionals~\cite{Nazarewicz:2013gda,Zhang:2016vcc,Lim:2019som,Gil:2018yah,Grasso:2018pen,Biswas:2021yge,Alford:2022bpp} or, more schematically, from meta-models \cite{Margueron:2017eqc,Margueron:2017lup,Zhang:2018vrx} parametrized by nuclear 
\begin{wrapfigure}{r}{0.61\textwidth}
	\centering
	\vspace{-6.75mm}
	\includegraphics[width=0.60\textwidth]{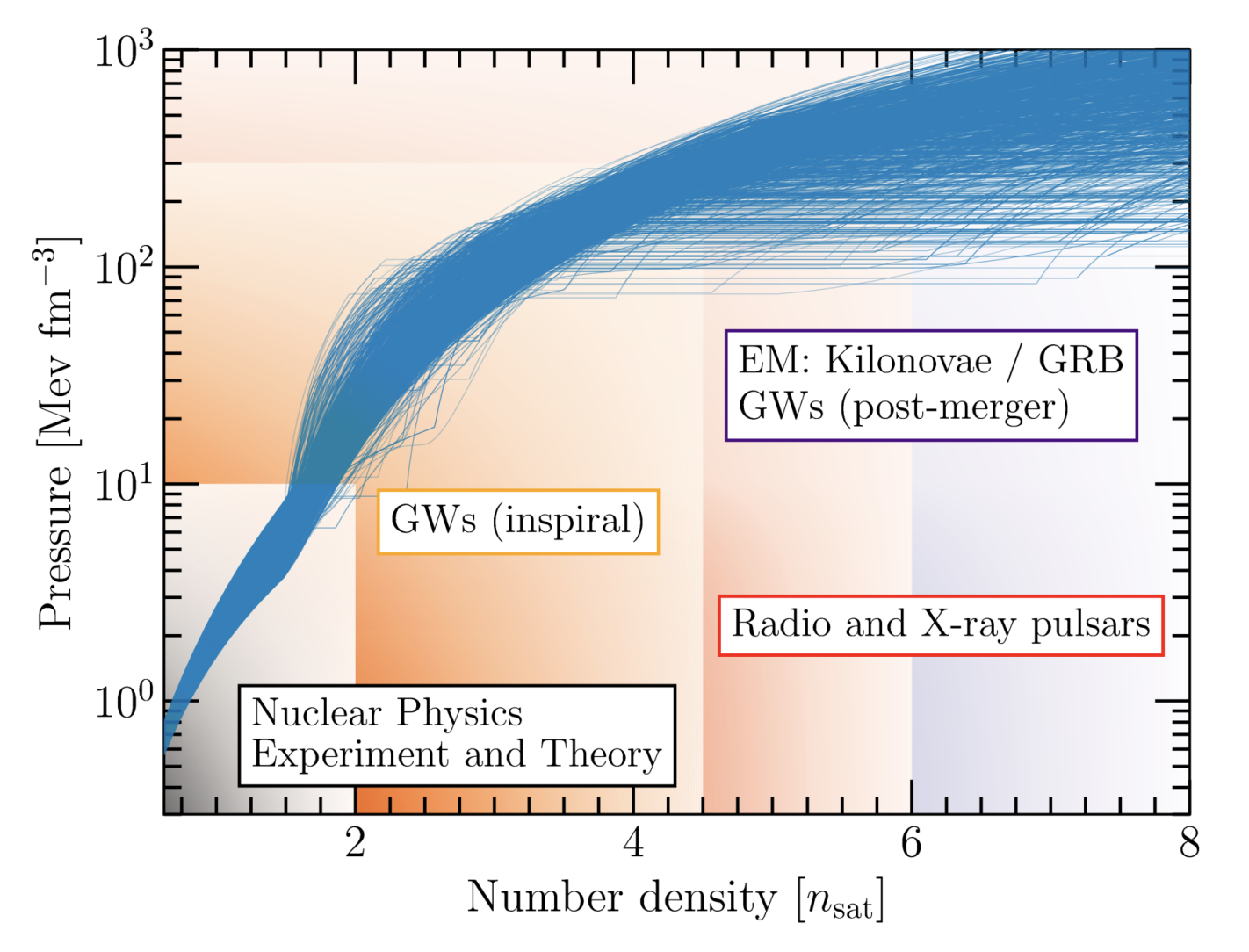}
	\vspace{-3.35mm}
	\caption{Impact of nuclear physics theory and experiment, and different astrophysical measurements on constraining the cold neutron-star EOS. Blue lines show a family of EOS that are constrained by chiral EFT at low densities. At higher densities, the EOS can then be constrained using GWs from inspirals of neutron star mergers, data from radio and X-ray observations of pulsars, and electromagnetic signals associated with neutron star mergers. The indicated boundaries between regions affected by these measurements are not strict and depend on the EOS and properties of the astrophysical system. Figure from ~\cite{Pang:2022rzc}.
	}
	\vspace{-7mm}
	\label{fig:cold-neutron_star_EOS_impacts}
\end{wrapfigure}
matter parameters, which can be used to make contact with heavy-ion collisions \cite{Huth:2021bsp}. \textit{Ab initio} calculations take into 
account more fundamental properties of the nuclear force (see Section~\ref{sec:microscopic_calculations_of_the_EOS}), but prohibit the calculation of large ensembles of EOSs spanning the nuclear parameter space. Meta-models allow rapid computation of such large ensembles, but 
encode mainly bulk properties of nuclear matter, which excludes them from being used to model finite nuclei. Density 
functionals represent a compromise, allowing both rapid computation of EOSs and use in finite nuclear models, and 
thus are more suited to combining nuclear experimental and astrophysical information. Many of these models can be 
smoothly extrapolated from the saturation-density to arbitrarily high density, in which case astronomical observations can be used to constrain the saturation-density nuclear matter parameters and their density dependence~\cite{Zhang:2018vrx,Gil:2020wqs,Zhang:2021xdt}. This extrapolation, however, is model-dependent, as different density functionals have different dependence on density. Additionally, this extrapolation might not be physically well-founded.

As densities inside neutron stars can reach up to several times nuclear saturation density, at some (as-yet not determined) density a description in terms of purely nucleonic degrees of freedom is expected to break down. Heavy-ion collisions can help us constrain that point, and the nature of any phase transitions that occur above saturation density. The nuclear EOSs can be then combined with models describing the EOS at higher densities. Models that explicitly include a 
range of possible high-density degrees of freedom, such as hyperons and quarks, can be constructed; the predicted neutron star compositions are then dependent on the particular model used. Another approach is to use more general models that give up the explicit dependence on the underlying degrees of freedom, thus losing information on, e.g., appearance of exotic particles at high densities, in favor of spanning the full space of physically consistent EOSs, reducing the model dependence of inferences from astrophysical observations~\cite{Tews:2019cap}. These schemes include piecewise polytropes~\cite{Read:2008iy,Ozel:2009da,Hebeler:2010jx,Steiner:2010fz}, line segments~\cite{Steiner:2017vmg,Al-Mamun:2020vzu}, speed-of-sound models \cite{Tews:2018kmu,Tews:2018iwm,Greif:2018njt,Raaijmakers:2021uju,Annala:2019puf}, spectral models \cite{Lindblom:2010bb} 
and non-parametric models generated from Gaussian processes (GPs) \cite{Landry:2018prl,Landry:2020vaw,Essick:2019ldf,Essick:2020flb,Legred:2021hdx} or machine learning techniques~\cite{Ferreira:2019bny}. If these more general approaches are used down to the nuclear saturation density, extra modeling is required to connect them to the microscopic nuclear EOS and nuclear observables~\cite{Essick:2021ezp}.

Once the EOS is specified, the solution of the Tolman--Oppenheimer--Volkoff equations and their extensions including rotation, determining the structure of a neutron star through balancing the attractive force of gravity and the repulsion coming from the EOS, provide predictions for bulk properties of the neutron star such as radii, tidal deformabilities, moments of inertia, and break-up frequencies of neutron stars as a function of their mass. 
All of these properties can be compared with multi-messenger observations, including gravitational waves and electromagnetic signals from neutron-star mergers and isolated neutron stars~\cite{Dietrich:2020efo}.

The systematic construction of neutron star EOS models and statistical inference of EOS parameters from data is an endeavor that is just over a decade old \cite{Read:2008iy,Ozel:2009da,Hebeler:2010jx,Steiner:2010fz}. 
This effort has matured in the current era of multi-messenger astronomy with a large push to explore the model-dependence of EOS inferences~\cite{Greif:2018njt,Legred:2022pyp} and ways of connecting the EOS with astrophysical and nuclear data~\cite{Annala:2019puf,Dietrich:2020efo,Raaijmakers:2021uju,Miller:2021qha,Essick:2021kjb,Huth:2021bsp}. Different choices of which observables to include or infer can be made. For example, astrophysical observations can be used to infer the EOS, which can then be connected to nuclear models to inform their parameters and predict nuclear observables. Conversely, nuclear observables can be used to infer nuclear parameters, which can then inform the neutron star models and predict astrophysical observables. The future lies in combining more and more sets of data of both types to understand nuclear and neutron star models better.

Exciting progress has been made in gathering astrophysical data to constrain our dense matter theories (see Fig.\ \ref{fig:cold-neutron_star_EOS_impacts} for an illustration of density regions affected by different observables). Neutron-star data from the last 5 years identified the heaviest neutron star known to date with a mass of $2.08(7)M_{\odot}$ \cite{Fonseca:2021wxt,NANOGrav:2019jur} (where $M_{\odot}$ is the solar mass), excluding EOSs which cannot reach that limit, while the kilonova AT2017gfo, associated with GW170817, has placed an upper limit on the maximum mass to be on the order of $2.3M_{\odot}$~\cite{Margalit:2017dij,Rezzolla:2017aly}. 
The detection of GW170817 by the LIGO-Virgo Collaboration has enabled us to place constraints on the tidal deformability of this system, $\tilde{\Lambda}_{\rm GW170817}\leq 720$~\cite{De:2018uhw,LIGOScientific:2018cki}.
Neutron Star Interior Composition Explorer Mission (NICER) has provided two mass-radius measurements by observing X-ray emission from several hot spots on the neutron star surface, finding a radius of $13.02^{+1.24}_{-1.06}$km for a star with mass $1.44^{+0.15}_{-0.14}M_{\odot}$ (PSR J0030+0451) and $13.7^{+2.6}_{-1.5}$km for a star with mass $2.08(7)M_{\odot}$ (PSR J0740+6620) in the analyses of Refs.~\cite{Miller:2019cac,Miller:2021qha,Riley:2019yda,Riley:2021pdl}. X-ray observations of the temperature of the neutron star in the Cas A supernova remnant have revealed core cooling on the timescale of years, hinting at the possible superfluid properties of the core \cite{Shternin:2021fpt}. 

These observations have enabled meaningful constraints on the EOS to be set and have already allowed us fascinating glimpses into the possible properties of high-density matter. For example, perturbative QCD predicts that the speed of sound squared approaches the conformal limit of $c_{\rm s}^2 =1/3$ from below as the density becomes arbitrarily high. Meanwhile, inferences of the neutron star EOS from observational data indicate that the speed of sound rises in the core to significantly above $1/3$~\cite{Bedaque:2014sqa,Alford:2015dpa,Tews:2018kmu,Fujimoto:2019hxv, Marczenko:2022jhl}. Consequently, this suggests that the speed of sound has a non-trivial behavior with increasing density \cite{Kurkela:2009gj,Tews:2018kmu,Tan:2021ahl}, first rising above $1/3$ to reach a maximum at moderate densities, then decreasing, and finally approaching $1/3$ from below at asymptotically high densities. Several microscopic mechanisms for this behavior have been suggested, including the appearance of quarkyonic matter~\cite{McLerran:2018hbz,Sen:2020qcd}, hadron-quark continuity within a topological Skyrmion model \cite{Rho:2022wco}, or superconducting quark matter~\cite{Ayriyan:2021prr,Contrera:2022tqh,Ivanytskyi:2022bjc}. At the same time, tentative evidence for a softening of the EOS in neutron star cores, based on an analysis using a novel speed-of-sound interpolation, has likewise been suggested~\cite{Annala:2019puf}.

If we want to leverage the substantial data we have on neutron star cooling and dynamical evolution, additional EOS quantities need to be supplied consistently for each EOS model, such as the effective masses (see also Section \ref{sec:density-dependence_effective_masses}) and superfluid neutron and proton gaps, essential for modeling thermal and dynamical properties of neutron stars. For example, the mutual friction of the core -- the strength of the coupling between the charged particles (electrons, protons) and superfluid neutrons -- depends on the effective neutron mass and the proton fraction \cite{Andersson:2008tp}, which both also correlate with the symmetry energy \cite{Li:2018lpy}. A consistent extraction of both symmetry energy parameters and effective masses from heavy-ion collision data is therefore required.

In contrast to efforts devoted to systematic, statistically meaningful inferences of the EOS in the cores of neutron stars, modeling the neutron star \emph{crust} is still in its infancy: The first calculations of large ensembles of systematically parametrized crust models and their use in statistical analysis have only been carried out recently \cite{Carreau:2019zdy,Balliet:2020nsh,Thi:2021hai,Newton:2021rni,Neill:2020szr}. However, much more nuclear experimental data can be brought to directly bear on crust physics, and we have entered an era where we can access information about the crust with unprecedented fidelity. For example, we have now observed the same neutron-star crust as it first cooled, then became heated by accreted matter, and then cooled again~\cite{Wijnands:2003tx,Cackett:2006qk,Cackett:2008tq,Cackett:2013wva,Parikh:2018pbx}. We have followed a pulsar through a glitch -- a sudden change in the rotation period of the pulsar -- and glitch recovery with a resolution of a few seconds~\cite{Ashton:2019hqd}. 
These observations have provided very strong evidence that the crust is solid, that there exist superfluid neutrons in the inner crust which can be decoupled from the nuclei in the crustal lattice, and that nuclear reactions from accreted material sinking into the crust provide deep crustal heating \cite{Cackett:2008tq,Brown:2009kw,Wijnands:2012tf}.

Additionally, models of the neutron star crust predict that, prior to the transition to homogeneous matter, isolated nuclei in the crust fuse to form cylindrical, planar, and more exotic shapes, termed ``nuclear pasta'', that can affect neutron-star observations~\cite{Sotani:2011nn,Gearheart:2011qt,Schneider:2013dwa,Schuetrumpf:2015nza}. This crust-core boundary region, often referred to as the mantle, is likely a complex fluid. 
Density functional theory and molecular dynamics calculations of these structures reveal a complex energy landscape with many coexisting shapes, and correspondingly complex mechanical and transport properties~\cite{Watanabe:2000rj,Berry:2015dos,Chugunov:2010ac,Caplan:2016uvu,Caplan:2018gkr,Pethick:1998qv,Pethick:2020aey}, which are strongly influenced by the EOS at around 0.5$n_0$ through the pressure, proton fraction, and surface energy of the structures. These properties can also be studied in multifragmentation reactions, which probe, among others, the competition between nuclear surface energy and Coulomb energy at sub-saturation density \cite{Botvina:2006xc,Souza:2008pj,Ogul:2010si}.

Inhomogeneous matter in the crust of a neutron star, including the dripped neutrons expected in the inner crust, can be modeled using a variety of nuclear theory techniques. These usually involve calculations within a single, repeating unit (Wigner--Seitz cell) of matter, typically containing a single nucleus~\cite{Chamel:2008ca,Lim:2017luh,Tews:2016ofv}. The compressible liquid drop model (CLDM) treats the nuclear matter inside and outside of nuclei as homogeneous and described by the bulk matter EOS, while the surface energy is specified by a separate function with additional parameters \cite{Baym:1971ax,Watanabe:2000rj,Forbes:2019xaz,Newton:2011dw,Tews:2016ofv}. 
The surface parameters and those that define the dimensions of the cell and nucleus are minimized to obtain the ground state. The Thomas-Fermi model employs the local density approximation, modeling matter with a specified form of the inhomogeneous nuclear matter density in the unit cell; here, the parameters of the density distribution are varied to obtain the ground state configuration~\cite{Oyamatsu:2006vd}. Microscopic approaches to describing inhomogeneous nuclear matter, in which individual neutrons and protons are the degrees of freedom, include quantum Hartree-Fock or Relativistic Mean Field models \cite{Magierski:2001ud,Gogelein:2007pb,Newton:2009zz,Schuetrumpf:2016kzq,Fattoyev:2017zhb,Newton:2021vyd}, and semi-classical molecular dynamics approaches \cite{Lopez:2014dga,Caplan:2018gkr}.

There is a great need for nuclear physics input into models of the neutron star crust, which analyses of heavy-ion collision data can provide. For example, the thickness, mass and moment of inertia of the crust depend on the higher-order symmetry energy parameters $L$, $K_{\rm sym}$, and $Q_{\rm sym}$~\cite{Newton:2014iha,Carreau:2019zdy,Thi:2021hai}.
Thus measurements of the symmetry energy parameters up to third order in heavy-ion collision experiments are essential to understand the properties of the crust. The symmetry energy, effective masses, and surface energies of nuclear clusters strongly affect the proton fraction on either side of the crust-core transition density, the extent of nuclear pasta near the crust-core boundary, the mechanical and transport properties, the thermal conductivity and specific heat, the electrical conductivity, and the shear modulus of the crust~\cite{Oyamatsu:2006vd,Chamel:2008ca,Newton:2011aa,Fattoyev:2017zhb}. Nuclear experiment can thus constrain neutron star crust models, and astrophysical observables associated with the crust can measure nuclear observables as well as measurements of neutron star bulk properties. For example, the symmetry energy can be constrained by combining nuclear data with crust and core observables, e.g., through a potential multi-messenger measurement of the resonant frequency of crust-core interface oscillations~\cite{Neill:2020szr}.

\subsubsection{Challenges and opportunities}

The next decade will provide a wealth of new data on neutron stars, as the LIGO-VIRGO-KAGRA detectors are expected to observe many new binary neutron-star mergers, some of them with electromagnetic counterparts~\cite{Baibhav:2019gxm,Colombo:2022zzp,Patricelli:2022hhr}. As NICER continues to measure more neutron star masses and radii, next-generation X-ray timing missions such as \mbox{Strobe-X}~\cite{STROBE-XScienceWorkingGroup:2019cyd} and radio telescopes such as the Square-Kilometer Array will increase the number of pulsars we see and are able to measure by an order of magnitude. Long-timescale observations of individual pulsars (using radio timing) and persistent gravitational waves from deformations of neutron stars will lead to measurements of their moments of inertia. These new data points might enable us to pin down the nuclear matter EOS, to discover or rule out the existence of phase transitions to exotic forms of matter in the cores of neutron stars, and to reliably constrain microscopic interactions between fundamental particles. Further ahead, next-generation detectors, such as the Einstein Telescope~\cite{Maggiore:2019uih} and the Cosmic Explorer \cite{Finstad:2022oni}, will be able to observe the neutron-star inspiral phase as well as the onset of tidal effects with high signal-to-noise ratio, with direct consequences for the resolving the interior structure of neutron stars and probing fundamental properties of matter at highest densities.

Although model-agnostic extrapolations to higher densities such as through the use of polytropes~\cite{Read:2008iy,Hebeler:2009iv}, speed of sound schemes~\cite{Tews:2018kmu,Greif:2018njt,Tan:2020ics}, Gaussian processes~\cite{Essick:2019ldf,Landry:2020vaw} and spectral methods~\cite{Lindblom:2010bb}, combined with robust data analysis, will eventually allow us to pin down the dense-matter EOS, they cannot answer the question about the relevant microscopic degrees of freedom at high densities. Hence, it is crucial to develop improved microscopic models with well-quantified uncertainties in this regime. At the same time, creating ensembles of outer core and crust models that allow for inclusion of astrophysical and nuclear data requires underlying nuclear models to have enough freedom to explore a large region of parameter space, and allow fast computation of relevant quantities that also capture the essential physics. Currently, it is energy density functionals like Skyrme, Gogny, and Relativistic Mean Field models that provide these properties. Consequently, progress could be made by making a stronger connection between these models and microscopic approaches, e.g., connecting energy-density functionals to \textit{ab initio} calculations allowing a more direct link to $\chi$EFT~\cite{Lim:2017luh,Marino:2021xyd,Yang:2021akb}. In the same spirit, EFT calculations of the EOS can be used as a ``low-density limit'' to calibrate higher-density models for neutron stars and heavy-ion collisions.

The crust can be modeled consistently with nucleonic matter in the core using density functional theory to model both. When choosing a model, a compromise must be made between accurate modeling of microscopic quantum effects, such as shell effects in the nucleus and surrounding neutron gas, and the computational expediency required to construct large ensembles of crust models needed for statistical inference. For example, quantum shell effects strongly determine the evolution of the mass and charge number of nuclei with density, alter the effective mass of dripped neutrons, and drive the complex energy landscape of nuclear pasta. Fully microscopic quantum calculations include shell effects self-consistently, but are computationally expensive.
The CLDM approach can be used to construct large numbers of crust models, but requires shell effects to be added by hand. 
Future work needs to develop schemes of incorporating such microscopic effects in large ensembles of crust models. The method that may allow that is the Extended Thomas-Fermi method, incorporating shell effects through the Strutinsky Integral (see, e.g., \cite{Shelley:2021pnx}).

Models should also incorporate nuclear pasta, as its extended structures may contribute to the mechanical and thermal properties of matter at the crust-core boundary. It is computationally demanding to model transport and mechanical properties of the crust microscopically or in simulations~\cite{Sagert:2022gwu}, particularly in the nuclear pasta phases, and it is unrealistic to include these quantities in large ensembles of crust models. Simpler schemes that extrapolate the mechanical and transport properties across the parameters space based on microscopic models could be developed. Also, representative crust models inferred from data can be used to calculate these crust properties.

Furthermore, when older neutron stars accrete matter in the crust, the matter gets gradually pushed down into the core  and replaced by the accreted matter. The temperatures in the crust are well below the nuclear potential energies, so the replacement crust cannot easily attain nuclear statistical equilibrium. Ensembles of \emph{accreted} crust models are yet to be constructed, but are necessary to correctly account for deep crustal heating and therefore to fully utilize the observations of cooling of accreted crusts in low mass X-ray binaries. 

There is also a need for a balance between accuracy and precision. 
A model can be accurate but not precise (predicting the correct value of a physical quantity but having large error bars), or precise but not accurate (predicting very small error bars, but not predicting the correct value of some physical observable). Individual crust models can be created from mass models that are precisely fit to data and which predict precise values for, e.g., the symmetry energy parameters. However, to make accurate inferences of nuclear matter parameters from astrophysical observables, and to include their experimentally measured ranges, ensembles of models spanning the parameter space should be employed. Both strategies are important, and the precision-fit models can act as benchmarks against which we assess the outcomes of statistical inferences.

\newpage
\section{Heavy-ion collision experiments}
\label{sec:laboratory_experiments}

Establishing the equation of state (EOS) of nuclear matter has been a major focus of heavy-ion collision experiments. While very low energy collisions can probe nuclear matter at densities smaller than the saturation density $n_0$, highly-compressed nuclear matter is produced in the laboratory by colliding heavy nuclei at relativistic velocities. At even higher energies, in the ultra-relativistic regime, quarks in the colliding nuclei become almost transparent to each other and therefore escape the collision region, which means that matter measured at midrapidity is characterized by a nearly-zero net baryon number. Heavy-ion collision experiments at top beam energies at the Relativistic Heavy Ion Collider (RHIC) and Large Hadron Collider (LHC) provided convincing evidence that at high temperatures and near-zero baryon density, nuclear matter becomes a quark-gluon plasma (QGP) \cite{Heinz:2000bk,BRAHMS:2004adc,PHENIX:2004vcz,PHOBOS:2004zne,STAR:2005gfr,Muller:2012zq,Pratt:2015zsa}, a deconfined but strongly-interacting state composed of color charges, confirming Lattice QCD (LQCD) calculations of the EOS at zero density~\cite{HotQCD:2018pds,Aoki:2006br,Borsanyi:2020fev}.

While the region of the QCD phase diagram explored in ultra-relativistic heavy-ion collisions is relatively well understood, the EOS of dense nuclear matter at moderate-to-high temperatures and moderate-to-high baryon densities is not known well due to the break-down of first-principle approaches in this regime. Answering pressing questions about the QCD EOS in this region, such as whether the quark-hadron transition becomes of first-order at high densities or what is the minimal energy required to produce the QGP, is the driving force behind Phase II of the Beam Energy Scan (BES) program at RHIC, the HADES experiment at GSI, and the future Compressed Baryonic Matter (CBM) experiment at the Facility for Antiproton and Ion Research (FAIR), Germany.

This renewed interest in the nuclear matter EOS at high densities, accessible in heavy-ion collisions at intermediate energies, coincides with an increased effort to constrain the EOS of neutron-rich matter, probed in studies of neutron stars and neutron star mergers (see Section~\ref{sec:neutron_star_theory} as well as recent white papers on \textit{QCD Phase Structure and Interactions at High Baryon Density: Continuation of BES Physics Program with CBM at FAIR}~\cite{Almaalol:2022xwv} and \textit{Dense matter theory for heavy-ion collisions and neutron stars}~\cite{Lovato:2022vgq}). Moreover, studies show that heavy-ion collisions in this regime and neutron star mergers probe similar temperatures and baryon densities \cite{Hanauske:2019qgs, Most:2022wgo}. However, while matter created in collisions of heavy-ions has comparable numbers of protons and neutrons, matter inside neutron stars is neutron-rich. Establishing the much needed connection between the studies of the nuclear EOS as probed in heavy-ion collisions and as inferred from neutron star observations is possible by leveraging the experimental capabilities of the newly commissioned Facility for Rare Ion Beams (FRIB), where energetic beams of proton- and neutron-rich nuclei can be produced. Heavy-ion collision experiments at FRIB can put tight constraints on the dependence of the nuclear matter EOS on the relative proton and neutron abundances \cite{FRIB400}, and thus enable a description of both dense nuclear and dense neutron-rich matter within a unified framework.

Indeed, if we assume that the core of a neutron star is composed of mostly uniform nucleonic matter, then nuclear matter and neutron stars should be described by a common EOS, specifying the relationship between the pressure and the temperature, density, and isospin content. The theoretical construct of symmetric nuclear matter consisting of equal amounts of neutrons and protons has been successful to derive properties of symmetric matter such as the saturation density and binding energy, however, an additional term in the EOS is needed to describe nuclear matter with unequal neutron-proton composition. This second term depends on the asymmetry~$\delta$, defined as $\delta = (n_{n} - n_{p})/n_B$, where $n_n$, $n_p$, and $n_B$ are the neutron, proton, and total baryon densities, respectively. Consequently, one can view the asymmetry as the neutron excess fraction. 
Mathematically, the energy per nucleon can be then expressed as a sum of two terms: $\epsilon(n_n,n_p) =\epsilon_{\rm{SNM}}(n) + S(n) \delta^2$. Here, the first term represents the energy per nucleon of symmetric nuclear matter, while 
\begin{wrapfigure}{r}{0.6\textwidth}
  \centering
  \vspace{-1mm}
    \includegraphics[width=0.59\textwidth]{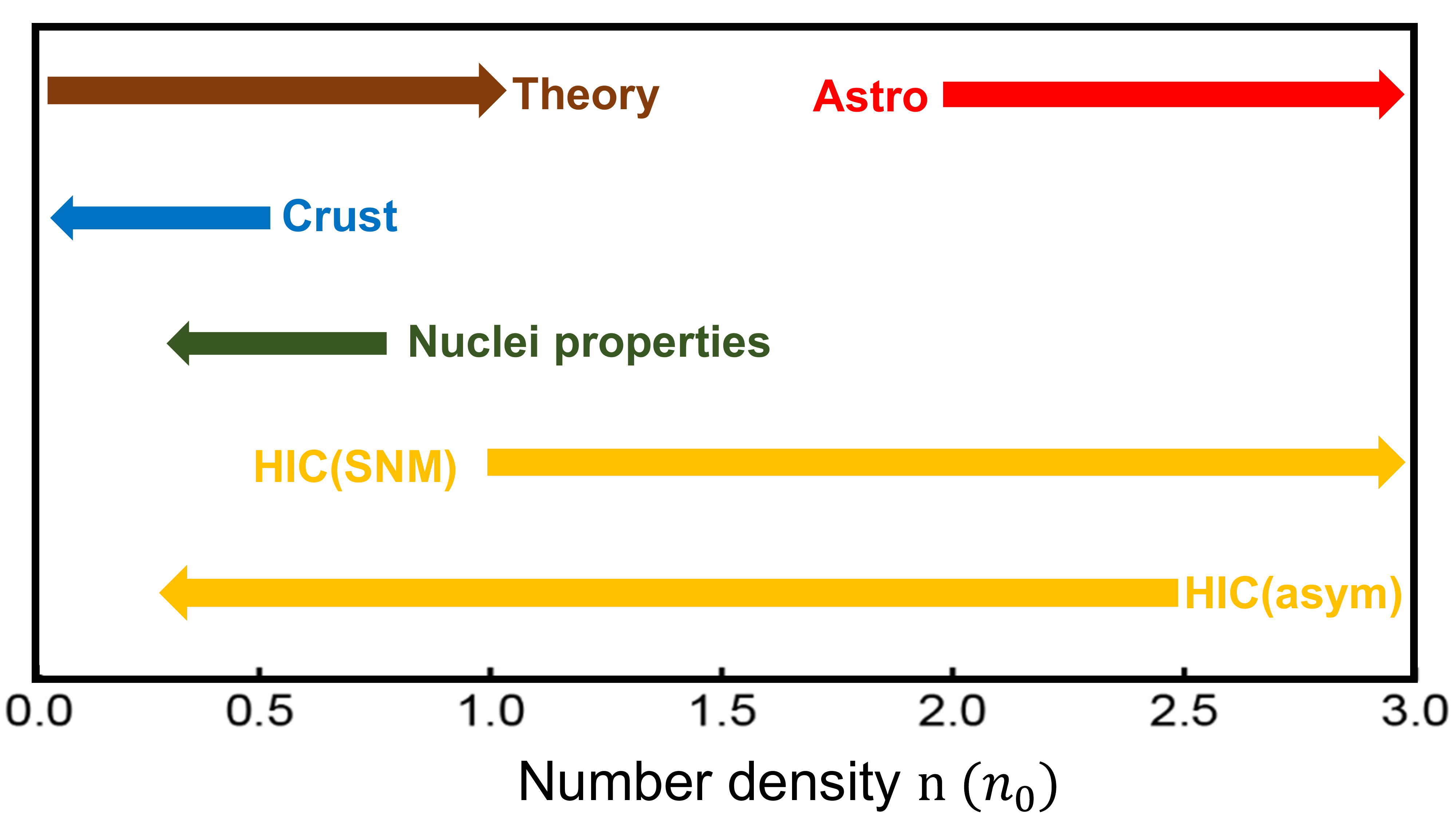}
    \vspace{-2mm}
  \caption{This schematic plot illustrates the approximate density ranges that are explored in the studies of chiral effective field theory, nuclei properties, heavy-ion collision experiments, and observations of neutron stars and their crusts in astronomy.
  }
  \vspace{-4mm}
  \label{fig:constraints}
\end{wrapfigure}
the second term accounts for the correction needed when $\delta \neq 0$. Therefore, $\delta$ is a crucial parameter that distinguishes neutron stars (with $\delta \gtrsim 0.8$) from most nuclei (with $\delta \lesssim 0.25$). Given the relatively small values of the asymmetry $\delta$ for nuclei, in heavy-ion collision 
experiments it is easier to constrain the coefficients of the EOS of symmetric matter, $\epsilon_{\rm{SNM}}(n_B)$. In contrast, the energy contribution from the asymmetric term, also known as the symmetry energy, constitutes a small fraction of the total energy of a nucleus even for neutron-rich heavy radioactive isotopes ($ < 5\%$ in the liquid drop model), and its determination requires precise measurements. Furthermore, because the isospin effects in any observable tend to diminish with temperature, it may be difficult to measure the symmetry energy at very high densities, which require high-energy heavy-ion reactions. Therefore, symmetry energy is best probed in heavy-ion collisions of highly asymmetric isotopes at low to intermediate energies.

Fig.\ \ref{fig:constraints} shows schematically the baryon density regions explored by different areas in nuclear physics studies. Recent breakthroughs in astronomical observations with state-of-the-art instruments led to the first detection of a binary neutron-star merger and the unprecedented radii measurements of neutron stars with accurately known masses (see Section \ref{sec:neutron_star_theory}). The neutron star mass-radius relationship provides an insight into the EOS at high densities above twice saturation density ($\gtrsim 2n_0$), as represented by the red arrow (labeled ``Astro'') in the upper right corner. Laboratory experiments, especially those using heavy-ion collisions, are essential to provide information on the dependence of the EOS on density and the asymmetry (see also Section \ref{sec:model_simulations_of_HICs}). High-energy heavy-ion collisions can provide insight into the symmetric nuclear matter EOS as represented by the gold right-pointing arrow (labeled ``HIC(SNM)''), while current probes of the symmetry energy are more suited for measurements of lower energy heavy-ion reactions ($\lesssim 600\ A\rm{MeV}$) as represented by the left-pointing gold arrow (labeled ``HIC(asym)''). Many properties of nuclei, such as masses and radii, have been shown to be mainly sensitive to densities around $(2/3)n_0$, however, with a careful selection of nuclear observables, the symmetry energy has been probed over densities of $0.3n_0<n_B<n_0$ using Pearson correlation methods \cite{Danielewicz:2016bgb, Zhang:2015ava} (green left-pointing arrow). Recent advances in chiral effective field theory (see Section \ref{sec:microscopic_calculations_of_the_EOS}) enabled extrapolations of the EOS to be extended up to $\approx 1.5n_0$ \cite{Drischler:2020hwi}, but the uncertainty increases exponentially with density for densities that are higher than $n_0$. It is not clear what is the maximum density up to which such extrapolations can succeed. Finally, one of the most interesting regions is at very low densities ($\lesssim 0.5n_0$), corresponding to the crust of a neutron star where matter is not uniform (see Section~\ref{sec:neutron_star_theory}). There, matter changes with increasing density from a Coulomb-dominated lattice to nuclear pasta and, ultimately, to uniform matter. The density and nature of these transformations are again dictated largely by the EOS. 

Measurements made in heavy-ion collisions at intermediate energies, probing high densities or, equivalently, small nucleon separations, will yield key insights into the nature of the nuclear force, including the density-dependence of the nuclear symmetry energy. Experimental efforts to determine the EOS for symmetric matter and the symmetry energy are described in Section \ref{sec:experiment_snm} and \ref{sec:experiment_asym}, respectively. 

Please note that all beam energies $\ekin$ quoted in this section are the single-beam kinetic energies per nucleon, in units of $A\rm{MeV}$ or $A\rm{GeV}$. (Alternatively, $\ekin$ is also sometimes denoted by other authors as $E/A$, with units of MeV or GeV). Additionally, while many results are reported in terms of their constraints on the incompressibility $K_0$, one should refrain from interpreting them as constraining the behavior of the EOS around the saturation density (see Section \ref{sec:selected_constraints} for more details).

\subsection{Experiments to extract the EOS of symmetric nuclear matter}
\label{sec:experiment_snm}

Heavy-ion collision experiments worldwide have extensively studied the EOS of symmetric nuclear matter at supra-saturation densities over the past four decades. Experiments based at the Schwerionensynchrotron-18 (SIS-18) ring accelerator at the GSI Helmholtz Centre for Heavy Ion Research (GSI) have probed Au+Au collisions at energies between $\ekin = 0.09$--$1.5\ A\rm{GeV}$ ($\snn = 1.92$--$2.52\ \rm{GeV}$), corresponding to fireball densities $1$--$2.5n_0$. Further experimental efforts with Au+Au collisions were carried out at higher energies, $\ekin = 2$--$10\ A\rm{GeV}$ ($\snn = 2.70$--$4.72\ \rm{GeV}$), at the Alternating Gradient Synchrotron (AGS) at the Brookhaven National Laboratory (BNL) to probe fireball densities $2.5$--$5n_0$. Complementing the densities reached at AGS-BNL is the Beam Energy Scan (BES) program of the Solenoidal Tracker at RHIC (STAR) experiment at RHIC in BNL, where high-statistics Au+Au collisions were performed at energies between $\ekin = 2.9$--$30.0\ A\rm{GeV}$ ($\snn = 3$--$7.7\ \rm{GeV}$) in the fixed-target mode. A selection of constraints on the EOS extracted from the above experiments is shown in Fig.~\ref{fig:joint_constraints_EOS}. Below, we describe the observables studied to extract the symmetric nuclear matter EOS, experiments probing the aforementioned density ranges, and inferences for the hadronic transport codes.

\subsubsection{Measurements sensitive to the EOS}

Collisions of heavy nuclei at relativistic energies lead to a rapid compression and heating of matter trapped in the collision region, followed by its dynamic expansion and cooling (see Fig.~\ref{fig:Collision}). The EOS governs both the compression as well as the expansion of the hot and dense nuclear matter, which in turn affect measured particle distributions. For example, a stiffer EOS (characterizing matter that is more incompressible) leads to a relatively smaller compression and, consequently, smaller heating, but a faster transverse expansion. The smaller temperatures reached in the fireball lead to smaller thermal dilepton and photon yields (see, e.g., \cite{Huovinen:1998tq,Rapp:2014hha,Seck:2020qbx,Savchuk:2022aev}), while the faster expansion manifests itself in relatively higher mean transverse momenta (see, e.g., \cite{Steinheimer:2022gqb}) and a shorter lifetime of the fireball, the latter of which can be probed by measuring femtoscopic correlations \cite{Bertsch:1989vn,Pratt:1990zq,Li:2022iil}.

The EOS also plays a large role in the interplay between the initial geometry of the system, the expansion of matter originating from nucleons trapped in the collision zone (participants), and the propagation of nucleons which are either still incoming into the collision region or whose trajectories do not directly cross the collision region (spectators). In systems colliding at beam energies for which the speed of the fireball expansion is comparable with the speed of the spectators, the resulting complex dynamical evolution affects the transverse expansion of the system and, therefore, the angular particle distributions in the transverse plane $dN/d\phi$. In particular, moments of the angular momentum distribution, known as the collective flow coefficients and defined as $v_n = \int d\phi ~ \cos (n\phi) ~ (dN/d\phi)  /  \int d\phi ~  (dN/d\phi) $ (see, e.g., \cite{Poskanzer:1998yz,Bilandzic:2010jr}), reflect the collective motion of the system and are highly sensitive to the EOS, as shown in numerous hydrodynamic~\cite{Stoecker:1980vf,Ollitrault:1992bk,Rischke:1995pe,Stoecker:2004qu,Brachmann:1999xt,Csernai:1999nf,Ivanov:2014ioa} and hadronic transport \cite{Hartnack:1994ce,Li:1998ze,Danielewicz:2002pu,LeFevre:2015paj,Wang:2018hsw,Nara:2021fuu} models. At the same time, flow observables can be measured with high precision, making them primary observables used to constrain the EOS.

In off-central collisions, the initial collision zone has an approximately elliptical shape, and the pressure gradients within the collision zone are larger along its short axis. If the spectator nucleons move out of the way before the fireball expands, the pressure gradients in the collision zone lead to particle distributions around midrapidity which have maxima coincident with the reaction plane (``in-plane'' emission). If, however, the spectators stand in the way of the fireball expansion, this leads to a preferential emission along the long axis of the collision zone (``out-of-plane'' emission, also referred to as ``squeeze-out'' due to the role that the spectators play in the expansion). The preferential emission in either in-plane or out-of-plane direction is described by the second Fourier coefficient of flow $v_2$, also known as the elliptic flow, which is positive in the former case and negative in the latter case (see the lower panel of Fig.~\ref{fig:v1_v2_world_data}). The magnitude of the elliptic flow, as well as the energy at which $v_2$ changes sign, are intrinsically connected to the stiffness of the EOS: for example, a stiffer EOS results in both a faster expansion and a more forceful blocking by spectators, which leads to a larger squeeze-out and a more negative~$v_2$.

The rapidity-dependence of the first Fourier coefficient of flow, the directed flow $v_1$, is also sensitive to the EOS as it measures the degree of spectator deflection due to the interaction with the collision zone \cite{Voloshin:2008dg}. In the center-of-mass frame, the spectators from a nucleus moving in the positive beam direction will be deflected to one side, while the spectators from the other nucleus, moving in the negative beam direction, will be deflected to the opposite side, resulting in a positive $v_1$ at positive rapidities and a negative $v_2$ at negative rapidities (here, the sign of $v_1$ is a matter of convention; see \cite{Oliinychenko:2022uvy} for a more detailed explanation). The magnitude of the directed flow in each region and, therefore, its slope at midrapidity are directly related to the EOS: for example, a softer EOS leads to a smaller deflection and a smaller slope of $v_1$ at midrapidity, where in particular a sufficiently soft EOS can even lead to a negative slope of $v_1$ \cite{Stoecker:2004qu,Zhang:2018wlk}. We note that spectators are necessary to obtain substantial magnitudes of the slope of the directed flow, as can be seen by its small values at high collision energies (see the upper panel of Fig.~\ref{fig:v1_v2_world_data}).

Beyond the collective flow phenomena, the EOS also has an effect on hadron production. In particular, much attention has been given to production of hadrons in heavy-ion collisions at energies below the nominal production threshold in $NN$ reactions (“sub-threshold” production), which requires multiple sequential hadron-hadron collisions to occur. The probability of these collisions is significantly higher in high-density regions, and consequently the yield of sub-threshold probes is expected to be substantially enhanced if higher densities are reached in the collision. Of particular importance for the EOS studies is sub-threshold production of $K^+$ mesons, which undergo few final-state interactions with the nuclear medium and therefore mostly leave the fireball unperturbed, making them a sensitive probe of the highest densities reached and, consequently, of the nuclear EOS~\cite{Aichelin:1985rbt}.

Similarly, two-particle correlations at small-relative momentum have shown strong sensitivity to the EOS. Correlations due to final-state interactions are stronger when particles are emitted closer together in space or time, as is the case for stiffer EOSs for which collisions are more explosive. Information about the space-time extent of the system can be obtained through measurements of femtoscopic radii $R_{\txt{long}}$, $R_{\txt{out}}$, $R_{\txt{side}}$ \cite{Heinz:1999rw}, where in particular the combination $R_{\txt{out}}^2 - R_{\txt{side}}^2 $ has been shown to be proportional to the duration of particle emission \cite{Bertsch:1989vn,Pratt:1990zq}. Such femtoscopic correlations played a central role in the Bayesian analysis of high-energy measurements which constrained the EOS at small baryon densities~\cite{Sangaline:2015isa,Pratt:2015zsa}. Recent calculations have demonstrated that the same ideas can constrain the EOS of baryon-dense matter in intermediate-energy collisions~\cite{Li:2022iil}.

\subsubsection{Experiments probing densities between 1--2.5\texorpdfstring{$n_0$}{TEXT}}
\label{snm-low}

\begin{figure*}[!t]

\includegraphics[width=0.49\linewidth]{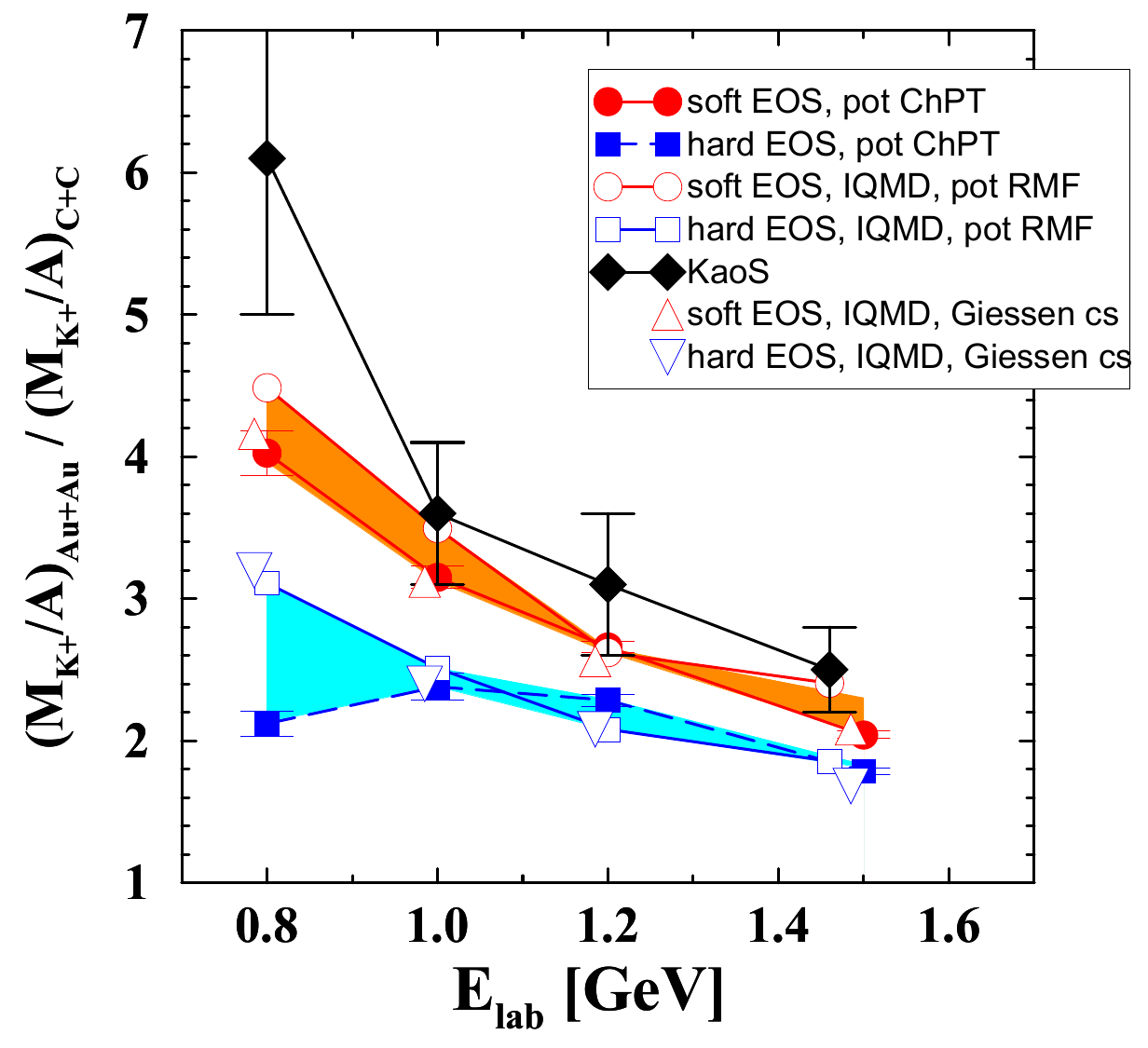}
\includegraphics[width=0.49\linewidth]{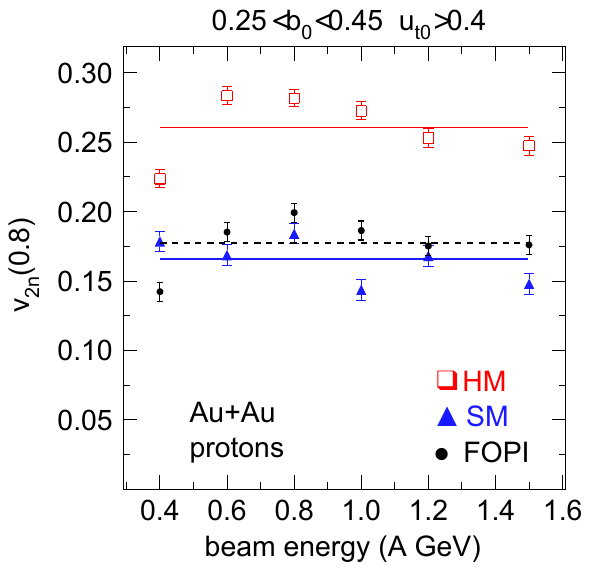}
\caption{\textit{Left panel:} Beam energy dependence of $K^+$ yield ratios in inclusive Au+Au collisions and C+C collisions between $\ekin = 0.8$--$1.5\ A\rm{GeV}$ ($\snn = 2.24$--$2.52\ \rm{GeV}$). A comparison of \texttt{RQMD} \cite{Fuchs:2000kp} and \texttt{IQMD}~\cite{Hartnack:2005tr} model calculations indicates a soft EOS ($K_0 = 200\ \rm{MeV}$, red symbols) instead of a hard EOS ($K_0 = 380\ \rm{MeV}$, blue symbols) when compared to KaoS data \cite{KAOS:2000ekm} (black symbols). Figure from \cite{Fuchs:2005zg}. \textit{Right panel:} Beam-energy dependence of the elliptic flow for protons in Au+Au collisions at $\ekin = 0.4\ {A}\rm{GeV}$ ($\snn = 2.07\ \rm{GeV}$) (black symbols) as measured by the FOPI experiment \cite{FOPI:2011aa}. Comparison to \texttt{IQMD} transport calculations with momentum dependence prefers a soft EOS (blue triangles) over a hard EOS (red squares), yielding $K_{0} = 190 \pm 30\ \rm{MeV}$. Figure from \cite{LeFevre:2015paj}.}
\label{fig:KaoS-FOPI}
\end{figure*}

As described above, sub-threshold particle yields can be used as probes of the EOS. In particular, due to their low in-medium cross-section, $K^+$ mesons produced at energies lower than the production threshold of $\ekin = 1.58\ \txt{GeV}$ ($\snn = 2.55\ \rm{GeV}$) can carry unperturbed information on the fireball density and the stiffness of the EOS \cite{Fuchs:2005zg}. 
The Kaon Spectrometer (KaoS) Experiment~\cite{Senger:2022ihi} at SIS18 in GSI studied the subthreshold production of $K^+$ mesons at beam energies between $\ekin = 0.6$--$2.0\ A\rm{GeV}$ ($\snn = 2.16$--$2.70\ \rm{GeV}$), and established it as a sensitive probe to the underlying EOS of the hot and dense nuclear matter. To reduce the experimental and model uncertainties, the production of $K^+$ mesons in a heavier Au+Au system was compared with the production in a lighter C+C system \cite{KAOS:2000ekm}. Analyzing the experimental results together with transport model calculations in the \texttt{RQMD} \cite{Fuchs:2000kp} and \texttt{IQMD} \cite{Hartnack:2005tr} model enabled extraction of the EOS of symmetric nuclear matter characterized by an incompressibility of $K_{0} = 200\ \txt{MeV}$ (see also Fig.\ \ref{fig:joint_constraints_EOS}). Both models included effects due to the momentum-dependence of the EOS by including $K^{+/-}N$ potentials, i.e., a repulsive mean field for $K^+$ and an attractive mean field for $K^-$, which are required to reproduce the $K^+$ and $K^-$ emission pattern~\cite{Uhlig:2004ue} (see Fig.\ \ref{fig:KaoS-FOPI}).

Collective behavior in heavy-ion collisions is likewise a very sensitive probe of the underlying EOS and has been extensively studied since its discovery by the Plastic Ball spectrometer at the Bevalac in Lawrence Berkeley National Laboratory \cite{Gustafsson:1984ka, Gutbrod:1988hh}. In particular, the elliptic flow $v_2$ is highly sensitive to both the initial geometry of the collisions and pressure gradients experienced throughout the evolution of the created systems \cite{Stoecker:1980vf, Stoecker:1986ci}. The Four Pi (FOPI) Experiment at SIS18 in GSI carried out extensive measurements of the beam energy dependence of the elliptic flow of protons and light fragments (such as deuterons, tritons, and $^3$He) over the entire range of SIS18 energies, $\ekin = 0.09$--$1.5\ {A}\rm{GeV}$ ($\snn = 1.92$--$2.52\ \rm{GeV}$) \cite{FOPI:2004bfz,FOPI:2011aa}. The nuclear EOS extracted from a comparison to \texttt{IQMD} simulations \cite{LeFevre:2015paj} is characterized by an incompressibility $K_0=190\pm30 \ \rm{MeV}$ when momentum-dependent interactions are taken into consideration. This constraint is consistent with the KaoS incompressibility inferences and suggests a soft EOS for symmetric nuclear matter at $1$-$2.5n_0$ (see Fig.~\ref{fig:KaoS-FOPI} and also Fig.\ \ref{fig:joint_constraints_EOS}).

\subsubsection{Experiments probing densities above 2.5\texorpdfstring{$n_0$}{TEXT}}
\label{snm-high}

Pioneering proton directed and elliptic flow measurements were performed in Au+Au collisions for beam energies $\ekin = 2$--$10\ A\rm{GeV}$ ($\snn = 2.70$--$4.72\ \rm{GeV}$) by the E895~\cite{E895:2000maf, E895:1999ldn} and E877~\cite{E877:1997zjw} experiments at AGS-BNL. Notably, it was observed that around $\ekin \approx 4\ A\rm{GeV}$ ($\snn \approx 3.3\ \rm{GeV}$), the proton $v_2$ changes from a preferential out-of-plane emission, reflecting a complex interplay between the spectators, the expanding collision zone, and the EOS, to an in-plane emission (see the lower panel of Fig.~\ref{fig:v1_v2_world_data}). The experimental results were used in a comparison with the \texttt{pBUU} transport model to extract the EOS for densities between $2$--$5n_0$, which constrained the EOS to those described by values of the nuclear incompressibility between $K_0 = 210$--$300\ \rm{MeV}$, ruling out extremely soft and extremely hard EOSs~\cite{Danielewicz:2002pu} (see Fig.\ \ref{fig:joint_constraints_EOS}). This rather broad constraint on $K_0$ reflects the fact that the experimental results for the collective flow could not be reproduced with one EOS.

\begin{figure*}[!t]
\includegraphics[width=0.99\linewidth]{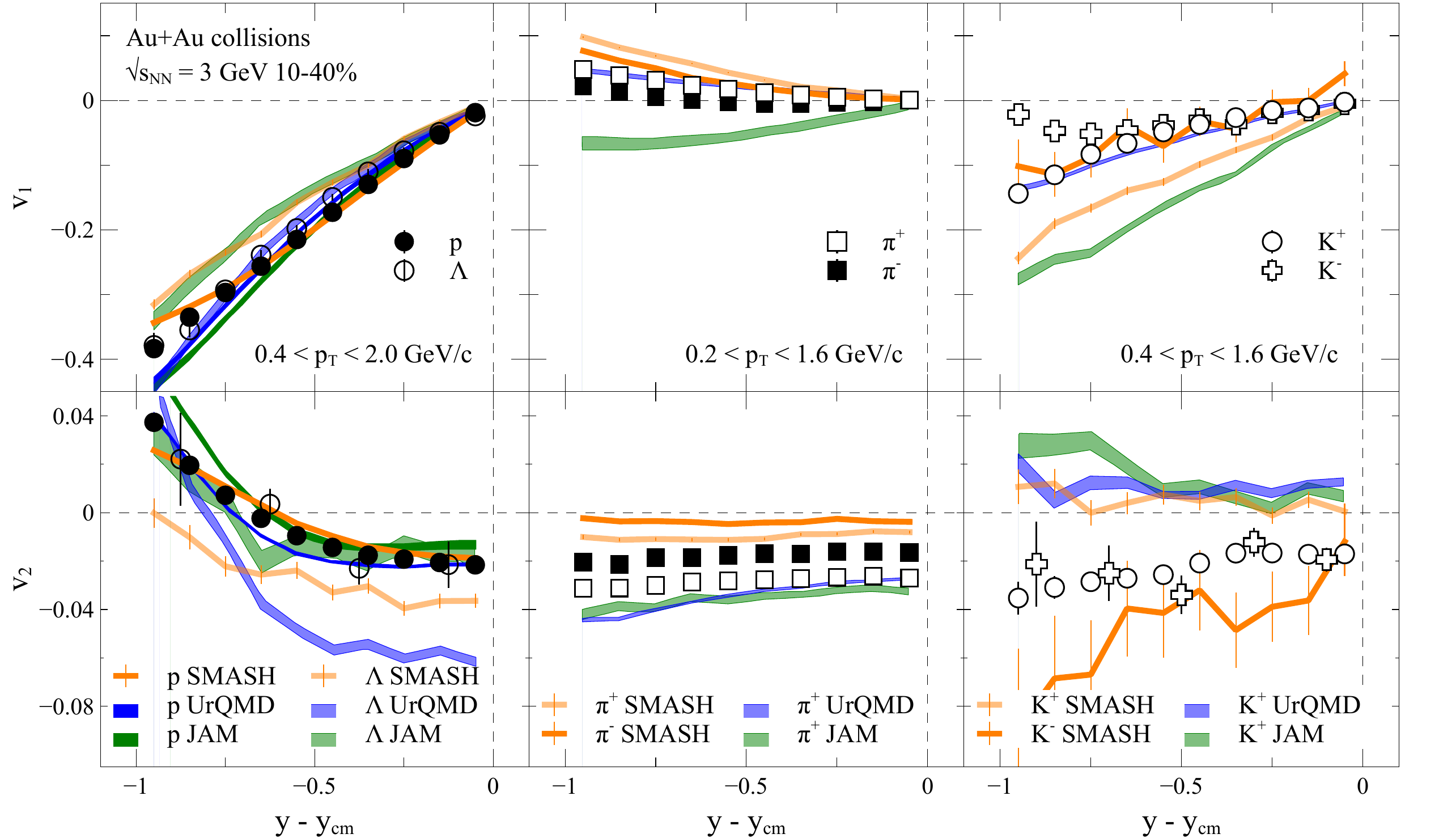}
\caption{Directed ($v_1$, \textit{top}) and elliptic ($v_2$, \textit{bottom}) flow of protons and lambda baryons (\textit{left panels}), pions (\textit{middle panels}), and kaons (\textit{right panels}) as a function of rapidity. Measurements from STAR \cite{STAR:2021yiu} (symbols) were performed with Au+Au collisions at $\snn=3\ \rm{GeV}$ ($\ekin = 2.9\ A\rm{GeV}$) and 10–40\% centrality. Results from \texttt{UrQMD} (blue bands), \texttt{JAM} (green bands), and \texttt{SMASH} (orange bands) hadronic transport models were obtained using a relatively hard EOS at moderate densities (characterized by $K_0 = 300$ in \texttt{SMASH} and $K_0 = 380\ \rm{MeV}$ in \texttt{JAM} and \texttt{UrQMD}), with the EOS used in \texttt{SMASH} becoming significantly softer at high densities (see \cite{STAR:2021yiu, Oliinychenko:2022uvy} for more details).}
\label{fig:STAR}
\end{figure*}

The STAR Experiment at RHIC-BNL with its Beam Energy Scan (BES) program \cite{BES_II_WhitePaper, Bzdak:2019pkr} performed Au+Au collisions for $\snn=3$--$200\ \rm{GeV}$. In terms of the freeze-out temperature and chemical potential, $(T_{\rm{fo}},~\mu_{\rm{fo}})$, this allowed STAR to comprehensively scan the QCD phase diagram from (80, 760)~MeV to (166, 25)~MeV, respectively. Probing the phase diagram at high densities was possible at RHIC in part due to STAR's capability to shift from a standard collider to a fixed-target (FXT) mode, which was used to scan through the lower energies $\snn=3$--$13.7\ \rm{GeV}$ ($\ekin = 2.9$--$98.0\ A\rm{GeV}$), thereby establishing a substantial overlap with the previously discussed AGS experiments \cite{Tribedy_QM22}. Recently, STAR measured collective flow ($v_1, v_2$) in collisions at $\snn = 3.0\ \txt{GeV}$~\cite{STAR:2021yiu} and $\snn = 4.5\ \rm{GeV}$ \cite{STAR:2020dav}. A comparison of results from the $\snn = 3.0\ \txt{GeV}$ data (see Fig.\ \ref{fig:STAR}) with \texttt{UrQMD} and \texttt{JAM} simulations indicates a relatively hard EOS (characterized by $K_{0} = 380\ \rm{MeV}$) \cite{STAR:2021yiu}; similarly, a recent Bayesian analysis of the STAR flow data based on a flexible parametrization of the EOS used in the \texttt{SMASH} transport code results in a relatively hard EOS at moderate densities (characterized by $K_0 = 300 \pm 60\ \rm{MeV}$) with a substantial softening at higher densities \cite{Oliinychenko:2022uvy}. However, both \texttt{UrQMD} and \texttt{SMASH} do not currently include momentum-dependent interactions, which are crucial for a correct description of the transverse-momentum-dependence of the elliptic flow \cite{Danielewicz:2002pu}. Moreover, while the above models reproduce the proton $v_1, v_2$ well, none of the models can simultaneously describe the flow of Lambda baryons and mesons (see Fig.~\ref{fig:STAR}).

\subsubsection{Challenges and opportunities}
\label{snm-challenges}

\textbf{Experiments probing densities between 1--2.5\texorpdfstring{$n_0$}{TEXT}}

\begin{figure*}[!b]
\includegraphics[width=0.40\linewidth]{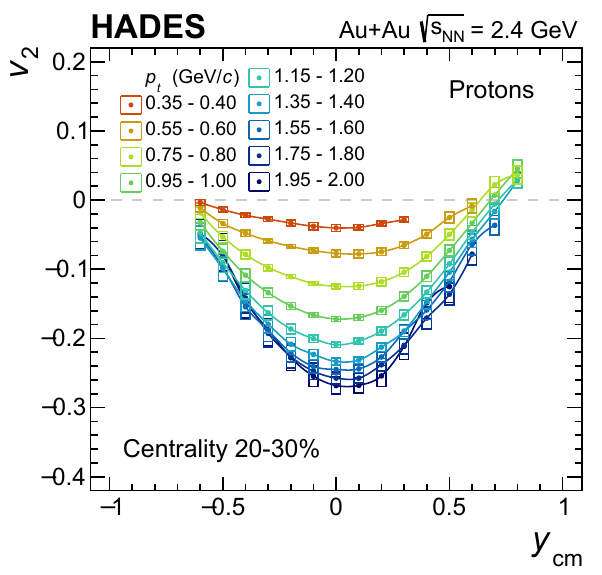}
\includegraphics[width=0.55\linewidth]{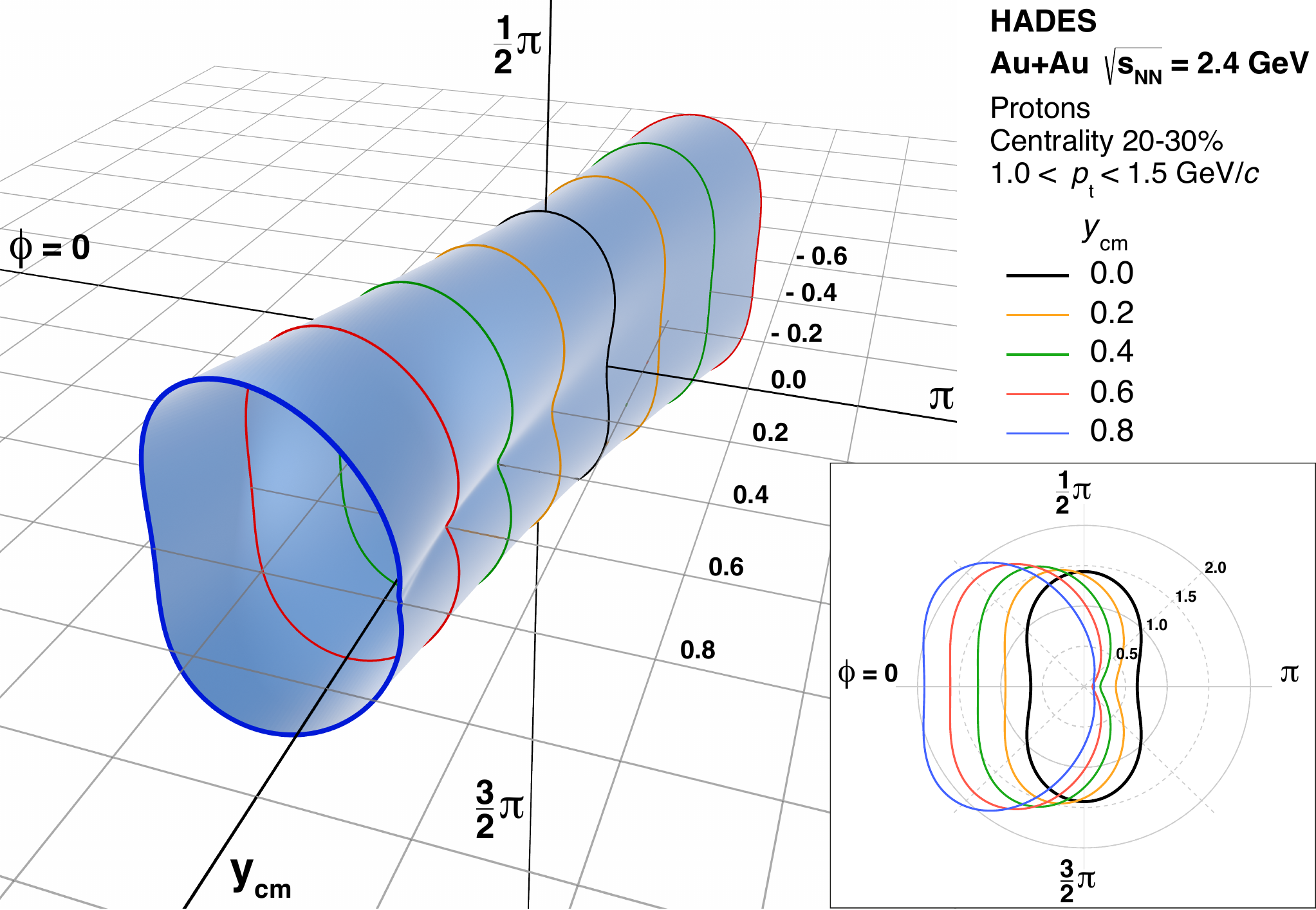}
\caption{\textit{Left:} Rapidity-dependence of proton elliptic flow ($v_2$) in semi-central Au+Au collisions at $\ekin = 1.23\  A\rm{GeV}$ ($\snn = 2.42\ \txt{GeV}$) for various $p_T$ bins (see legend) as measured by the HADES experiment. Figure from \cite{HADES:2022osk}. \textit{Right:} 3D-picture of the proton angular emission pattern in momentum space (flow coefficients from $v_1$ to $v_6$) for HADES data in semi-central Au+Au collisions at $\ekin = 1.23\  A\rm{GeV}$ ($\snn = 2.42\ \txt{GeV}$). Figure from \cite{HADES:2020lob}.}
\label{fig:HADES}
\end{figure*}

The High Acceptance Di-Electron Spectrometer (HADES) Experiment \cite{HADES:2009aat} at SIS-18 in GSI has performed collective flow measurements in Au+Au collisions at $\ekin = 1.23\ A\rm{GeV}$ ($\snn = 2.42\ \txt{GeV}$). The high acceptance and high statistics of HADES measurements allow one to perform multi-differential studies of flow harmonics, ranging from $v_1$ up to $v_6$, which in turn enables reconstruction of a full 3D-picture of the emission pattern in the momentum space \cite{HADES:2020lob, HADES:2022osk} (see Fig.~\ref{fig:HADES}). In addition to the collective flow measurement capabilities, HADES can also precisely measure the dielectron excess yield, which was used to extract the fireball temperature, finding it to be 71.8 $\pm$ 2.1 MeV/$k_B$ \cite{HADES:2019auv}. These precise measurements of the fireball temperature and the underlying dielectron spectra allow HADES to investigate the presence of a first-order phase transition at SIS-18 energies and look for signs of a potential change of degrees of freedom \cite{Seck:2020qbx}.

During its 2024 beam campaign, HADES will be in a unique position to measure the fireball caloric curve and the beam energy dependence of the collective flow from Au+Au collisions at $\ekin = 0.4$--$0.8\ A\rm{GeV}$ ($\snn = 2.07$--$2.24\ \rm{GeV}$). Furthermore, there are ongoing efforts to establish systematic consistency between results from FOPI and HADES, including understanding various detector-related effects. This is highlighted by the observed discrepancy in pion multiplicities between FOPI and HADES, which could be partially explained by different methods used by the respective experiments to estimate the number of participant nucleons \cite{HADES:2020ver}.

The abundance of available and future data presents an opportunity to benchmark transport model simulations with measurements from KaoS, FOPI, and HADES experiments by enabling systematic studies of the symmetric nuclear matter EOS. A recent comparison between FOPI measurements and \texttt{dcQMD} transport code \cite{Cozma:2021tfu}, using the transverse rapidity \cite{FOPI:2010xrt} and flow spectra~\cite{FOPI:2011aa} of protons and light clusters at $\ekin = 0.15$--$0.80\ A\rm{GeV}$ ($\snn = 1.95$--$2.24\ \rm{GeV}$), has further tightened the constraints on the nuclear EOS at the probed densities to one characterized by an incompressibility $K_0 = 236\pm6~\rm{MeV}$. The \texttt{dcQMD} analysis for the FOPI data is planned to be extended up to $\ekin = 1.5\ A\rm{GeV}$ ($\snn=2.52\ \rm{GeV}$), probing densities above $2n_0$, by taking into account an improved description of reaction dynamics through using more accurate approximations for 3-body terms in the interaction and considering multi-pion decay channels for the resonances~\cite{Cozma_INT22-84W}.

Moreover, perfect-fluid hydrodynamic calculations for binary-neutron-star mergers and heavy-ion collisions at SIS-18 energies show that comparable temperatures ($T \approx 50$~MeV) and densities ($n_B \approx 2n_0$) are reached in both systems~\cite{Hanauske:2019qgs, Most:2022wgo}. This has led to increasing efforts to use the existing constraints on the EOS of symmetric nuclear matter from KaoS and FOPI experiments in a multi-physics effort to constraint neutron star properties \cite{Huth:2021bsp, Ghosh:2021bvw, Ghosh:2022lam}. Such multi-physics constraints are discussed in detail in Section \ref{sec:combined_constraints}.

\vspace{5mm}
\textbf{Experiments probing densities above 2.5\texorpdfstring{$n_0$}{TEXT}}

\begin{figure*}[!t]
\includegraphics[width=0.49\linewidth]{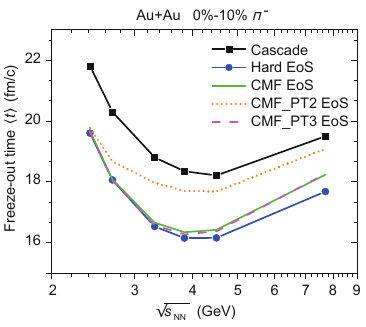}
\includegraphics[width=0.49\linewidth]{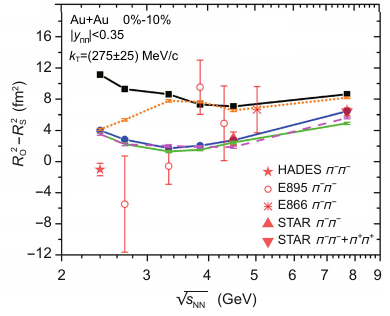}
\caption{\textit{Left:} Beam energy dependence of the $\pi^{-}$ freeze-out time as extracted from \texttt{UrQMD} simulations for different EOSs. The CMF$\_$PT2 and CMF$\_$PT3 EOS soften at low and high baryon density, respectively, by introducing a first-order phase transition, and the pure cascade mode can be considered as an extremely soft EOS. Figure from \cite{Li:2022iil}. \textit{Right:} Comparison of the beam energy dependence of $R_{\rm{out}}^2-R_{\rm{side}}^2$ for $\pi^{-}$ extracted from experiments and \texttt{UrQMD} simulations with different EOSs. Figure from \cite{Li:2022iil}.}
\label{fig:hbt}
\end{figure*}

While collective flow can be used to deduce the geometry of the colliding system and its properties in an indirect way (see Figs.~\ref{fig:STAR} and~\ref{fig:HADES}), a more direct method -- femtoscopy -- can provide a direct handle on the space-time evolution of the fireball \cite{Lisa:2005dd}. Here, the time of the particles' emission $\Delta \tau$ is also a probe of the underlying EOS, with larger values of $\Delta \tau$ corresponding to a softer EOS~\cite{Li:2022iil} (see Fig.\ \ref{fig:hbt}). Access to this information is provided by measurements of femtoscopic radii $R_{\rm{long}}$, $R_{\rm{out}}$, $R_{\rm{side}}$ \cite{Pratt:2008sz}, where the relation between $R_{\rm{out}}$ to $R_{\rm{side}}$ is strongly correlated with~$\Delta \tau$. The sensitivity of pion emission to the EOS has already been studied \cite{Li:2022iil, Batyuk:2017smw}, however, experimental uncertainties are still too big to make precise comparisons with model calculations. Ongoing studies of proton femtoscopy at STAR are expected to bring new, substantial references for such investigations of the EOS. 

In addition to studying the EOS of dense nuclear matter, the STAR BES program also aims to search for a potential first-order phase transition from hadronic to partonic phase at higher baryonic densities. This search can provide an input on collision energies at which hadronic transport models should take into consideration new degrees of freedom. Among the explored observables, number-of-constituent-quark (NCQ) scaling was used as an important evidence of creation of QGP at the highest RHIC energy of $\snn=200\ \rm{GeV}$ ($\ekin = 21,300.0\ A\rm{GeV}$)~\cite{STAR:2015gge}. Recent results point to the breaking of the NCQ scaling in Au+Au collisions at $\snn = 3\ \rm{GeV}$ ($\ekin = 2.9\ A\rm{GeV}$) \cite{STAR:2021yiu}. Other observables that can hint at the possible existence of a first-order phase transition include the thermodynamic susceptibilities of pressure, which are predicted to fluctuate in the vicinity of a critical point and manifest as a specific behavior of higher-order moments of conserved quantities (such as baryon number, strangeness, and electrical charge) with the beam energy \cite{Stephanov:2008qz, Stephanov:2011pb}. STAR~\cite{STAR:2020tga, STAR:2021iop} and HADES \cite{HADES:2020wpc} have observed tentative non-monotonic behavior in the beam-energy-dependence of the fourth-order net-proton cumulant (a proxy for the net-baryon number cumulant) in Au+Au collisions at $\snn = 7.7$--$27.0\ \rm{GeV}$ ($\ekin = 2.9$--$390\ A\rm{GeV}$).

The experimental effort to uncover the symmetric nuclear matter EOS will be further strengthened by the Compressed Baryonic Matter (CBM) experiment \cite{CBM:2016kpk, Friman:2011zz} at the currently-under-construction Facility for Antiproton and Ion Research (FAIR) in Darmstadt, Germany. The CBM experiment, which at the time of writing is expected to become operational in 2028-29, aims to use nucleus-nucleus collisions to precisely explore the QCD phase diagram with Au+Au collisions in the energy range of $\snn = 2.9$--$4.9\ \rm{GeV}$ ($\ekin = 2.6$--$10.9\ A\rm{GeV}$). Other particle beams, such as $Z = N$ species and protons can also be used at $\ekin = 15\  A\rm{GeV}$ ($\snn = 5.63\ \txt{GeV}$) and $\ekin = 30\  A\rm{GeV}$ ($\snn = 7.73\ \txt{GeV}$), respectively. This will be enabled by using primary heavy-ion beams from the Schwerionensynchrotron-100 (SIS-100) ring accelerator operating at an intensity of $10^9$~ions/s \cite{Durante:2019hzd}. CBM will operate at unprecedentedly high peak interaction rates of up to 10 MHz, which will be further complemented by a novel trigger-less data acquisition scheme and online event selection. This will allow CBM to perform systematic, multi-differential measurements of the dependence of observables on the beam energy and system size. The most promising observables to explore are: (i) event-by-event fluctuations, (ii) thermal radiation (photons and dileptons), (iii) (multi-)strangeness, (iv) hypernuclei, and (v) charm production (recent physics performance results can be found in \cite{Agarwal:2022ydl, CBM_NuPECC2024}). Moreover, the HADES Experiment will be moved to the SIS-100 beamline in the CBM experimental cave to complement the overarching CBM physics program in 2031 \cite{Gasik_KHuK2022}. The HADES detector, given its large polar angle acceptance ($18^{\circ}\leq\theta\leq85^{\circ}$), will perform reference measurements for CBM at lower SIS-100 energies. This will be done with light collision systems, e.g., proton beams and heavy-ion beams with moderate particle multiplicities (such as Ni+Ni or Ag+Ag collisions) \cite{CBM:2016kpk}. Altogether, CBM represents an opportunity to link the physics programs at SIS-18 and RHIC, thereby leading to a continuation of the Beam Energy Scan program (see also the white paper on \textit{QCD Phase Structure and Interactions at High Baryon Density: Continuation of BES Physics Program with CBM at FAIR}~\cite{Almaalol:2022xwv}).

Overall, STAR-FXT and CBM-FAIR are capable of performing high-statistics multi-differential measurements of the relevant EOS observables. However, a successful inference of the EOS depends on comparisons to transport simulations. Although many transport codes are available for describing heavy-ion collisions in different energy ranges and extracting the underlying EOS (see~\cite{TMEP:2022xjg} for a review), currently none of the available codes can reproduce all proposed experimental observables (see, e.g., Fig.\ \ref{fig:STAR}). A meaningful description of experimental data in the STAR-FXT and CBM-FAIR range will require transport codes to incorporate physics allowing reproducing all of the above-mentioned key measurements and more, see Section \ref{sec:model_simulations_of_HICs}.

\subsection{Experiments to extract the symmetry energy}
\label{sec:experiment_asym}

The energy contribution from the isospin dependence term, also known as the symmetry energy, is a small fraction of the total energy of a nucleus even for neutron-rich heavy radioactive isotopes ($ < 5\%$ in the liquid drop model). However, due to the large isospin asymmetry in neutron stars, the density dependence of the symmetry energy is very important, determining many neutron star properties, including their size and the cooling pathways \textit{via} neutrino emission.
While experimental inferences of the symmetry energy pose significant challenges, researchers have developed methods to elucidate the relatively small effects that the asymmetry has on isospin-dependent observables, e.g., by measuring ratios of neutron and proton observables or charged pion observables. Experimental as well as theoretical systematic errors are further minimized by taking double ratios of the same observable using two reactions that differ mainly in the neutron/proton content, as in the measurement of isoscaling. To reach the widest range of asymmetry between reactions, intense radioactive beams are necessary.
Large experiments designed to measure symmetry energy can require large collaborations. However, small-scale experiments can likewise have an impact on some of the outstanding problems. Consequently, many groups contribute to the diverse experimental results.

\subsubsection{Experiments that probe low densities}
\label{sec:experiments_at_very_low_densities}

\begin{wrapfigure}{l}{0.5\textwidth}
  \centering
  \vspace{-4.35mm}
    \includegraphics[width=0.5\textwidth]{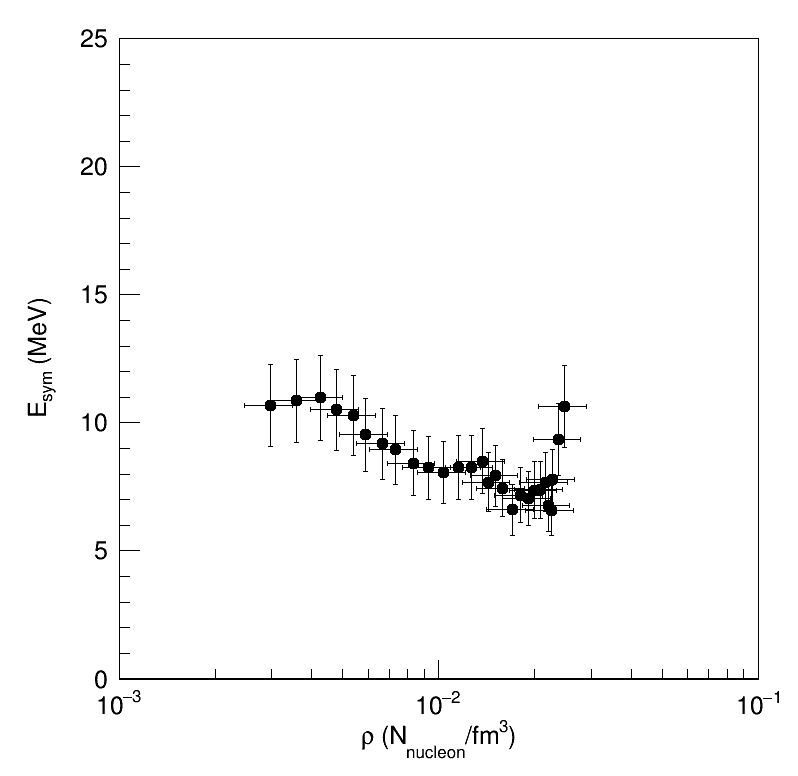}
    \vspace{-10mm}
  \caption{The symmetry energy of clustered matter at very low densities. Figure from ~\cite{Wada:2011qm}.
  }
  \vspace{-7mm}
  \label{fig:Wada2012}
\end{wrapfigure}

At beam energies below $\ekin = 100\ A\rm{MeV}$ ($\snn = 1.93\ \rm{GeV}$), the colliding nuclei overlap briefly and then expand, with most of the detected particles being emitted during the expansion stage. The rates of emission of neutrons and protons during the expansion are influenced by the symmetry energy. Some nucleons emerge within fragments or clusters that are 
formed and emitted throughout the reactions. Nearly all theory studies require the symmetry energy to be zero at zero density. 
However, before matter reaches zero density, at low densities of 
$0.002 \leq n/n_0 \leq 0.02$, many nucleons combine into clusters and preserve the information about the symmetry energy at those low densities. Following the work presented in Ref.~\cite{Natowitz:2010ti}, based on the EOS developed in Ref.~\cite{Typel:2009sy}, clustering is shown to have a significant impact on the symmetry energy below 0.03 nucleons/fm$^3$~\cite{Wada:2011qm,Hagel:2014wja}, see Fig.~\ref{fig:Wada2012} (we note here that this conclusion depends on the definition of the symmetry energy). Overall, the presence of clusters changes the characterization of the symmetry energy. Nonetheless, low-density clusterization is an important ingredient in supernova matter and for the EOS in the neutrino sphere. It is also relevant to the nature of proto-neutron star matter as it cools and the crust crystallizes~\cite{Pais:2015xoa}.

\subsubsection{Measurements to extract symmetry energy up to 1.5$n_0$}

In the past decade, many studies have been conducted to extract the symmetry energy and 
symmetry pressure \cite{Lynch:2021xkq}, focusing mostly at low densities. Since the nuclear EOS should give a 
good description of the properties of the nuclei, 
including the masses or binding energies, the large nuclear mass data base provides a great resource to determine the symmetry energy at about (2/3)$n_{0}$ from (1) masses of double magic nuclei using Skyrme density functional \cite{Brown:2013mga}, (2) nuclear masses using density functional theory  \cite{Kortelainen:2011ft}, and (3) the energies of isobaric analogue states \cite{Danielewicz:2016bgb}.

By nature, a heavy nucleus has excess neutrons which are needed to overcome the Coulomb repulsion from the protons inside the nucleus. The symmetry-energy forces the excess neutrons 
to the surface. This surface layer of excess neutrons is referred to as a neutron skin. Thickness of this neutron skin reflects the symmetry pressure, or equivalently the slope of the symmetry energy at the saturation density. The long-awaited measurements of the neutron skin of the $^{208}$Pb nucleus inferred from parity violation in electron scattering were recently published by the PREX collaboration \cite{PREX:2021umo,Reed:2021nqk}. The measured value of $0.283 \pm 0.071\ \rm{fm}$, corresponding to the symmetry pressure of $2.38\pm0.75$ MeV/fm$^{3}$ at $(2/3)n_0$, is rather large and disagrees with most theoretical predictions. Subsequently, the PREX/CREX collaboration measured the neutron skin of $^{48}$Ca to be $0.121 \pm 0.026\ \rm{(exp)} \pm 0.024\ \rm{(model)} \ \rm{fm} $ \cite{CREX:2022kgg}, which is much thinner than the $^{208}$Pb skin. However, the $^{48}$Ca value is much closer to the theoretical predictions. The discrepancies between the two results and the expectations from models have not been resolved, and the CREX collaboration has not released official values for the slope of the symmetry energy or symmetry pressure, even though there have been many attempts by outside groups to resolve the apparent discrepancies between the two skin measurements \cite{Reinhard:2022inh,Zhang:2022bni,Papakonstantinou:2022gkt}.

In the last few years, an alternative way to measure skin thickness has been proposed \cite{Brown:2017xxo}. In the limit of an exact charge symmetry, the proton radius of a given nucleus is identical to the neutron radius of its mirror partner. Thus the neutron skin for a given nucleus may be determined from the difference in proton radii measured in these mirror pairs. In reality, there are relativistic and finite-size corrections, as well as corrections from the Coulomb force which breaks the isospin symmetry. In principle, these corrections can be calculated within the energy density functional theory. While larger neutron skins are expected in heavier nuclei due to the larger neutron excess, making them a better probe, proton-rich mirrors of heavy nuclei is typically far beyond the limits of existence. Thus this technique is limited to species of relatively low mass and isospin. Even with the use of high-intensity isotope beams near the proton driplines, it is still a challenge to do such experiments. The most recent result with this technique is from the $^{54}$Ni-Fe mirror pair~\cite{Pineda:2021shy}.

Complementary to structure experiments, heavy-ion collisions have probed the symmetry energy and pressure over a wide density range. At incident energies below $\ekin = 100\ A\rm{MeV}$ ($\snn \leq 1.93\ \rm{GeV}$), low densities (estimated to be around $(1/3)n_{0}$) are reached when matter expands after the initial impact and compression of the projectile and target. Therefore, the corresponding experimental observables primarily reflect the symmetry energy at sub-saturation densities~\cite{Tsang:2008fd,Morfouace:2019jky}. 
The transport of neutrons and protons allows systems with isospin gradients to equilibrate, where the degree of equilibration depends on the strength of the potential experienced by the nucleons and the duration of transport. The technique of equilibration chronometry allows the visualization of the time evolution of the neutron excess. Signatures of neutron-proton equilibration obeying first-order kinetics are observed both in experimental data~\cite{McIntosh:2019tec,Jedele:2017xow,RodriguezManso:2017emd,Hannaman:2020trw} and in transport calculations ~\cite{Harvey:2020fvx}. Since the equilibration depends on the neutron and proton chemical potentials, this technique offers new experimental data to constrain the sub-saturation EOS through comparisons with simulations~\cite{Tsang:2004zz,Tsang:2008fd,TMEP:2022xjg}.

Isoscaling was first observed in central $^{124}$Sn+$^{124}$Sn and central $^{112}$Sn+$^{112}$Sn collisions at beam energy $\ekin = 50\ A\rm{MeV}$ ($\snn = 1.90\ \rm{GeV}$)~\cite{Tsang:2001dk, Tsang:2001jh}. Isoscaling describes a simple scaling law governing the ratios of isotope yields from two systems which differ mainly in their neutron-proton composition. It arises from the differences in the neutron and proton chemical potentials of the two reactions and is, therefore, sensitive to the symmetry energy.  The isospin diffusion, derived from the isoscaling observable, reflects the driving forces arising from the asymmetry term of the EOS~\cite{Tsang:2004zz,Liu:2006xs,Sun:2010km,Tsang:2008fd} and provides a measurement of the symmetry energy at around $(1/3)n_0$~\cite{Lynch:2021xkq}.

Other observables used to study the symmetry energy with light charged particles include both $n/p$ and $t$/$^{3}$He ratios and their double ratios obtained from two reactions with different isospin content \cite{Coupland:2011px,Zhang:2014sva,Li:2002qx,Li:2000bj,TMEP:2022xjg}. Due to the difficulties in measuring neutrons, neutron data is not widely available. However, recent isoscaling measurements have allowed the construction of “pseudo neutrons”, that is a reconstruction of neutron yields from light particle ratios such as $t$/$^3$He~\cite{Chajecki:2014vww}. In particular, this method allows for a reconstruction of low-energy neutrons. However, due to the lack of high-energy charged particles data, it is a challenge to reconstruct high-energy neutron spectra in this way. Therefore, to study the symmetry energy at supranormal densities, neutron arrays constructed with new advanced materials will be needed in the next generation of experiments.

In experiments utilizing central $^{124}$Sn+$^{124}$Sn and central $^{112}$Sn+$^{112}$Sn collisions at $\ekin=120\ A\rm{MeV}$ ($\snn = 1.94\ \rm{GeV}$)~\cite{Coupland:2014gya}, the spectra of neutrons emitted to 90 degrees in the center-of-mass frame are compared to the corresponding proton spectra.
Transport calculations predict that if the effective masses of neutrons and protons satisfy $m_n^* < m_p^*$, then fast neutrons coming from the compressed participant region experience a more repulsive potential and a higher acceleration than do fast protons at the same momentum, resulting in an enhanced ratio of neutron over proton ($n/p$) spectra at high energies. In contrast, calculations for $m_n^* > m_p^*$ predict that the effective masses enhance 
the acceleration of protons relative to neutrons, resulting in a lower $n/p$ spectral ratio. Bayesian analysis of the experimental results~\cite{Morfouace:2019jky} compared to \texttt{ImQMD} calculations shows that the values of the first two Taylor expansion coefficients of the symmetry energy, $S_0$ and $L$, depend on both the symmetry energy and to the effective mass splitting. More examples of Bayesian analyses used to simultaneously constrain multiple parameters will be discussed in Section \ref{sec:combining_different_constraints}, where methods to extract multiple transport model input parameters are discussed.

\subsubsection{Selected constraints on the symmetry energy around 1.5$n_0$}
\label{selected_constraints_experiment}

Current constraints on the symmetry energy above saturation are obtained with large uncertainties, mainly at densities around $1.5n_0$. This is the area of future opportunities, and we discuss this in more detail here to illustrate the complexity of the experiments and analyses as well as the 
central role played by transport models.

The nucleon elliptic flow is sensitive to the pressure generated in nuclear collisions and, therefore, to the EOS. Since a higher symmetry pressure will yield a larger magnitude of the elliptic flow at midrapidity for neutrons than for protons, comparisons of the neutron and proton elliptic flows provide sensitivity to the density-dependence of the symmetry energy~\cite{Li:2002qx}. The neutron and hydrogen elliptic flow from Au+Au collisions at a beam energy of $\ekin = 0.4\ A\rm{GeV}$ ($\snn = 2.07\ \rm{GeV}$) were 
measured in the FOPI-LAND and Asymmetric-Matter EOS (ASY-EOS) experiments, using 
\begin{wrapfigure}{l}{0.45\textwidth}
  \centering
  \vspace{-1.75mm}
    \includegraphics[width=0.45\textwidth]{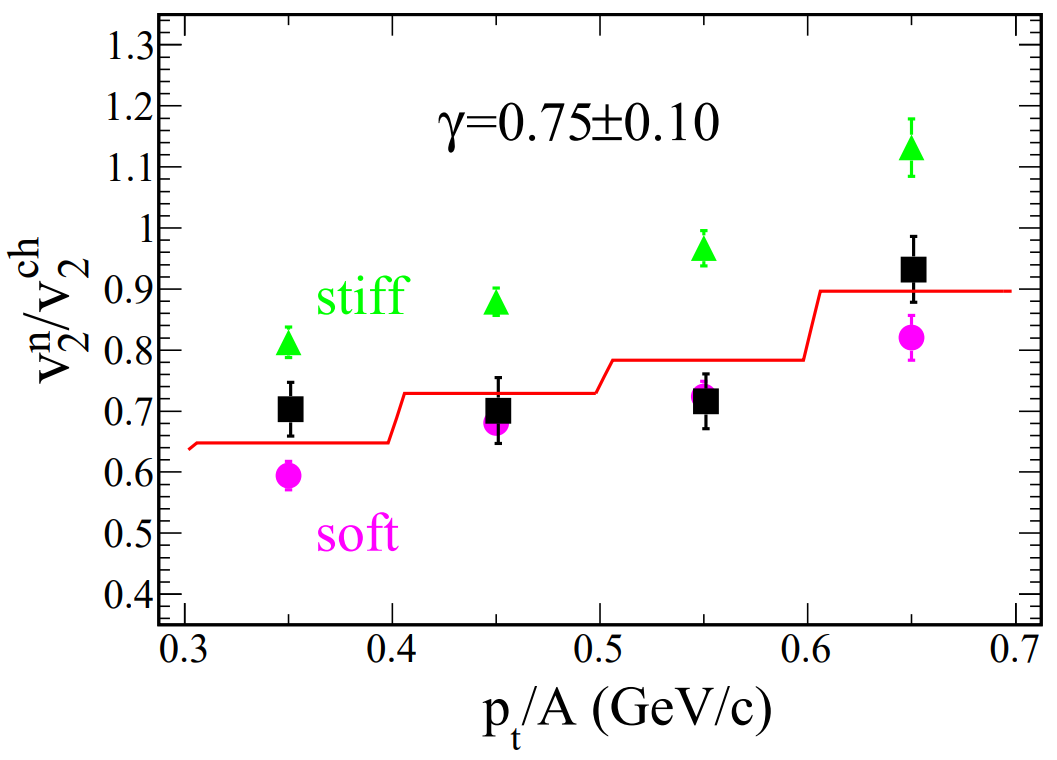}
    \vspace{-10mm}
  \caption{Ratio of the elliptic flows of neutrons over charged particles $v^{n}_2/v^{ch}_2$ as a function of transverse momentum per number of constituent nucleons $p_t/A$ in Au+Au collisions at $\ekin = 0.4\ A\rm{GeV}$ ($\snn = 2.07\ \rm{GeV}$). The comparison between ASY-EOS measurements (square black symbols) and \texttt{UrQMD} transport model calculations with a soft (pink dots) and hard (green triangles) symmetry potentials shows a preference for a soft symmetry energy; solid red line indicates $\gamma_{\rm{asy}} = 0.75 \pm 0.10$. Figure from~\cite{Russotto:2016ucm}.
  }
  \vspace{-7mm}
  \label{fig:ASY-EOS}
\end{wrapfigure}
the Land Area Neutron Detector (LAND) for the measurement of the neutron flow. A~comparison of data to \texttt{UrQMD} simulations, shown in Fig.~\ref{fig:ASY-EOS}, was used to extract the dependence of the symmetry energy on density, parametrized as proportional to $(n_B/n_0)^{\gamma_{\txt{asy}}}$, and the symmetry energy slope parameter~$L$. The FOPI-LAND experiment reported $\gamma_{\rm{asy}} = 0.9 \pm 0.4$ and $L = 83 \pm 26 \ \txt{MeV}$~\cite{Russotto:2011hq}, whereas the ASY-EOS obtained $\gamma_{\txt{asy}} = 0.72 \pm 0.19$ and $L = 72 \pm 13\ \rm{ MeV}$~\cite{Russotto:2016ucm}, indicating a moderately soft symmetry energy (see Fig.~\ref{fig:ASY-EOS} and also Fig.~\ref{fig:joint_constraints_EOS}). The analysis also illustrates the dependence of $S_0$ and $L$ on other input parameters of the EOS, such as $\gamma_{\txt{asy}}$. A subsequent comparison of data with \texttt{dcQMD} model~\cite{Cozma:2017bre} gives a value of $L=85 \pm 32$ MeV at $n=1.5n_0$.

In addition to the ASY-EOS experiment, another effort that explores this density region is the SAMURAI Pion-Reconstruction and Ion-Tracker (S$\pi$RIT) experiment, performed with radioactive tin isotopes at RIKEN, Japan. For constraining the symmetry energy at supra-saturation densities, pion yield ratios are considered as a unique observable since they do not form composite particles with other particles. This makes their yields independent of clusterization processes which can affect the symmetry energy (see Section~\ref{sec:experiments_at_very_low_densities}). Furthermore, pion observables are predicted to be sensitive to the nuclear EOS at high densities due to their unique production mechanism: Above $\ekin = 200\ A\rm{MeV}$ ($\snn = 1.97\ \rm{GeV}$), some of the interactions occurring in central collisions are energetic enough to form excited $\Delta(1232)$ baryon resonances (through the $NN\leftrightarrow N\Delta$ scattering process), which then promptly decay into pions and nucleons. The high production threshold of the $\Delta(1232)$ resonance ensures that pions originate from the early stages of the reaction, and therefore from regions characterized by a high density. The S$\pi$RIT collaboration measured charged pion emission from systems characterized by a wide range of asymmetry~\cite{SpRIT:2020blg} by colliding tin isotope beams of $^{108,112,124,132}$Sn with isotopically enriched targets of $^{112,124}$Sn.

\begin{wrapfigure}{r}{0.40\textwidth}
  \centering
  \vspace{-2.5mm}
    \includegraphics[width=0.40\textwidth]{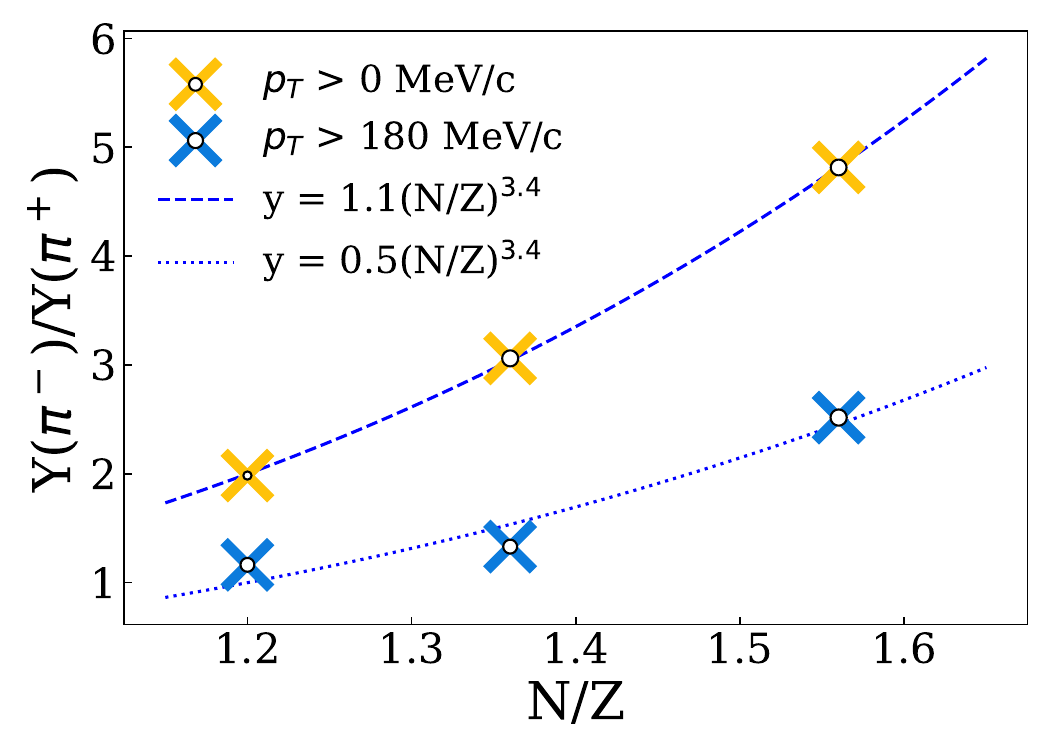}
    \vspace{-10mm}
  \caption{Ratios of yields of $\pi^-$ over yields of $\pi^+$ in central ($b < 3~\rm{fm}$) events for pions with $p_z > 0$ in the center-of-mass frame, plotted as a function of $N/Z$. 
    The yellow crosses show yield ratios with no transverse momentum cut, while the blue crosses show yield ratios for $p_T > 180~{\rm{MeV}}/c$. The radius of the circle inside each cross represents the statistical uncertainty. The dashed and dotted blue line corresponds to the best-fitted power functions of $N/Z$ for $p_T > 0$ and $p_T > 180~{\rm{MeV}}/c$ pion ratios, respectively. Figure from~\cite{tsang2021constrain}.
  }
  \vspace{-4mm}
  \label{pion_ratio_NZ}
\end{wrapfigure}

The production of $\pi^-$ strongly depends on $n$-$n$ collisions in the high-density region, while $\pi^+$ production largely depends on $p$-$p$ collisions (the production of $\pi^-$ and $\pi^+$ is equally likely in $n$-$p$ collisions). It follows that the relative production of $\pi^-$ and $\pi^+$ depends on the relative numbers of neutrons and protons and, therefore, is sensitive to the symmetry energy in the high-density region. Assuming a $\Delta$-resonance model for pion production, one would expect that the pion yield ratio $Y(\pi^-)/Y(\pi^+)$ follows a $(N/Z)^2$ dependence~\cite{FOPI:2010xrt, Li:2002qx}. 
However, the measured total pion yield ratio follows $N/Z$ with a best-fitted power index of $3.4$, as shown in Fig.~\ref{pion_ratio_NZ}, where yield ratios without a transverse momentum cut are depicted by yellow crosses with circle markers. The radius of the circle in the center of each cross represents the experimental uncertainty, showcasing very good experimental accuracy of the measurement in which systematic errors are reduced by taking pion yield ratios. Moreover, comparisons of systems with different $N/Z$ measured in the same experiment reduces systematic errors \cite{estee2020charged}. The discrepancy between the theoretical expectation and 
experimental data indicates the presence of dynamical factors beyond a simple $\Delta$-resonance model, 
while the large measured exponent suggests that the ratios are strongly affected by the symmetry energy. When a transverse momentum cut of 
\mbox{$p_T > 180~{\rm{MeV}}/c$} is imposed, the result (represented in Fig.~\ref{pion_ratio_NZ} by blue crosses with circle markers) still shows the same $(N/Z)^{3.4}$ dependence, suggesting that effects due to the symmetry energy persist in high-momentum pions. Interestingly, current transport models do not seem to be able to reproduce the strong $N/Z$ dependence~\cite{SpRIT:2020blg}.

\begin{figure}[!t]
    \centering
    \includegraphics[width=0.99\linewidth]{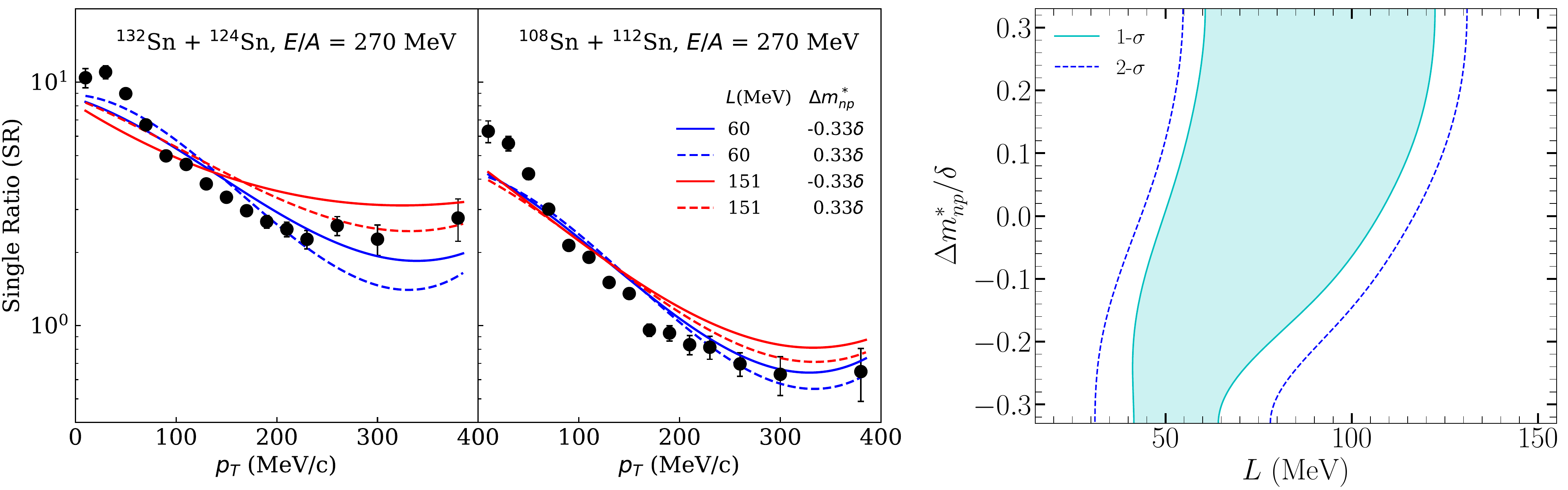}
    \caption{\textit{Left panel:} Single pion spectral ratios for $^{132}$Sn+$^{124}$Sn (\textit{top}) and $^{108}$Sn+$^{112}$Sn (\textit{bottom}) reactions compared with four selected \texttt{dcQMD} predictions \cite{Cozma:2021tfu}. Figure from ~\cite{SRIT:2021gcy}. \textit{Right panel:} Correlation constraint between $L$ and $\Delta m_{np}^*/\delta$, extracted from pion single ratios at $p_T > 200~{\rm{MeV}}/c$ in collisions of both neutron-deficient $^{108}$Sn+$^{112}$Sn and neutron-rich $^{132}$Sn+$^{124}$Sn systems. The light blue shaded region (dashed blue lines) corresponds to 68\% (95\%) confidence interval. Figure from ~\cite{SRIT:2021gcy}.
    }
    \label{fig:single_ratio_and_L_M_fit}
\end{figure}

While Fig.~\ref{pion_ratio_NZ} shows the total yield, the left panel of Fig.~\ref{fig:single_ratio_and_L_M_fit} focuses on the $p_T$-dependence of the single ratio spectrum ${\txt{SR}}
(\pi^-/\pi^+) = [dN(\pi^-)/dp_T]/[dN(\pi^+)/dp_T]$ for two extreme cases: reactions of neutron rich ($^{132}$Sn+$^{124}$Sn) and of near-symmetric ($^{108}$Sn+$^{112}$Sn) systems. The data is compared with the \texttt{dcQMD} model~\cite{Cozma:2014yna,Cozma:2017bre}, a Quantum Molecular Dynamic transport model that includes total energy conservation and other advanced features.
To extract the EOS, the \texttt{dcQMD} model was used to predict single ratios with 12 different parameter sets in the $(L, ~\Delta m_{np}^*)$ space, forming a regular lattice; here, $L$ is the slope of the symmetry energy and $\Delta m_{np}^*$ is the neutron-proton effective mass splitting.  The value of $L$ in the lattice is either 15, 60, 106, or $151~$MeV and $\Delta m_{np}^*/\delta$ is either -0.33, 0, or 0.33. All other input parameters in the \texttt{dcQMD} have been fixed by comparing to FOPI data, as well as by comparing the predictions to the total yield of the charged pions and the average $p_T$ obtained from the pion spectra. Details of the comparison can be found in Ref.\ \cite{SRIT:2021gcy}. The left panel in Fig.~\ref{fig:single_ratio_and_L_M_fit} shows a few selected calculations and the measured single ratios. 
The $(L, \Delta m_{np}^*)$ values for the solid blue line are $(60, -0.33\delta)$, for the dashed blue line are $(60, 0.33\delta)$, for the solid red line are $(151, -0.33\delta)$, and for the dashed red line are $(151, 0.33\delta)$. Coulomb effects dominate the low $p_T$ region, causing a steep rise in the measured ratios at $p_T < 200~{\rm{MeV}}/c$. 
All calculations at $p_T < 200~{\rm{MeV}}/c$ disagree with data, which could be caused by inaccuracies in the simulation of Coulomb interactions or of the pion optical potential above the saturation density. At $p_T > 200~{\rm{MeV}}/c$, the Coulomb and pion potential effects diminish and the ratios should be good probes of the symmetry energy effects.

The predicted single ratios at $p_T > 200~{\rm{MeV}}/c$ are interpolated with 2D cubic splines over the $(L, \Delta m_{np}^*)$ space, and the interpolated predictions are then compared to experimental measurements through a chi-square analysis. The resultant multivariate constraint on $L$ and $\Delta m_{np}^*$
is shown in the right panel of Fig.~\ref{fig:single_ratio_and_L_M_fit}, where the green shaded region is the $1\sigma$ confidence interval and the area enclosed by the two blue dashed curves is the $2\sigma$ confidence interval. 
The correlation between $\Delta m_{np}^*$ and $L$ occurs because both parameters influence the nucleon momenta; $L$ influences the momenta through its isospin-dependent contribution to the nucleon potential energy, and $\Delta m_{np}^*$ influences the momenta \textit{via} its isospin-dependent impact on the nucleonic kinetic energy. Either increasing $L$ or decreasing $\Delta m_{np}^*$ will increase the energies of neutrons relative to protons. This increases the numbers of $n$-$n$ collisions relative to $p$-$p$ collisions at energies above the pion production threshold and enhances the production of $\pi^{-}$ relative to that for $\pi^{+}$.

\subsubsection{Challenges and opportunities}

\textbf{Experiment and theory}

Currently, there are few experiments that aim at inferring the symmetry energy and symmetry pressure from heavy-ion collisions probing densities of 1--2$n_0$. Furthermore, the available constraints have very large uncertainties, especially for the symmetry pressure.  It is worth noting that heavy-ion collision experiments do not measure the symmetry energy or pressure directly, but rather they depend on comparisons with transport model simulations that describe the dynamics of the collisions \cite{TMEP:2022xjg}. The large uncertainties in available constraints mainly arise from the intrinsic uncertainties of the transport models and the accuracy of determining the parameters used as an input in these models. For example, a general feature of low-energy heavy-ion collisions is that more nucleons are emitted in light clusters than are emitted as free neutrons and protons, while the reverse is true of most transport model simulations of these reactions. Theoretical approaches to this issue have been proposed (see Section \ref{sec:transport_opportunities}), but are rarely implemented to model the coalescence of nucleons in the medium into the observed distribution of clusters, and therefore it is not clear to what extent these approaches are valid. The current inaccuracy in cluster production complicates and limits the scientific conclusions that can be drawn by comparing data to transport theory, and therefore improving the accuracy of cluster production in transport theory would be a very significant achievement, enabling more stringent constraints on the symmetry energy.

It is important to quantify major sources of systematic uncertainties in the transport model implementations and in the model parameters. Due to the quality as well as technical details of solutions adopted in different models, it may not be realistic to establish all uncertainties for all transport models. Nonetheless, developing methods to validate transport models and performing these validations remains a primary goal for the Transport Model Evaluation Project (TMEP) collaboration, and it is essential to extracting reliable constraints on the EOS from heavy-ion collisions (see Section \ref{sec:model_simulations_of_HICs}).

The current capabilities at FRIB, using beam energies up to $\ekin = 200\ A\rm{MeV}$ ($\snn = 1.97\ \rm{GeV}$), allow for exploration of densities up to 1.5$n_0$, and the neutron excess can be varied over a wide range by changing the composition of the rare isotope beams and targets, allowing to more closely recreate the matter found in extreme astrophysical environments (e.g., neutron stars). From the dense collision region in heavy-ion collisions, pions and free nucleons are emitted with high transverse momentum. The relative yields of these particles, especially as a function of energy, as well as particle elliptic flow contain information about the dense collision zone and thus can be used to constrain the EOS that governs supra-saturation matter. Individual efforts based on small-scale experiments, which are the strength of the field, have provided a diversity of results. Nevertheless, in order to take advantage of multiple-parameter Bayesian analyses, described below, and given the tight allotment of the expensive (and coveted) beam time, future experiments at FRIB should utilize detectors that provide large coverage with the ability to measure multiple observables simultaneously. In particular, a large coverage of neutron detectors for EOS experiments would be indispensable. However, neutrons are notoriously difficult to detect, which calls for research into constructing a neutron array with advanced materials and technologies. Furthermore, development of a time projection chamber (TPC) detector is essential to measure both pions and charged particles. When the construction of the High Rigidity Spectrometer (HRS) is completed, a similar set up as used in the S$\pi$RIT experiment can be employed for EOS studies \cite{Barney:2020mxk}. Before the completion of the HRS, a simpler set up can be used, utilizing high resolution silicon detector arrays and large area neutron arrays coupled with a lower resolution 4$\pi$ detector array to determine the reaction plane and the collision geometry, necessary for experiments constraining the EOS. Pursuing these experimental needs is necessary to maintain the U.S. leadership in the EOS research at FRIB.

Reaching higher densities requires the energy upgrade to $\ekin = 400\ A\rm{MeV}$ ($\snn = 2.07\ \rm{GeV}$). With the capability for producing high-intensity rare isotope beams with a wide range of asymmetries, FRIB400 is essential for the U.S.\ effort to lead in the determination of the density-dependence of the symmetry energy \cite{FRIB400}.

The beams available at FRIB, being complementary to those that can be accelerated at European facilities, may represent a unique opportunity to conduct nuclear transport investigations also by the international nuclear physics community. As described above, the development of detector arrays with high isotopic resolution over a wide dynamic range, from light particles to heavy fragments, provides the prospect of measuring observables (especially in the context of isospin diffusion and drift as well as in collective motion phenomena) that can amplify the sensitivity to the symmetry energy.
Coupled with its capability to use high-quality radioactive beams, FRIB may represent a focus of interest for a wider community, stimulating the need for international discussions and collaborations in the coming years. Such an interest may concern also theoretical physicists that have been collaborating with FRIB colleagues within the TMEP initiative, aimed at improving investigations of the isospin-dependent EOS with comparisons to experimental observables (see also Section~\ref{sec:model_simulations_of_HICs}).

\vspace{5mm}
\textbf{Multiple Parameter Bayesian analysis}

The EOS is only one of many input parameters in transport models used to simulate heavy-ion collisions. Often, multiple measurements probing different parts of the collisions are needed to constrain other parameters of these models, such as the momentum-dependence of the isovector mean-field potential, or the in-medium isospin-dependent cross sections. However, constraining transport model parameters with experimental results is a delicate endeavor. The outcomes of nuclear collisions are influenced by a multitude of processes, and therefore the experimentally 
measured final stage observables can depend simultaneously on values of multiple parameters. However, carefully chosen observables may only be sensitive to just a few specific parameters. The full extent of the dependence of a given transport model on input parameters can only be tested empirically after performing a complete series of simulations of heavy-ion collisions.

Bayesian statistical methods provide means to quantify the relation between observable values and physical parameters. They also provide a systematic way of constraining multiple nuclear properties and utilizing prior knowledge from different experiments, prior constraints from other sources, and results from new experimental measurements. For example, in the $n/p$ ratio experiment mentioned above, measuring the yield ratios of neutron and protons spectra, a Bayesian analysis comparing the experimental results~\cite{Morfouace:2019jky} to \texttt{ImQMD} calculations determines both $\Delta m_{np}^*$ and the relationship between $S_0$ and $L$, even though the uncertainties are large. More precise measurements in the future will enable a better resolution.

\begin{figure}[!t]
    \vspace{-0mm}
    \centering
    \includegraphics[width=0.99\linewidth]{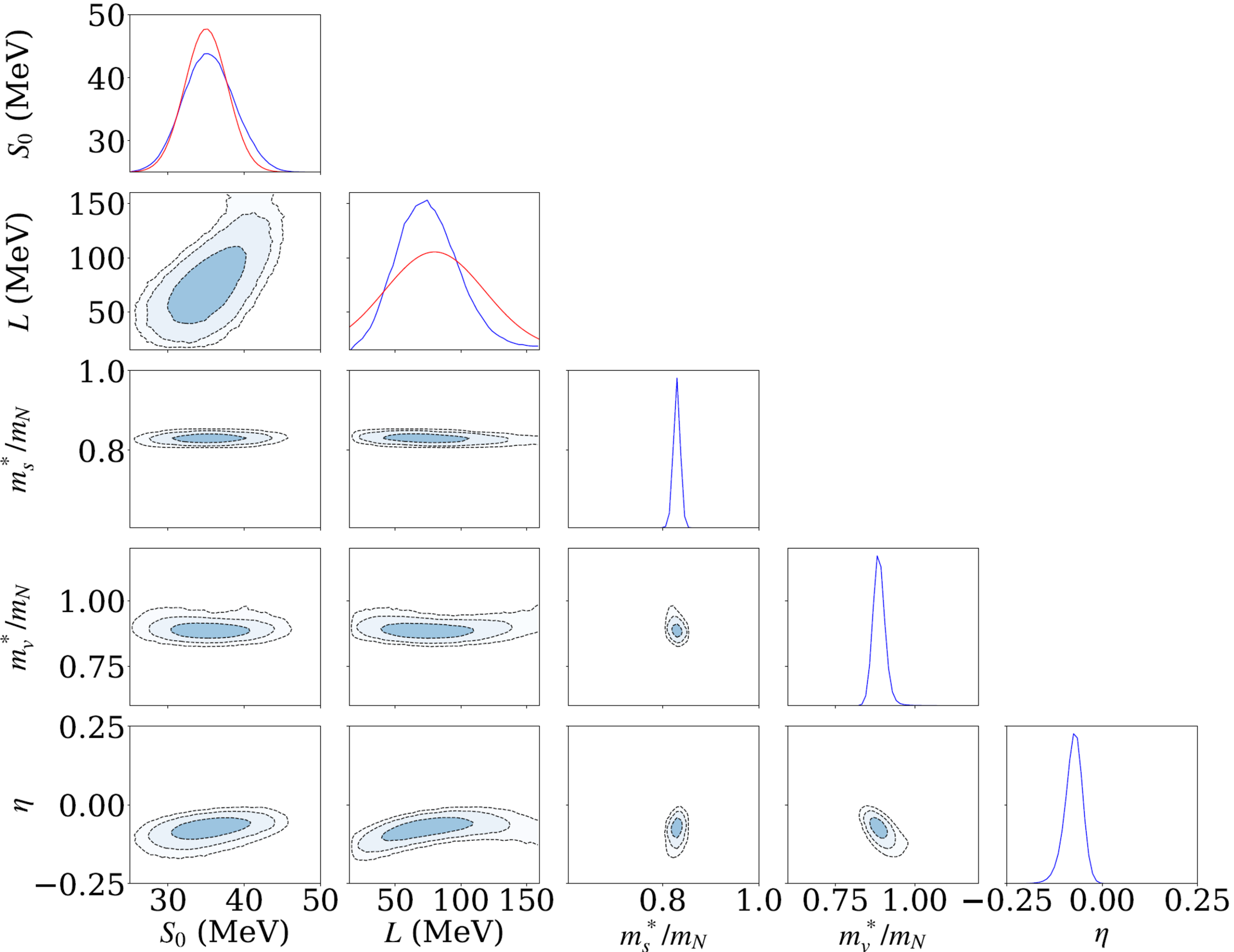}
    \caption{Posterior distribution obtained from a Bayesian analysis of \texttt{ImQMD} simulation results and experimental data from S${\pi}$RIT experiments~\cite{tsang2021constrain}. Eight available observables are used for Bayesian analysis. The values for median and 68\% confidence interval of the marginalized distribution are tabulated on the upper right-hand side of the figure. Figure from~\cite{tsang2021constrain}.}
    \label{fig:ExpPosterior}
\end{figure}

In the long term, it is important to develop Bayesian analyses of multiple observables to determine multiple parameters simultaneously. As an example,  in the S$\pi$RIT experiment many observables have been measured with four reaction systems. Eight observables in total, including the directed flow, elliptical flow, and the stopping observable from different reactions, are fitted simultaneously by varying five transport model input parameters (two pertaining to the shape of the symmetry energy term in the nuclear matter EOS, two pertaining to the nuclear effective masses, and one pertaining to  the nuclear in-medium cross-section). The posterior distribution shows a weak constraining power on the symmetry energy terms, but a strong sensitivity to effective masses and in-medium cross-section~\cite{tsang2021constrain}, see Fig.~\ref{fig:ExpPosterior}.

The posterior parameter distributions are generated from repeated sampling of transport model predictions for hundreds of thousands of times, each with different parameter values. If carried out directly, this process would consume an unreasonable amount of computational resources. This can be alleviated with an effective, efficient, and capable model emulator which emulates the behavior of transport models at all points of the allowed parameter space from predictions at just a few tens of parameter values. Gaussian processes are readily available and commonly used in emulators, but the procedures for tuning hyperparameters vary across analyses. Numerous heuristics and cost functions are proposed for the optimization of hyperparameters, and one can also marginalize over all nuisance parameters with a Markov chain Monte Carlo.

\section{The equation of state from combined constraints}
\label{sec:combined_constraints}

\begin{table}[h]
\scriptsize
\vspace{-3mm}
\centering
\setlength\extrarowheight{5pt}
\begin{tabular}{|l|l|} 
\hline
\bf{Nuclear} & \bf{Neutron star} \\
\hline
Isospin diffusion in HICs  & Masses and radii \\
Dipole polarizability & Tidal deformability \\
Spectral ratios of light clusters \hspace{5mm}  & Moment of inertia \\
Nuclear masses and radii  & Gravitational binding energy \\
Isobaric analog states  & Cooling of young neutron stars\\
$n/p$ ratios in HICs & Bulk oscillation modes \\
Neutron skins & Crust cooling \\
Mirror nuclei & Pulsar glitches \\
Giant resonances & Lower and upper limits on neutron star spin periods \hspace{5mm}\\
Flow of particles in HICs & Torsional crust oscillations \\
Charged pion ratios in HICs  & Crust-core interface modes \\
\hline 
\end{tabular}
\caption{Illustrative list of nuclear and astrophysical observables.}
\label{tab:observables}
\end{table}

\begin{wrapfigure}{r}{0.60\textwidth}
  \centering
  \vspace{-2mm}
    \includegraphics[width=0.60\textwidth]{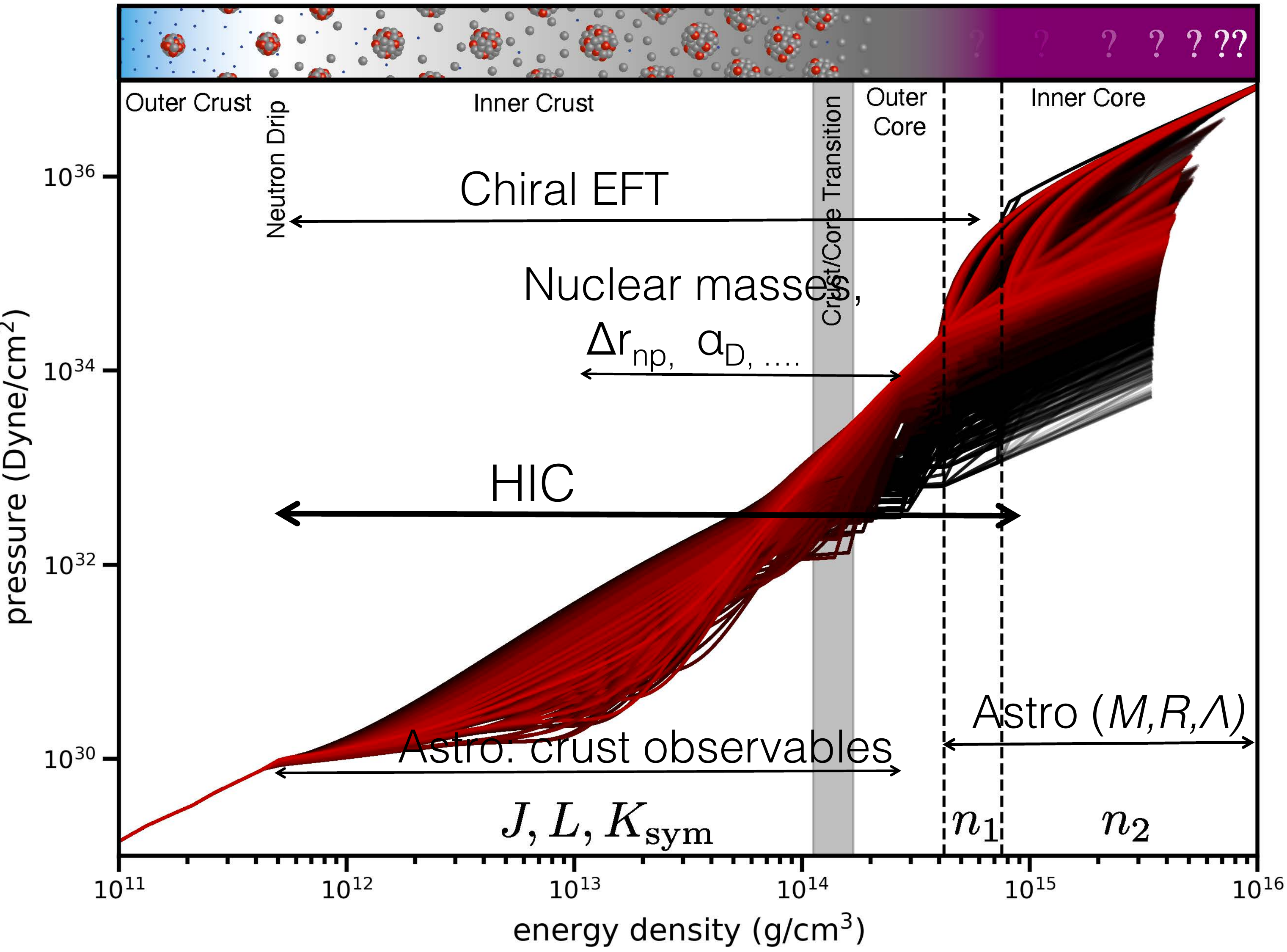}
    \vspace{-6.0mm}
  \caption{An ensemble of EOSs that range over crust and core uncertainties consistently. Density ranges over which different nuclear and astrophysical observables probe the EOS are indicated. Figure modified from~\cite{Neill:2022psd}.
  }
  \vspace{-4.0mm}
  \label{fig:crust_core_EOSs}
\end{wrapfigure}

There have been many attempts to extract the equation of state (EOS) as a function of density from both nuclear experiments and astronomical observations. In Table~\ref{tab:observables}, we provide an illustrative list of relevant experimental and observational measurements. Importantly, these observables probe the EOS at different densities: a few probe the EOS near the saturation density~$n_0$, but many probe densities that are significantly higher or lower. For example, nuclear structure typically probes densities that are somewhat lower than $n_0$, while analyses of heavy-ion collisions or properties of neutron stars probe larger density ranges, as schematically illustrated in Figs.\ \ref{fig:constraints} and \ref{fig:crust_core_EOSs}.

Comparing constraints based on different measurements allows one to test their consistency and ultimately find tight constraints on the EOS over the full range of densities that can be probed either by experiments or by astronomical 
observations. Techniques of Bayesian inference 
or Pearson correlation analyses are well-suited to this endeavor and can provide more readily useful and testable information on the density-dependence of the EOS than, e.g., statistical comparisons or combining the Taylor expansion coefficients (such as $S_0$, $L$, and $K_{\rm{sym}}$) obtained from individual analyses. Key to this approach is the determination of the density that each experimental observable most accurately probes. Away from that density, weaker constraints on the EOS are possible, but the analysis is more complex.

In this section, we review the variety of observables that have been used to place constraints on 
the EOS; heavy-ion collision experiments, which produce many of these constraints, are described in Section~\ref{sec:laboratory_experiments}. We then discuss recent attempts at combining various constraints that result in meaningful EOSs with quantified uncertainties.

\subsection{Constraints}
\label{sec:Constraints}
As discussed in Section \ref{sec:laboratory_experiments}, experiments are often designed to explore certain aspects of the EOS. Accordingly, we classify the constraints obtained from laboratory measurements as sensitive to either the symmetric nuclear matter EOS or the symmetry energy. In addition to the experimental inferences, constraints on the EOS can be also obtained from neutron star observations as well as 
from chiral EFT theory at low densities. The list of constraints discussed here is not exhaustive. Rather, it represents a slice of widely acknowledged constraints at the moment of writing. We note that some of the constraints reviewed here have already been presented in Sections~\ref{sec:theory} and \ref{sec:laboratory_experiments}, to which we refer when appropriate.

\vspace{5mm}
{\textbf{Symmetric matter constraints from laboratory experiments}}

Some properties of the symmetric nuclear matter are fairly well-known near $n_0$. For example, the generally accepted values of $n_0$ and binding energy at saturation $E_{0}$ are $0.16\ \rm{fm}^{-3}$ and $-16\ \rm{MeV}$~\cite{Drischler:2017wtt, Drischler:2015eba}, respectively, to within 4\%. The incompressibility parameter, $K_{0}$, has been determined from giant monopole resonance (GMR) experiments \cite{Youngblood:1999zza} to be $231\pm5\ \rm{MeV}$. However, subsequent GMR measurements of the Sn isotopes cast larger uncertainties on $K_{0}$ \cite{Dutra:2012mb}. While these larger uncertainties are consistent with values of $K_{0}$ determined from heavy-ion collision experiments \cite{Fuchs:2001gv,KAOS:2000ekm,LeFevre:2015paj}, we note here that these experiments derive their constraints on $K_{0}$ based on density functionals that are parametrized with $K_0$, but used to describe the high-density behavior of the EOSs (i.e., these experiments do not probe the incompressibility at saturation; see also a similar discussion in Section \ref{sec:selected_constraints}). 
Measurements of the collective flow from high energy Au+Au collisions have constrained the EOS for symmetric nuclear matter at densities spanning $(1$--$4.5)n_0$ \cite{Danielewicz:2002pu,Lynch:2009vc,LeFevre:2015paj, Oliinychenko:2022uvy} (see the left panel in Fig.~\ref{fig:joint_constraints_EOS}), as described in Sections \ref{sec:selected_constraints} and \ref{sec:experiment_snm}.

\vspace{5mm}
{\textbf{Symmetry energy constraints from laboratory experiments}}

In the past decade, many studies have been conducted to extract the symmetry energy and the 
symmetry pressure, and some of the widely-known constraints are plotted in the right panel of Fig.~\ref{fig:joint_constraints_EOS}, which includes both the usual EOS constraint bands as well as symbols located at densities which a novel analysis in Ref.~\cite{Lynch:2021xkq} identified as the most sensitive densities for a given measurement. At $(2/3)n_{0}$, precise symmetry energy constraints have been obtained from studies on nuclear masses using Skyrme density functional forms for the EOS. These are labeled in the right panel of Fig.\ \ref{fig:joint_constraints_EOS} as ``mass(Skyrme)'' \cite{Brown:2013mga} and ``mass(DFT)'' \cite{Kortelainen:2011ft}, respectively. In this density region there are also precise constraints obtained from the energies of isobaric analogue states \cite{Danielewicz:2016bgb}, indicated in the right panel of Fig.\ \ref{fig:joint_constraints_EOS} by a data point labeled as ``IAS''. The dipole polarizability $\alpha_D$, marked in the right panel of Fig.\ \ref{fig:joint_constraints_EOS} as ``$\alpha_D$'' and reflecting the response of a nucleus to the presence of an external electric field, also helps to constrain the symmetry energy at low densities. Constraints on the symmetry pressure $P_{\rm{sym}}$, which is proportional to the derivative of the symmetry energy with respect to density, have been recently provided by the measurements of the neutron skin of $^{208}$Pb in the Lead Radius EXperiment (PREX and PREX-II)~\cite{Abrahamyan:2012gp,PREX:2021umo,Reed:2021nqk} and of the neutron skin of $^{48}$Ca in the Calcium Radius EXperiment (CREX)~\cite{CREX:2022kgg,Reinhard:2022inh,Zhang:2022bni}, both at Jefferson Lab, which use parity-violating weak neutral interactions to probe the neutron distribution in $^{208}$Pb and $^{48}$Ca. A range of other scattering experiments have measured the neutron skins of a number of neutron-rich isotopes and likewise used them to constrain the symmetry energy \cite{Chen:2010qx,Xu:2020fdc,Newton:2020jwn}. Giant dipole resonances and polarizabilities \cite{Klimkiewicz:2008lqv,Reinhard:2010wz,Tamii:2011pv,Birkhan:2016qkr} in neutron-rich isotopes provide another source of information about the symmetry energy \cite{Carbone:2010az,Hashimoto:2015ema,Roca-Maza:2015eza,Piekarewicz:2021jte,Reinhard:2021utv,Zhang:2015ava,Daoutidis:2011zz}, as do mirror nuclei~\cite{Brown:2017xxo,Pineda:2021shy}. At densities far from $(2/3)n_{0}$, heavy-ion collisions have been used to probe the symmetry energy, as is described in Sections \ref{sec:selected_constraints} and \ref{selected_constraints_experiment}, and shown in the right panel of Fig.~\ref{fig:joint_constraints_EOS}.

\vspace{5mm}
{\textbf{Constraints from astronomical observations}}

The bulk properties of neutron stars (such as their maximum mass, radii, tidal deformabilities, moments of inertia, limits on the rotation frequency, and binding energy) depend strongly on the 
distribution of matter throughout the star, therefore providing a measure of the EOS integrated over the range of densities present in the star. 
The mass-radius relationship has a one-to-one correspondence to the neutron star EOS \cite{Lindblom1992}, and it is known that the radius, the tidal deformability, 
and the moment of inertia provide the strongest constraints on the EOS above 2$n_0$ \cite{Miller:2019nzo,Legred:2021hdx}, while the maximum measured mass of neutron stars constrains the EOS at the highest densities. 
Together, the tidal deformability measured in GW170817, the mass of J0740+6620, and the two mass-radius measurements of NICER, discussed in Section~\ref{sec:neutron_star_theory}, form the current gold standard in measuring the neutron star properties using astronomical observations.

A number of astronomical observables also probe the neutron star crust physics, which results in constraints on the pure neutron matter EOS, and in particular on the symmetry energy. This is because the neutrons provide the hydrostatic pressure that supports the inner crust, and the interplay between these neutrons and the lattice of nuclei that makes up the crust determines the crust-core boundary as well as the possible nuclear pasta shapes that appear near that boundary. The crust physics also depends, more weakly, on the symmetric matter EOS. The nuclear EOS at subsaturation densities, down to where the neutron drip begins ($n_B \approx$ 10$^{-4} n_0$), is therefore an essential ingredient in crust models.

Due to the complexity of crust physics, extracting rigorous EOS constraints from observations of crust-associated neutron star behavior is in its early stages, and it is an area where substantial progress can be made over the next decade. Here we list some constraints on the symmetry energy as an illustration of this potential, but, at the same time, we note that they are very tentative and do not have well-quantified errors; indeed, some of them are mutually exclusive, emphasizing the need to make progress in applying microscopic nuclear physics models to these observations.

Constraints on the symmetry energy and its slope can be obtained from studying the following phenomena: A study of the cooling of the neutron star in the Cas A supernova remnant \cite{Newton:2014iha}, which has been observed to cool on a timescale of decades, implies that the neutron star core may have superfluid properties \cite{Shternin:2010qi,Page:2010aw,Shternin:2021fpt}. Studying the temperatures of the population of neutron stars whose surface X-ray emission is observable leads to constraints on the neutron star masses and radii and the composition of the core \cite{Beloin:2018fyp}. Constraints from quasi-periodic oscillations in the X-ray tail of gamma-ray flares from soft gamma-ray repeaters \cite{Steiner:2009yg,Sotani:2011nn,Sotani:2012qc}, which could be a signature of torsional oscillations of the crust. Potential measurements of the crustal moment of inertia from glitches -- sudden changes in rotation frequency -- of radio pulsars and some X-ray pulsars~\cite{Link:1999ca,Andersson:2012iu,Chamel:2012ae,Piekarewicz:2014lba,Hooker:2013fda,Newton:2015wka}, which can constrain properties of the crust. Limits on the longest and shortest observed periods of neutron stars probe physics such as the magnetic field evolution in the crust~\cite{Pons:2013nea} and the development of rotation-induced instabilities in oscillations such as $r$-modes~\cite{Wen:2011xz,Vidana:2012ex}. During the last few seconds of an inspiral prior to the merger of two neutron stars or a neutron star with a black hole, tidal forces may shatter the crust, causing a gamma-ray flare: in this scenario, coincident timing of the flare with the gravitational wave signal measures the resonant frequency of crust-core interface modes and sets constrains on the symmetry energy~\cite{Tsang:2011ad,Neill:2020szr}.
The cooling of the crusts of quiescent low-mass X-ray binaries promises to provide a source of constraints on the composition and size of the neutron star crust and the extent of nuclear pasta phases therein~\cite{Brown:2009kw,Horowitz:2014xca,Wijngaarden:2019tht,Lalit:2019okh}. The expected accurate measurement of the moment of inertia of pulsar J0737-3039a \cite{Kramer:2021jcw} will set constraints on the EOS competitive with the current radius constraints~\cite{Lattimer:2004nj,Fattoyev:2010tb,Steiner:2014pda,Raithel:2016vtt,Greif:2020pju}. The heat capacity of a neutron star core can be measured by using inferences of the core temperature of transiently-accreting neutron stars, and strongly suggests that a core dominated by a color-flavor-locked quark phase is ruled out \cite{Cumming:2016weq}. Transiently-accreting neutron stars are also observed to have efficient cooling in the core, constraining superfluid gap models and the symmetry energy~\cite{Mendes:2022gbq}. With next-generation gravitational wave observatories, bulk neutron star oscillations excited during a binary inspiral~\cite{Hinderer:2016eia,Steinhoff:2016rfi,Ho:1998hq} could be detected \cite{Pratten:2019sed,Williams:2022vct}, where the frequencies and eigenmodes of such oscillations depend on the EOS and the structure of the star. In particular, g-mode oscillations depend on the proton fraction gradient in high-density regions~\cite{Reisenegger_gModes_1992}, thus probing the symmetry energy at high baryon density.

\vspace{5mm}
{\bf{Constraints from nuclear theory}}

In recent years, many-body nuclear theory such as chiral effective field theory (\ceft), discussed in Section \ref{sec:microscopic_calculations_of_the_EOS}, has made significant progress to be considered as the canonical nuclear matter EOS at low densities with rigorous uncertainty quantification~\cite{Hebeler:2015hla,Lynn:2019rdt,Drischler:2021kqh,Drischler:2021kxf}. 
Even though the theory is developed mainly for densities below saturation, it has been extended to $2n_0$, and it is a popular constraint for studies that focus mainly on astronomical observations.

\subsection{EOS obtained by combining various constraint sets}
\label{sec:combining_different_constraints}

Each of the nuclear and astrophysical observables discussed above provides vital information about the EOS over some density range, that can be combined with other constraints to globally constrain the density-dependence of the EOS from sub-saturation to supra-saturation densities. Such analysis techniques are relatively new, but several of such global constraints now exist, and a selection of studies is briefly described below to illustrate their potential.

Beloin \textit{et al.}~\cite{Beloin:2018fyp} used relativistic mean-field models of the nuclear interaction to model the structure and cooling of neutron stars, consistently combining nuclear data, neutron star mass-radius measurements, and neutron star cooling measurements within a Bayesian framework.

\begin{wrapfigure}{l}{0.43\textwidth}
  \centering
  \vspace{-7.0mm}
    \includegraphics[width=0.43\textwidth]{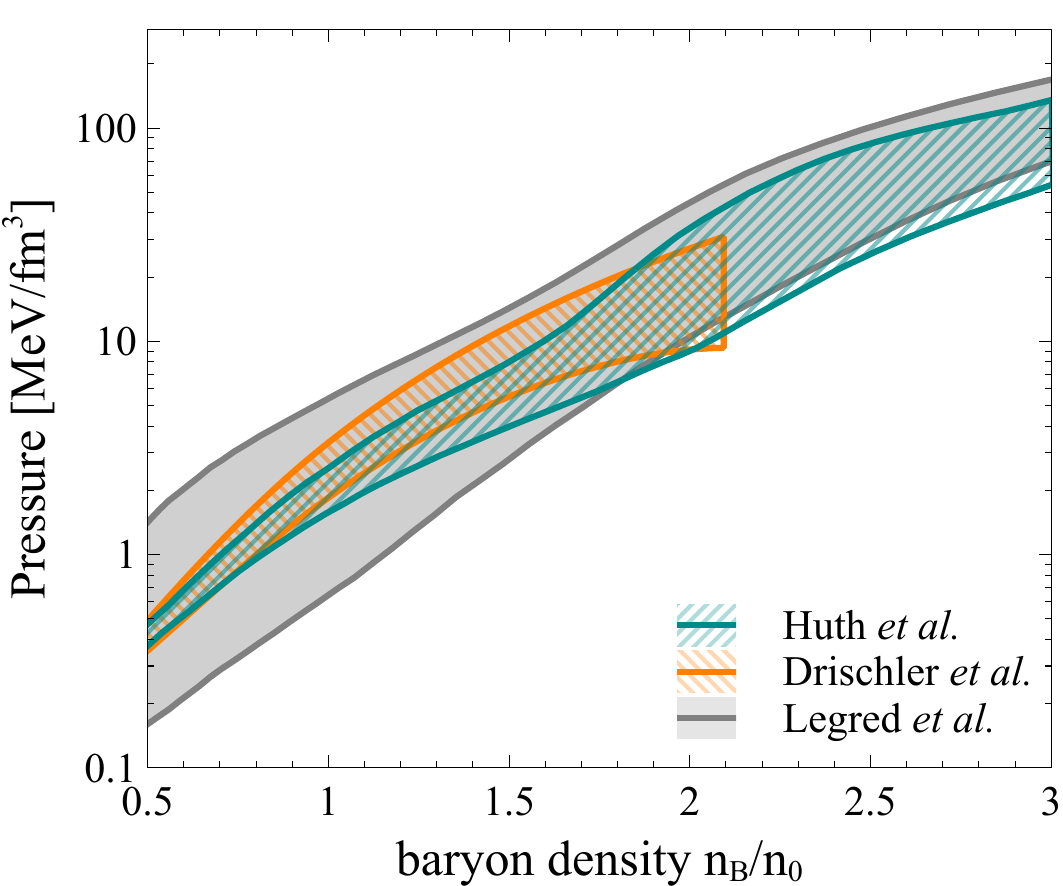}
    \vspace{-9.35mm}
  \caption{The pressure of neutron star matter as a function of number density $n_B$, as obtained by Huth \textit{et al.}~\cite{Huth:2021bsp}, Drischler \textit{et al.}~\cite{Drischler:2020fvz}, and Legred \textit{et al.}~\cite{Legred:2021hdx} at 95\%, 95\%, and 90\% confidence interval, respectively. 
  }
  \vspace{-5mm}
  \label{fig:NSM}
\end{wrapfigure}

Legred \textit{et al.}~\cite{Legred:2021hdx} performed nonparametric EOS inference based on Gaussian processes, combining information from 
X-ray, radio, and gravitational wave observations of neutron stars. Their results are plotted in Fig.\ \ref{fig:NSM} and labeled as 
``Legred \textit{et al.}''. These Bayesian analyses incorporate astrophysical data and provide constraints on the neutron star EOS at higher densities.

Drischler \textit{et al.}~\cite{Drischler:2020fvz} performed a Bayesian analysis of correlated effective field theory truncation errors based on order-by-order calculations up to next-to-next-to-next-to-leading order in the $\chi$EFT expansion. The neutron star matter pressure calculated with these EOS is shown in Fig.\ \ref{fig:NSM} and labeled as ``Drischler et al.''.

Huth \textit{et al.}~\cite{Huth:2021bsp} combined nuclear theory \textit{via} $\chi$EFT calculations (constraining the EOS below 1.5$n_0$), EOS inferences from heavy-ion collisions \textit{via} the FOPI (constraining the symmetric matter EOS up to 2$n_0$) and ASY-EOS (constraining the symmetry energy at around 1.5$n_0$) experiments, and astrophysical data on bulk neutron star properties (constraining the total neutron star EOS above~2$n_0$).  
The EOS models were extended to high densities using a speed-of-sound model. The results are shown in Fig.~\ref{fig:NSM} and labeled as ``Huth et al.''.

Yue \textit{et al.}~\cite{Yue:2021yfx} constructed neutron star models using a Skyrme energy-density functional, which allowed them to calculate the neutron skin of $^{208}$Pb and combine constraints from heavy-ion collisions, neutron skin measurements, and astrophysical observations within the same model.

Neill \textit{et al.}~\cite{Neill:2022psd} followed a similar strategy as the example above, using Skyrme models which were 
extended to the crust. This allowed them to combine neutron skin measurements, NICER/LIGO observations, a~crust observable (the resonant frequency of the crustal $i$-mode), and nuclear mass data to constrain both the core and crust properties as well as the EOS. By calculating all these quantities using the same underlying Skyrme energy density functional (and polytrope parametrizations at the highest densities), some poorly controlled modeling uncertainties were eliminated. This work demonstrated the complementarity of different observables: within the particular model used, nuclear masses constrain mainly the zeroth and first symmetry energy expansion coefficients, $S_0$ and~$L$, the crust observable has the largest impact on the inferred values of $L$ and the second expansion coefficient~$K_{\rm sym}$, and the neutron star radius and tidal deformability have the largest impact on the inferred values of $K_{\rm sym}$ and two polytrope parameters. Thus, when combined, different observables provide complementary information that can contribute to a complete picture of the EOS. The ranges of these overlapping data are depicted in Fig.~\ref{fig:crust_core_EOSs}.

Without crust observables, neutron star radii and tidal deformabilities tend to give weaker constraints on~$K_{\rm sym}$ 
and stronger constraints on $L$ (see the analysis of a large number of studies in~\cite{Li:2021crp}). However, the relative constraints on the symmetry energy parameters change when the used priors include the criterion that the crust is stable and incorporate potential data from crust observables~\cite{Neill:2022psd}. While this result is model-dependent and correlations with higher-order symmetry parameters need to be investigated, it demonstrates the way in which crust observables could significantly contribute to constraining the EOS, and motivates the need for improving models to consistently combine crust and core observables with nuclear data.

One of the defining strengths of the global constraint analysis is that one or more additional constraint(s) can be always included as long as an assessment of the corresponding statistical and systematic uncertainties is also provided. Moreover, the more data from nuclear and astrophysical observables can be meaningfully included in such EOS inferences, the greater the ability to deliver a robust EOS. Therefore, constraining the EOS by combining various inferences is highly promising and, furthermore, well-suited for the coming era of multi-differential observables from heavy-ion collisions and multi-messenger astronomical observations.

\newpage
\section{Connections to other areas of nuclear physics}

\subsection{Applications of hadronic transport}
\label{sec:applications_of_hadron_transport}

In addition to the use of transport codes to study fundamental nuclear physics, their ability to describe the transport and interactions of particles in a material also make them valuable for applications that benefit society. Examples include the design of nuclear physics experiments, detector development and simulations of detector performance, as well as medical applications and radiation shielding in accelerator and space exploration. Some of these uses are outlined here.

Transport models are widely used to simulate particle emission from nucleus-nucleus collisions. In these simulations, the four-momentum of every emitted particle is tracked, making it possible to generate double differential distributions, the particle spectra at various emission angles. These distributions are particularly important for applications.

Most transport models are optimized for describing physics in certain energy ranges.  The type of code required can be tailored to the desired application.  For example, some models perform best at energies of a few hundred MeV and below, which is near the peak of the cosmic ray flux~\cite{Slaba}, while others are more applicable for GeV-scale energies and above, in the tail of the cosmic ray flux. Over the entire experimental energy range, from intermediate energies through the highest collider energies, transport codes have been successfully employed to design complex detectors, optimize experimental setups, and carry out analyses of experimental data, including assessing the detector efficiencies and background contributions.

\subsubsection{Detector design}

In high-energy experiments, the code packages most commonly used for detector development and data analysis are \texttt{Geant3}~\cite{Brun:1987ma}, \texttt{Geant4}~\cite{GEANT4:2002zbu}, and \texttt{FLUKA}~\cite{Ahdida:2022gjl}. Most of these simulation packages use cascade codes, that is codes without mean-field potentials, to describe particle transport through matter. Modern transport codes that can cover a wide range of energies such as \texttt{PHITS} (Particle and Heavy-Ion Transport code System) \cite{Sato:2013,Sato:2015iqw} can, however, provide a more complex description of particle transport~\cite{TMEP:2022xjg}. 

\begin{figure}[!t]
    \centering
    \includegraphics[width=0.99\linewidth]{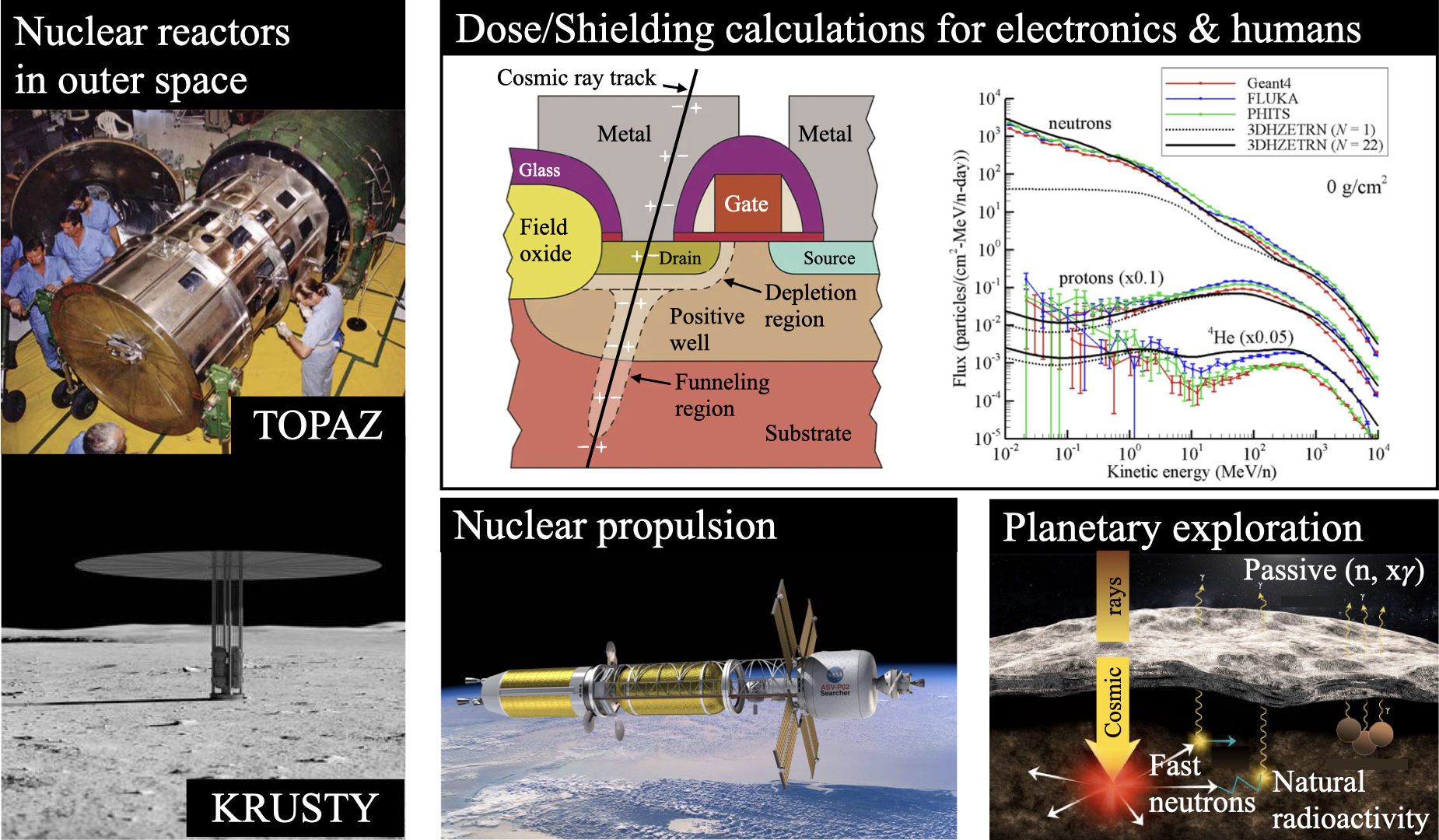}
    \caption{Some of the applications of hadronic transport calculations and nuclear data in space exploration research (\textit{counterclockwise from top left}): energy production in outer space, such as with the TOPAZ nuclear reactor \cite{INL_TOPAZ} or the proposed fission surface power system on the Moon KRUSTY \cite{NASA_KRUSTY}; nuclear thermal rocket propulsion \cite{NASA_NTR}; planetary exploration \cite{LLNL_planet_exploration}; and dose and shielding calculations of ions passing through electronics and humans (\textit{left:} a heavy-ion interaction with a shielding structure \cite{Cosmic_ray_in_shield}, \textit{right:} particle spectra calculated for an incident solar minimum GCR iron spectrum \cite{WILSON201546}).
    }
\label{fig:space_applic}
\end{figure}

Aside from heavy-ion collisions, transport simulations play an important role for a variety of fundamental physics experiments. For example, long-baseline neutrino experiments need to determine the incoming neutrino energy in order to extract the neutrino mixing parameters, CP violating phases, and neutrino mass ordering~\cite{Diwan:2016gmz}.  However, because the neutrino beam is generated from fixed-target proton-nucleus interactions producing secondary $\pi$ and $K$ mesons with neutrino decay products, there are large uncertainties on the energy of the interacting neutrinos. The neutrino energy must be reconstructed from the measurement of the final state~\cite{Mosel:2016cwa,Mosel:2017nzk} which is often modeled by simple Monte-Carlo cascade approaches.  Reliable transport descriptions could significantly improve these studies for the Deep Underground Neutrino Experiment (DUNE)~\cite{DUNE:2018tke}, as well as for the ongoing experiments NuMI Off-axis $\nu_e$ Appearance (NOvA)~\cite{NOvA:2021nfi} and Tokai to Kamioka (T2K)~\cite{T2K:2019bcf}. Other experiments that require transport simulations of backgrounds include dark matter searches~\cite{HPS:2018xkw}, semi-inclusive electron scattering such as ($e$,$e'p$) on nuclear targets at Jefferson Lab to search for color transparency, short-range correlations~\cite{Wright:2021dal}, and hadronization in a nuclear medium~\cite{CLAS:2021jhm}.

\subsubsection{Space exploration, radiation therapy, and nuclear data}

Transport models can also be used in applications relevant to space exploration to understand and mitigate the harmful effects of the space radiation environment on electronics and astronauts. Collisions of galactic cosmic rays (GCRs) with nuclei, whether in the Earth’s atmosphere or in 
the material of a spacecraft above it, can generate showers of particles, including pions, muons, neutrinos, electrons, and photons as well as protons and neutrons. GCRs cover a wide range of energies, from tens or hundreds of MeV up to the TeV scale, and ion species, spanning elements $1 \leq Z < 28$~\cite{Badhwar:1992}, making it challenging to determine all their potential effects in a given material. The penetrating power of the initial GCRs and the secondaries generated by their interaction with matter can have a serious impact on the safety and viability of space exploration. The 1\% 
of GCR primaries which are heavier than Helium nuclei can pose an especially serious problem, given that the damage they inflict scales as $Z^2$. The secondary particles generated from GCR interactions with spacecraft materials~\cite{Finckenor:2018} such as aluminum, polyethylene, and composites can harm astronauts and disrupt or even disable electronic systems. Moreover, spacecraft shielding designed to reduce the GCR flux is itself a target that can increase the secondary flux. Because of the wide variety of possible shielding materials and thicknesses, transport models are essential to determine the sensitivity of the secondaries (regarding both their flux and composition) to different shielding configurations, as well as the subsequent harmful impact of those secondaries on electronic systems~\cite{Hoeffgen:2020} and humans~\cite{Durante:2011}. A pictorial overview of applications transport modeling and nuclear data in space missions is given in Fig.~\ref{fig:space_applic}.

Due to the lack of data at the appropriate energies, simulations of space radiation effects have large uncertainties. The space research community has generally relied on phenomenological nuclear reaction models such as the Double Differential Fragmentation model (DDFRG)~\cite{Norbury:2021coa}, which consists of a sum of multiple exponential distributions with 
\begin{wrapfigure}{l}{0.59\textwidth}
  \centering
  \vspace{-7.5mm}
    \includegraphics[width=0.64\textwidth]{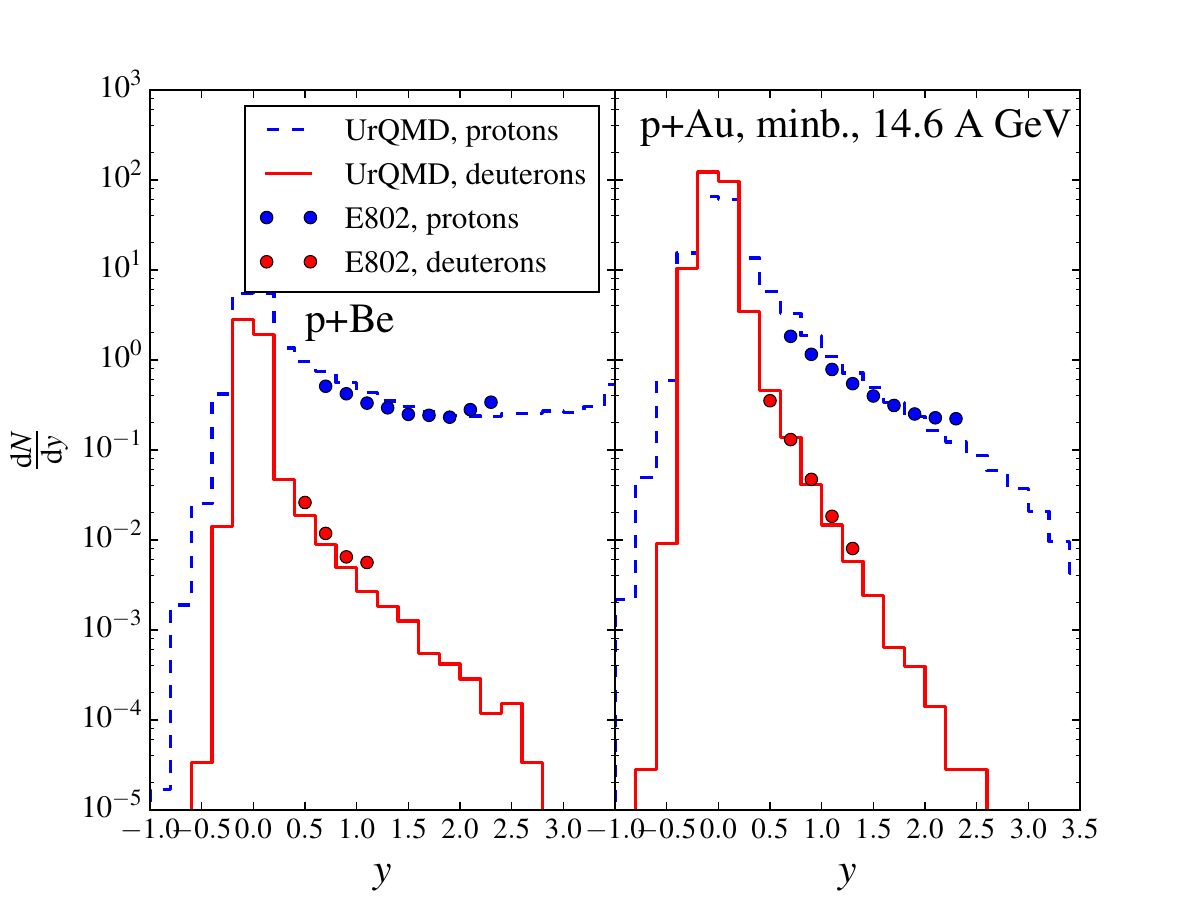}
  \vspace{-11.5mm}
  \caption{Rapidity distributions of protons and deuterons in minimum bias p+Be (\textit{left}) and p+Au (\textit{right}) collisions at a beam energy of $\elab = 14.6\ A\rm{ GeV}$ ($\snn = 5.4\ \rm{GeV}$). Blue dashed and red solid lines show results for protons and deuterons, respectively, obtained from the \texttt{UrQMD} model, compared to data from the E892 experiment (blue and red dots for protons and deuterons, respectively)~\cite{E-802:1991unu}. Figure from Ref.~\cite{Sombun:2018yqh}.
  }
  \vspace{-4mm}
  \label{fig:UrQMDfig}
\end{wrapfigure}
parameters fit to data. Many of these models employ abrasion-ablation models~\cite{Hufner:1975zz,Werneth:2021} (where abrasion and ablation refer to particle removal in ion-ion interactions and nuclear de-excitation following abrasion, respectively) or semi-empirical parametrizations, see Ref.~\cite{Luoni:2021bne}. Researchers modeling these interactions could benefit from transport codes discussed in this White Paper. The use of hadronic transport models such as the Ultrarelativistic Quantum Molecular Dynamics (\texttt{UrQMD}) code~\cite{Bass:1998ca,Bleicher:1999xi}, which was shown to correctly predict proton and deuteron yields from the BNL Alternating Gradient Synchrotron measurements of collisions of protons on Be and Au targets at $\elab = 15\ A\rm{GeV}$~\cite{Sombun:2018yqh,E-802:1991unu}  (\mbox{$\snn = 5.4\ \rm{GeV}$}), see Fig.~\ref{fig:UrQMDfig}, could significantly advance simulations of collisions relevant for space exploration. For further information about the needs for space applications, see Refs.~\cite{Kolos:2022stv,Smith:2022xyz}.

Similar transport modeling needs arise in charged particle therapy for medical applications such as cancer treatment. In this case, the ion beam is tuned to penetrate the tissue at the tumor location so that the Bragg peak, or maximal dose, is delivered to the tumor site while minimizing the spread of the charged particle beams into surrounding tissue due to target fragmentation and secondary scattering~\cite{Durante:2016}. Transport models can play an important role in improving the effectiveness and safety of charged particle therapy in cancer treatment~\cite{Durante:2016,Malouff:2020}. Moreover, if ions such as carbon are used instead of protons, the beam may also fragment and spread in the body. These interactions are also studied employing abrasion-ablation models. Better models of projectile fragmentation are needed to determine the effect of ion beams on normal tissue. Recently, the Stochastic Mean Field (\texttt{SMF})~\cite{Colonna:1998ase}  and Boltzmann--Langevin One Body (\texttt{BLOB})~\cite{Napolitani:2013ad} models have been coupled with Geant4 for studies related to radiation therapy~\cite{Mancini:2020}.

The nuclear information required for applications falls under the general umbrella term of ``nuclear data". The \texttt{Geant3}, \texttt{Geant4}, and \texttt{FLUKA} codes all utilize information taken directly from nuclear data libraries. However, standard nuclear databases cover almost exclusively neutron-induced reactions, while few charged-particle data are available. In addition, the energy range 
covered by these databases typically only extends to 20~MeV. In higher energy databases such as the 
GSI-ESA-NASA database \cite{Luoni:2021bne}, there are essentially no data for light ions beyond $\elab \approx 3\ A\rm{GeV}$ ($\snn \approx 3\ \rm{GeV}$) and scant data for heavy ions beyond a few hundred $A\rm{MeV}$~\cite{Luoni:2021bne}. Transport models such as Quantum Molecular Dynamics (\texttt{QMD})~\cite{Aichelin:1992ur} and \texttt{PHITS}~\cite{Sato:2013,Sato:2015iqw} have been used to simulate higher-energy collisions to fill the gaps in data. Experiments at nuclear accelerators are needed to verify these calculations.

The U.S. Nuclear Science Advisory Committee (NSAC) has been charged to ``assess challenges, opportunities and priorities for effective stewardship of nuclear data''. As part of the development of the Long Range Plan for nuclear science, town halls involving different sub-fields of the U.S. nuclear physics community have adopted nuclear data resolutions, including a recommendation to identify  cross-cutting opportunities with other programs. We suggest that one of these opportunities is the use of transport codes to advance and enhance high-energy applications, such as space research and advanced medical treatments.

\subsection{Hydrodynamics}
\label{sec:hydrodynamics}
%
 
Relativistic hydrodynamics (an early formulation of which was given by Landau and Lifshitz~\cite{LL6}) can be defined as the effective field theory~(EFT) describing fluids on energy scales much smaller than the fluid temperature~\cite{Kovtun:2012rj}. 
Hydrodynamic equations of motion encode the time evolution of hydrodynamic variables, such as fluid velocity, as well as conserved charges, such as baryon number or electric charge, and their associated currents in spacetime. 

Solving the hydrodynamic equations of motion requires the EOS as a crucial ingredient leading to a closed system of equations, which fully constrain the dynamics of the system (in other words, the EOS is needed to have as many equations as there are independent degrees of freedom). Thus, in turn, hydrodynamics can {\it in principle} be used to constrain the EOS.  
While this task has proven to be difficult, some of the most prominent attempts relevant for the quark-gluon plasma~(QGP) generated in heavy-ion collisions include those presented in Refs.~\cite{Pratt:2015zsa,Alba:2017hhe,Busza:2018rrf,Spieles:2020zaa,Monnai:2021kgu,An:2021wof}, while some attempts in the context of neutron star mergers can be found in Refs.~\cite{Baiotti:2016qnr,Font:2008fka,Oechslin:2006uk}.  
In modern relativistic viscous fluid dynamics applied to describing heavy-ion collisions, out-of-equilibrium quantities such as the shear stress tensor are considered as independent variables evolving according to the hydrodynamic equations of motion. Following the hydrodynamic phase the fluid is {\it particlized} and hadron transport codes are used to describe the particles at low temperatures~\cite{Petersen:2008dd}.
However, fundamental questions remain on the limitations of hydrodynamics at ever lower collision energies and higher baryon as well as isospin densities.

\subsubsection{Status} 
%
Hydrodynamics has had a great success describing nuclear matter generated in heavy-ion collisions over a wide range of energies, (see, e.g., \cite{Heinz:2013th,Luzum:2013yya,Gale:2013da,DerradideSouza:2015kpt}).
Remarkably, hydrodynamics applies to various system sizes accessible in heavy-ion collisions~\cite{Bozek:2011if,Kozlov:2014fqa,Dusling:2015gta,Weller:2017tsr,Song:2017wtw,Schenke:2020mbo,Sievert:2019zjr}, with ALICE, ATLAS, and CMS experiments showing collective fluid behavior in proton-ion~\cite{ALICE:2012eyl,ATLAS:2012cix,CMS:2015yux} and even proton-proton collisions~\cite{ALICE:2016fzo,ATLAS:2015hzw,CMS:2015fgy}, which was also successfully reproduced hydrodynamically~\cite{Kozlov:2014fqa,Bozek:2016kpf,Weller:2017tsr,Song:2017wtw,Schenke:2020mbo}. 
Collective behavior in small systems was also observed at RHIC by the PHENIX and STAR experiments~\cite{PHENIX:2013ktj,STAR:2022pfn}. 

It was realized early on that first-order hydrodynamics (in Landau or Eckart frame) is causality-violating and unstable~\cite{Hiscock:1985zz}. 
At this time, the standard solution to this problem is the M\"uller-Israel-Stewart (MIS) theory~\cite{Muller:1967zza,Israel:1976tn,Israel:1979wp}, or versions thereof~\cite{Denicol:2012cn,Baier:2007ix}, which are used in most hydrodynamic codes modeling heavy-ion collisions. 
In the MIS theory, transient modes are added as regulators ensuring a causal time evolution~\cite{Florkowski:2017olj}. 
The behavior of such transient modes depends on the way they are introduced~\cite{Denicol:2012cn,Florkowski:2017olj} (and, in particular, they may encode physical behavior when derived using a microscopic theory, e.g., kinetic theory~\cite{Israel:1979wp,Denicol:2012cn,Wagner:2022ayd}).
Since transient modes are generally not associated with any conserved quantities, their behavior is not what hydrodynamics aims to describe.
MIS thus relies on these transient modes to decay sufficiently fast for the observables to behave hydrodynamically. 
This poses a problem for MIS at early times in a heavy-ion collision, when the regulator transient modes are still present, because observables sensitive to the early times may reflect the physics of these regulators. 
In addition, the causality violation~\cite{Plumberg:2021bme} and stability~\cite{Almaalol:2022pjc} in these setups has to be monitored when modeling, for example, heavy-ion collisions. 

Alternatively, a more direct approach to constructing causal viscous hydrodynamics is based on the realization that hydrodynamics is causal when considered in a general frame (and not, e.g., Landau frame or Eckart frame). 
In that case it is not necessary to introduce any regulator or auxiliary fields, as the differential equations governing the hydrodynamic fields (temperature, fluid velocity, and chemical potential) are hyperbolic (i.e., there exists a solution for all times) and their time evolution is causal by construction. This leads to the Bemfica--Disconzi--Noronha--Kovtun (BDNK) theory~\cite{Bemfica:2017wps,Bemfica:2019knx,Kovtun:2019hdm,Hoult:2020eho,Bemfica:2020zjp,Abbasi:2022rum}, which is capable of, for example, modeling neutron star mergers~\cite{Bemfica:2020zjp,Pandya:2021ief}. 
BDNK also has a practical use in constructing manifestly causal numerical codes solving hyperbolic equations. 
Note, however, that BDNK is merely a causal formulation of hydrodynamics, and thus BDNK is still not expected to be a good approximation at early times.  
Furthermore, BDNK (just like MIS) contains a non-hydrodynamic sector; in fact, this statement is true for every causal stable formulation of dissipative hydrodynamics~\cite{Heller:2022ejw}.

Finally, a rigorous field-theory formulation of hydrodynamics was achieved, which expresses it as an EFT based on a generating functional~\cite{Jensen:2012jh,Banerjee:2012iz,Grozdanov:2013dba,Kovtun:2016lfw,Haehl:2015pja,Crossley:2015evo,Jensen:2018hse} (see Ref.~\cite{Liu:2018kfw} for a summary). This approach employs the  nonequilibrium Green's function (also known as the Schwinger--Keldysh) formalism of thermal field theory.  
As applications of this formulation, effects of stochastic interactions on hydrodynamic correlation functions~\cite{Jain:2020zhu} as well as a theory of non-linear diffusion were derived, taking into account large hydrodynamic fluctuations (for example, leading to the dependence of transport coefficients on fluctuations of the hydrodynamic fields)~\cite{Vogel:2007yq,Chen-Lin:2018kfl,Roch:2020zdl,Abbasi:2022aao}. 

As an invaluable theoretical and practical tool, Kubo formulae~\cite{Kubo:1957mj} are used to relate macroscopic transport coefficients (such as shear or bulk viscosity) to the correlation functions of the underlying (potentially strongly-interacting) microscopic theory. In particular, transport coefficients for QCD and similar theories can be consistently incorporated into relativistic hydrodynamics and magnetohydrodynamics as well as used in numerical hydrodynamic codes. In the limit of an infinite ’t~Hooft coupling, correlation functions were computed and Kubo formulae used to derive transport coefficients of $\mathcal{N} = 4$ SYM theory using the gauge/gravity correspondence  (also known as AdS/CFT or holography)~\cite{Policastro:2001yc, Policastro:2002se, Policastro:2002tn, Kovtun:2004de, Teaney:2006nc, Baier:2007ix, Bhattacharyya:2007vjd, Buchel:2007mf, Haack:2008xx, Erdmenger:2008rm, Banerjee:2008th, Romatschke:2009kr, Kanitscheider:2009as, Bigazzi:2010ku, Arnold:2011ja, Finazzo:2014cna, Wu:2016erb}; moreover, finite coupling corrections are likewise known~\cite{Buchel:2008sh, Buchel:2008bz, Buchel:2008kd, Saremi:2011nh, Grozdanov:2014kva, Grozdanov:2016fkt,Grozdanov:2016vgg,Stephenson:2021xbx}. 
Transport coefficients have also been computed within the kinetic theory using the Boltzmann equation in the relaxation time approximation~\cite{Baier:2007ix, Romatschke:2011qp, Jaiswal:2014isa}, in perturbative QCD in the limit of a small gauge coupling as well as considering finite coupling corrections~\cite{Arnold:2003zc, York:2008rr, Romatschke:2009ng, Moore:2012tc}, from lattice QCD (SU(3) Yang-Mills theory on the lattice)~\cite{Nakamura:2004sy, Meyer:2007ic, Meyer:2011gj, Philipsen:2013nea, Borsanyi:2013bia, HotQCD:2014kol, Ding:2015ona,Borsanyi:2018srz}, and by applying non-perturbative quantum field theory approaches to compute correlators used then in Kubo formulae~\cite{Christiansen:2014ypa}. Kubo formulae were also derived for thermodynamic transport coefficients~\cite{Kovtun:2016lfw,Kovtun:2018dvd}, as well as for chiral hydrodynamics in strong external magnetic fields~\cite{Ammon:2020rvg} and for (chiral) magnetohydrodynamics~\cite{Grozdanov:2016tdf,Hernandez:2017mch,Das:2022fho,Landry:2022nog}.

\subsubsection{Range of applicability} 
Many factors influence whether a system may be described hydrodynamically. Most importantly, like any EFT, hydrodynamics requires a separation of scales between the microscopic physics and the scales on which the system is described. Let us focus on two remarkable results regarding the range of applicability of hydrodynamics:   \vspace{-2mm}
\begin{itemize}
\item[1)] The {\it unreasonable effectiveness of hydrodynamics} far away from equilibrium.  \vspace{-2mm}
\item[2)] The possibility to extend the range of applicability of hydrodynamics by considering its systematic extensions.  \vspace{-2mm}
\end{itemize}
%

Regarding 1), the applicability of hydrodynamics has been historically tied to a requirement of a near-equilibrium state, near-isotropy, and small gradients. Astonishingly, heavy-ion collisions, where neither of these three conditions is met, 
\begin{wrapfigure}{r}{0.6\textwidth}
	\centering
	\vspace{-5mm}
	\includegraphics[width=0.6\textwidth]{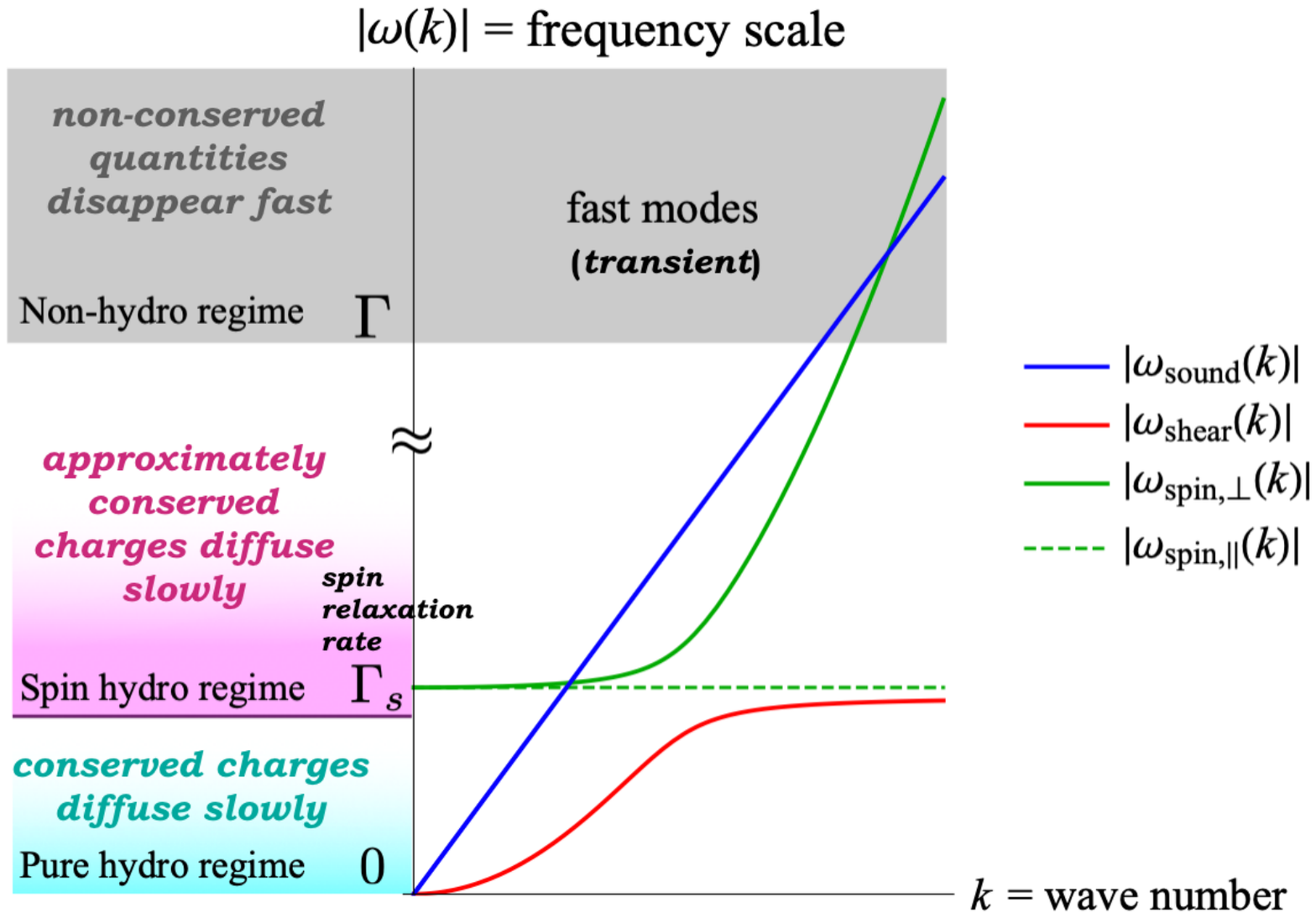}
	\vspace{-13.5mm}
	\caption{Example of an extension of the regime of applicability of hydrodynamics: spin hydrodynamics. While standard hydrodynamics is valid at small frequencies and momenta (labeled as ``pure hydro regime'', indicated by cyan blue region), in the presence of spin degrees of freedom spin hydrodynamics is valid in an extended regime (labeled as ``spin hydro regime'', indicated by a pink region). This is facilitated by adding the slowly relaxing spin modes (green curves) to the spectrum of standard hydrodynamic shear (red solid curve) and sound (blue solid curve) modes. HYDRO+ is constructed in a similar way by adding modes which relax slower and slower when approaching the critical point in the QCD phase diagram, bearing implications for the EOS~\cite{Rajagopal:2019xwg,Abbasi:2021rlp}. Figure adapted from~\cite{Hongo:2021ona}.
	}
	\vspace{-4mm}
	\label{fig:extendedHydroExample}
\end{wrapfigure}
were successfully described hydrodynamically, which is often referred to as the {\it unreasonable effectiveness of hydrodynamics}~\cite{Noronha-Hostler:2015wft}.  
As a possible explanation, hydrodynamic attractors were proposed~\cite{Heller:2015dha} in a conformal MIS theory, and subsequently studied for heavy-ion collisions~\cite{Romatschke:2017acs,Blaizot:2019scw,Kurkela:2019set,Giacalone:2019ldn,Almaalol:2020rnu,Blaizot:2020gql,Jankowski:2020itt,Ambrus:2021sjg,Blaizot:2021cdv,Chattopadhyay:2021ive,Jaiswal:2021uvv} at vanishing baryon densities, and more recently also at finite baryon densities~\cite{Dore:2020jye,Du:2020zqg,Chattopadhyay:2022sxk}. The underlying reason for the attractor behavior is proposed to be kinematic, {i.e.},~owed to a fast expansion in the boost-invariant plasma~\cite{Blaizot:2017ucy,Kurkela:2019set,Jaiswal:2019cju,Chattopadhyay:2019jqj,Chattopadhyay:2021ive,Jaiswal:2021uvv,Spalinski:2022cgj}, combined with exponential decay of non-hydrodynamic modes. Since systems cease to be boost invariant as the collision energy is lowered, the development of hydrodynamic attractors in nuclear matter at low to intermediate energies may or may not occur.

Inspired by the gauge/gravity correspondence~\cite{Heller:2013fn}, the position-space hydrodynamic expansion (in proper time) around the Bjorken flow within the MIS theory was shown to diverge factorially~\cite{Heller:2015dha} (the approach has been further generalized in~\cite{Heller:2020uuy,Heller:2021oxl}). In contrast, for the same factorially divergent holographic theory Fourier-transformed into the momentum space, there is a finite convergence radius limited by the branch point singularity closest to the origin in the complex momentum plane~\cite{Withers:2018srf,Grozdanov:2019kge,Heller:2020hnq,Heller:2020uuy,Heller:2021yjh}. Along this line of ideas based on convergence, the formulation of far-from-equilibrium hydrodynamics \textit{via} resummation was suggested for conformal MIS~\cite{Heller:2015dha}, which was later generalized 
to kinetic theory and holographic models~\cite{Romatschke:2017vte}; such a resummation scheme was already proposed earlier~\cite{Lublinsky:2007mm}. 
Note that in order not to rely on any gradient or inverse proper time expansion, relativistic hydrodynamics can also be phrased in terms of expansions in Knudsen number and inverse Reynolds number~\cite{Denicol:2012cn}.

Regarding 2), a standard method to extend the regime of validity of hydrodynamics is to add one or several mode(s). In fact, promoting the shear tensor to an auxiliary field (regulator) adds a mode to the spectrum of first-order hydrodynamic formulation yielding the MIS model. In another crucial example, critical fluctuations need to be taken into account near the critical point in the QCD phase diagram, and a set of slow modes can be added to the hydrodynamic modes yielding HYDRO+~\cite{Stephanov:2017ghc,Stephanov:2004wx}, which in turn bears implications for the EOS and the speed of sound~\cite{Rajagopal:2019xwg,Abbasi:2021rlp}. Furthermore, Lambda hyperon polarization data~\cite{STAR:2017ckg} 
indicates that the QGP is highly vortical and polarized~\cite{Karpenko:2016jyx,Becattini:2017gcx,Becattini:2021iol}, which motivated the inclusion of spin in various hydrodynamic descriptions~\cite{Florkowski:2017ruc,Montenegro:2020paq,Bhadury:2020cop,Weickgenannt:2020aaf,Fukushima:2020ucl,Li:2020eon,Gallegos:2021bzp,Hongo:2021ona,Gallegos:2022jow,Wang:2021wqq,Weickgenannt:2022zxs,Weickgenannt:2022qvh} (see, e.g., Fig.~\ref{fig:extendedHydroExample}). 
Within a different systematic extension of hydrodynamics, dynamical electromagnetic fields can be added, leading to versions of magnetohydrodynamics which couple the hydrodynamic conservation equations to Maxwell's equations~\cite{Grozdanov:2016tdf,Hernandez:2017mch,Armas:2018zbe}, and which can also include the chiral anomaly~\cite{Hattori:2017usa,Das:2022fho,Vardhan:2022wxz}, relevant for neutron stars~\cite{Vardhan:2022wxz}. 
Note that various implementations of different versions of magnetohydrodynamics exist, for example~\cite{Inghirami:2016iru,Rogachevskii:2017uyc,DelZanna:2018dyb,Nakamura:2022ssn,Chabanov:2022twz}. 
Finally, another natural extension of hydrodynamics is the simultaneous inclusion of multiple conserved charges, in particular, baryon number $B$, strangeness $S$, and electric charge $Q$~($BSQ$ charges)~\cite{Almaalol:2022pjc}. This renders transport coefficients matrix-valued, which means that gradients in one charge may lead to diffusion of another charge~\cite{Greif:2017byw,Fotakis:2019nbq}.

Modern hydrodynamics has been developed in close relation to the gauge/gravity correspondence. This development, which notably yielded the only consistent theoretical description of fluids with $\eta/s$ as low as found in heavy-ion data~\cite{Gyulassy:2004zy,Muller:2006ee,Luzum:2008cw}, began with the insight that a lower bound on entropy production per degree of freedom ($\eta/s$) for a certain class of theories is related to black branes in the Anti de Sitter (AdS) spacetime~\cite{Kovtun:2003wp}. (It~is important to stress that the value $\eta/s=1/(4\pi)$ is {\it not} a universal lower bound~\cite{Buchel:2008vz}. Instead, $\eta/s=1/(4\pi)$ is the lower bound for a certain class of theories which, roughly speaking, can be characterized as those gauge theories which have Einstein gravity as their gravity dual~\cite{Buchel:2003tz,Iqbal:2008by}.)
The fluid/gravity correspondence~\cite{Bhattacharyya:2007vjd} as a systematic construction tool led to the discovery of the chiral vortical effect~\cite{Erdmenger:2008rm,Son:2009tf} and the re-discovery of the chiral magnetic effect~\cite{Kharzeev:2004ey,Son:2009tf}.  
Holographic models are also suitable for exploration of plasmas at high densities~\cite{Son:2006em}, 
phase transitions (in particular a holographic version of the QCD critical point~\cite{DeWolfe:2010he}), 
neutron stars~\cite{Jarvinen:2021jbd}, taking into account finite coupling~\cite{Grozdanov:2016zjj,Grozdanov:2021gzh}, and for investigating the 
far-from-equilibrium regime of holographic plasmas~\cite{Janik:2006gp,Chesler:2010bi,Chesler:2016ceu}.

\subsubsection{Challenges and opportunities} 
Given the recent developments described above, there are strong reasons to assume that hydrodynamics either is valid or can be extended to be valid for the description of dense nuclear matter at intermediate energy scales, even in small systems with large gradients, far from equilibrium, and near the QCD critical point. 
Such (extended) versions of hydrodynamics may well overlap with the regime of validity of hadronic transport simulations, which needs to be studied. 
Here, in particular, further development of hybrid approaches using both hydrodynamics and hadronic transport will contribute to a better description of intermediate energy heavy-ion collisions.

The way ahead will require pushing forward the development of the rigorous theoretical formulation of hydrodynamics, as well as testing its applicability with exactly solvable models (e.g., constructed using the gauge/gravity correspondence) and, most importantly, against experimental data. 
By continuing the development of hydrodynamics in parallel with gauge/gravity models, the proposed versions of spin hydrodynamics can be tested and constructed rigorously using the correspondence; the same statement also applies to versions of magnetohydrodynamics. 
This approach can also reveal further effects of hydrodynamic attractors with implications for, for example, the stiffness of the EOS or the speed of sound~\cite{Cartwright:2022hlg}. 
In the context of (magneto)hydrodynamics, one may also explore the interplay of multiple conserved charge currents and anomalous currents, leading to novel transport phenomena~\cite{Kharzeev:2010gd,Neiman:2010zi,Jensen:2013vta,Ammon:2020rvg,Landry:2022nog,Das:2022fho}. 
For an efficient modeling of heavy-ion collisions (as well as neutron stars and neutron star mergers), the BDNK approach needs to be developed and implemented in standard codes for data analysis. 
At high densities, it becomes necessary to describe the propagation of multiple conserved charges, in particular, the $BSQ$ charges~\cite{Almaalol:2022pjc}. Consequently, the initial state used in numerical hydrodynamic simulations must be modified to include $BSQ$ degrees of freedom~\cite{Shen:2017bsr,Martinez:2019jbu,Carzon:2019qja}. Similarly, the EOS~\cite{Monnai:2019hkn,Noronha-Hostler:2019ayj} and the exact charge conservation when particles are formed (see, e.g., Ref.~\cite{Oliinychenko:2019zfk,Oliinychenko:2020cmr}) need to take into account $BSQ$ charges. 
At the same time, intermediate and low-energy versions of hydrodynamic codes need to be developed, extending initial efforts such as the multi-fluid model~\cite{Batyuk:2016qmb,Cimerman:2023hjw}. 
Beyond describing all conserved charges, theoretical consistency on one hand and the need to describe systems far-from-equilibrium on the other hand both necessitate a rigorous treatment of hydrodynamic fluctuations, which has been done using a deterministic approach to fluctuations \cite{An:2019osr,An:2019csj,An:2020vri,An:2022jgc}. As a viable future complementary approach, hydrodynamic fluctuations can be included using the nonequilibrium Green's function formulation of hydrodynamics~\cite{Liu:2018kfw}. 
These goals are in line with two recent white papers: {\it Snowmass Theory Frontier: Effective Field Theory Topical Group Summary}~\cite{Baumgart:2022yty} and {\it Snowmass White Paper: Effective Field Theories for Condensed Matter Systems}~\cite{Brauner:2022rvf}.

\newpage
\section{Exploratory directions}
\label{sec:exploratory_directions}

\subsection{Dense nuclear matter EOS meeting extreme gravity and dark matter in supermassive neutron stars}
\label{sec:extreme_gravity_and_dark_matter}

Do we need an independent determination of the nuclear EOS using terrestrial experiments in the era of high-precision multi-messenger astronomy? While it is often emphasized that combined data analyses of heavy-ion reactions and neutron star observations within a unified EOS theory framework are a powerful tool to study the EOS (see Section \ref{sec:combined_constraints}), the independent extraction of the nuclear EOS from heavy-ion reactions alone is fundamentally important. This assertion is motivated by a well-known degeneracy \cite{Shao:2019gjj} between the EOS of dense matter (including hadronic and/or quark matter, and dark matter) and strong-field gravity in studies aimed at understanding properties of super-massive neutron stars, the minimum mass of black holes, and properties of dark matter \cite{Bertone:2007ae, Kouvaris:2007ay, Bertoni:2013bsa, McKeen:2018xwc, Horowitz:2019aim, Singh:2022wvw, Das:2020vng, Karkevandi:2021ygv, Miao:2022rqj}.

In the \textit{Astro2020 Science White Paper on Extreme Gravity and Fundamental Physics} \cite{Sathyaprakash:2019yqt}, 
future gravitational wave~(GW) observations are envisioned to enable unprecedented and unique science related to \vspace{-2mm}
\begin{itemize}
\item \textit{The nature of gravity: Can we prove Einstein wrong? What building-block principles
and symmetries in nature invoked in the description of gravity can be challenged?} \vspace{-2mm}
\item \textit{The nature of dark matter: Is dark matter composed of particles, dark objects, or modifications of gravitational interactions?}
\vspace{-2mm}
\item \textit{The nature of compact objects: Are black holes and neutron stars the only astrophysical extreme compact objects in the Universe? What is the EOS of densest matter?}
\vspace{-2mm}
\end{itemize}
An independent determination of the EOS of dense nuclear matter from terrestrial experiments, which are free from gravitational effects, will address the question of whether exotic physics, such as modified gravity, is necessary to describe the behavior and phenomena of supermassive stars. Thus constraining the EOS from heavy-ion collision experiments will help realize the astrophysical science goals.

The fundamental questions listed above are among the eleven greatest physics questions for the new century identified by the U.S.\ National Research Council in 2003 \cite{Council2003}. While gravity was the first force discovered in nature, the quest to unify it with other fundamental forces remains elusive, partially because of its apparent weakness at short distances~\cite{Arkani-Hamed:1998jmv,Hoyle:2003dw}. Moreover, while Einstein's general relativity (GR) theory for gravity has successfully passed all observational tests so far, it is still not fully tested in the strong-field domain \cite{Psaltis:2008bb}. Searches for evidence of possible deviations from GR are at the forefront of several fields in natural sciences. It is fundamentally important to test whether GR will break down at the strongest possible gravitational fields reachable. For this goal, supermassive neutron stars are among the ideal testing sites \cite{Yunes:2009ch,Jimenez:2021wik}. However, as already mentioned above, their properties can be accounted for by either modifying gravity, adding dark matter, and/or adjusting the nuclear EOS. Thus, an independent inference of the nuclear EOS from terrestrial experiments is fundamentally important for breaking the degeneracy between the EOS of supermassive neutron stars and the strong-field gravity.

There are already some indications that the EOS of dense neutron-rich matter may play a significant role in  understanding the nature of gravity \cite{Wen:2009av,Krastev:2007en,Xu:2012wc,Jiang:2012hy}. Effects of the nuclear symmetry energy on the gravitational binding energy \cite{Newton:2009vz}, surface curvature, and red shift \cite{He:2014yqa}, which are normally used to measure the strength of gravity of massive stars in GR, as well as examples of mass-radius relations in several classes of modified gravity theories are reviewed briefly in Ref.~\cite{Li:2019xxz}. More precise information about the dense nuclear matter EOS from terrestrial experiments will enable further progress in this direction.

\subsection{Nuclear EOS with reduced spatial dimensions}

Nuclear systems under constraints, with high degrees of symmetry and/or collectivity, may be considered as effectively moving in spaces with reduced spatial dimensions. Historically, in developing modern methodologies, the spatial dimension $d$ has been considered to be either a continuous or a discrete variable.
Many exciting and fundamentally new experimental discoveries in reduced dimensions have been made in recent years in material sciences (e.g., the graphene \cite{Sarma2011} and topological insulator \cite{Hasan2010,Qi2011}) and cold atom physics (e.g., the superfluidity in a strongly correlated 2D Fermi gas \cite{Sobirey2021} and the generalized hydrodynamics in a strongly interacting 1D Bose gas~\cite{Malvania2021}).

The EOS of neutron-rich matter in spaces with reduced dimensions can be linked to that in the conventional 3-dimensional (3D) space by the $\epsilon$-expansion ($\epsilon=d-4$) \cite{Wilson:1973jj,MaSK1976,Wallace1976}. The latter is a perturbative approach that has been successfully used in treating second-order phase transitions and related critical phenomena in solid state physics and, more recently, in studying the EOS of cold atoms in 1D, 2D, and two-species Fermi and/or Bose gases with mixed dimensions \cite{Nishida:2006br,Nishida:2008kr}. The energy per nucleon $E(n_B,\delta,d)$ in cold nuclear matter of dimension $d$ at density $n_B$ and isospin asymmetry $\delta$ can be expanded around $n_B = n_0$, $\delta=0$, and $d=3$. In cold symmetric nuclear matter, the $E(n_B,\delta=0,d)$ is predicted to decrease with decreasing $d$, indicating that nuclear matter with a smaller $d$ tends to be more bound but, at the same time, saturates at a higher 3D-equivalent density. The symmetry energy was also found to become smaller in spaces with lower dimensions compared to the conventional 3D case \cite{Cai:2022gjr}.

Can we find or make 1D and/or 2D nuclear systems in our 3D world? Can nucleons in neutron-skins of heavy nuclei be considered as living in spaces with reduced spatial dimensions, and if so, can we discover the related effects in heavy-ion collisions? Can some of the objects (e.g., lasagna) in the predicted pasta phase \cite{Caplan:2016uvu,Lopez:2020zne,Newton:2014iha} of the neutron star crust be described as nuclear systems with 1D, 2D, or fractional dimensions?  What are the roles of the dimension-dependent EOS in multi-dimensional models of late stellar evolution? If 1D/2D simulations using 3D EOS do not lead to supernova explosions, what will happen if the corresponding 1D/2D EOSs are used instead? 
Answers to these questions may provide new perspectives on the EOS of neutron-rich matter in 3D and help solve some of the unresolved puzzles.

\subsection{Interplay between nucleonic and partonic degrees of freedom: SRC effects on nuclear EOS, heavy-ion reactions, and neutron stars}
\label{sec:SRC_effects_on_nuclear_EOS}

Short-range correlations (SRCs) in nuclei, that is correlations in the nuclear ground-state wave function, are mostly due to isosinglet neutron-proton pairs that have temporally fluctuated into a high-relative-momentum state with approximately zero total center-of mass-momentum and a spatial separation of about 1 fm \cite{Brueckner:1955zzd,Bethe:1971xm,Frankfurt:1981mk,Hen:2014nza,CiofidegliAtti:2015lcu,Hen:2016kwk}. 
Subnucleonic degrees of freedom are expected to play an important role in understanding SRC-related phenomena. Altered quark momentum distributions in nucleons embedded in nuclei with respect to those in free nucleons, known as the EMC effect, have been studied extensively since 1983 \cite{EuropeanMuon:1983wih}.  
SRCs have been proposed as one of the two leading causes of the EMC effect \cite{CiofidegliAtti:1991ae,CiofidegliAtti:1990dh}. Recent experiments found that the strength of SRCs and the EMC effect are strongly correlated \cite{Weinstein:2010rt,CLAS:2019vsb} and that they both depend strongly on the isospin asymmetry of the nuclei. Moreover, strong evidence was found that only the momentum distributions of quarks in SRC nucleon pairs in nuclei are modified with respect to free nucleons. Furthermore, the distributions of quarks in protons of neutron-rich nuclei are modified more than in neutrons, implying that, on average, $u$ quarks are modified more than $d$ quarks in neutron-rich nuclei \cite{CLAS:2019vsb}, in analogy to an earlier finding that SRCs make protons move faster than neutrons in neutron-rich nuclei \cite{CLAS:2018yvt}. These phenomena reflect profound QCD effects in the nuclear medium. 
Studying the flavor- and spin-dependence of nucleon structure functions is at the forefront of QCD and is a major science driver of future EIC experiments. An example of a correlation formed on short-range QCD length scales are quark-quark correlations known as diquarks. It was recently proposed that diquark formation across two nucleons \textit{via} the attractive QCD quark-quark potential is the underlying QCD-level source of SRCs in nuclear matter and the cause of the EMC effect~\cite{West:2020tyo}.

The SRC-related effects have consequences for the nuclear structure, high-energy quark distributions (the EMC effect), and high-density nuclear matter, including its EOS and in-medium nucleon-nucleon scattering cross sections. Better understanding of SRC effects in dense neutron-rich medium through heavy-ion reactions may have important ramifications. The profound consequences of SRCs on the nuclear matter EOS can be easily seen when one considers the well-known Hugenholtz--Van Hove (HVH) theorem, which was derived by assuming there are sharp Fermi surfaces for nucleons. The theorem provides intrinsic connections among the nuclear symmetry energy, momentum dependences of both isoscalar and isovector nucleon potentials, and the corresponding nucleon isoscalar effective mass and neutron-proton effective mass splitting in neutron-rich matter \cite{Xu:2010fh}. However, due to the SRCs nucleons do not have sharp Fermi surfaces, but extended high-momentum tails, as evidenced by many experiments at the Jefferson Laboratory (JLAB) and Brookhaven National Laboratory (BNL), see, e.g., Refs.\ \cite{Arrington:2011xs,Hen:2014nza,Hen:2016kwk,Arrington:2022sov} for recent reviews. Therefore, the above relations may be completely altered by the isospin-dependence of SRCs induced by the nuclear tensor force, which is much stronger in the symmetric nuclear matter than in pure neutron matter \cite{Rios:2009gb,Charity:2006zb,Rios:2013zqa}. Consequently, the composition of the symmetry energy itself (e.g., the ratio of its kinetic over potential parts) may also be very different from the one without considering the SRCs~\cite{Xu:2012hf}. Most of the parametrizations of the nuclear symmetry energy used so far in both nuclear physics and astrophysics adopt the kinetic symmetry energy predicted by the free Fermi gas model. However, such parametrizations neglect SRC effects that may lead to a reduced or even negative kinetic symmetry energy. This effect originates in the fact that SRCs are dominated by isosinglet neutron-proton pairs. As the system becomes more neutron-rich, an increasingly larger fraction of protons, compared to neutrons, are found in the high momentum tail \cite{Hen:2014nza,Hen:2016kwk,CLAS:2019vsb}. Consequently, the kinetic symmetry energy is reduced compared to the free Fermi gas model prediction \cite{Xu:2012hf,Vidana:2011ap,Carbone:2011wk,Carbone:2013cpa,Rios:2013zqa,Hen:2014yfa,Li:2014vua,Cai:2015xga}.

Interesting indications have been found very recently of SRC effects on the cooling of protoneutron stars, the formation of baryon resonances, dark matter, and nuclear pasta as well as on tidal deformation and mass-radius correlation in neutron stars \cite{Souza:2020gjs,Souza:2020eyq,Lourenco:2021dvh,Hong:2022noq,Pelicer:2022euf,Dutra:2022mxl,Lu:2021xvj,Lu:2022ngd}, and also on several features of nuclear matter and heavy-ion reactions  \cite{Yong:2017zgg,Guo:2021zcs,Hagel:2021vjh,Wang:2017odj,Cai:2022cre,Burrello:2022tjw,Cai:2022grw,Zhang:2022tsw}. However, much more work remains to be done to systematically and consistently address the SRC-related issues in hadronic transport simulations (see Section \ref{sec:model_simulations_of_HICs}).
Investigations of SRC effects on the nuclear EOS using heavy-ion collisions at FRIB and FRIB400 will complement the ongoing and planned SRC research programs at JLAB, GSI, and EIC at BNL. Together, these efforts will reveal new knowledge about the spin-isospin dependence of three-body and tensor forces in dense neutron-rich matter. At short distances, these forces are mostly due to the $\rho$-meson exchange \cite{Brown:1990kj}. The in-medium $\rho$-meson mass, determined by QCD, may be significantly different from its free-space value \cite{Brown:1990kj,Brown:1991kk,Rapp:1997ii}. Such modification in the $\rho$-meson mass has been found to significantly affect the high-density behavior of the nuclear symmetry energy \cite{Xu:2009bb,Xu:2012hf}. 
However, effects of the QCD quark-quark potential and the modified tensor or three-body force on in-medium nucleon-nucleon cross sections remain to be explored.

\subsection{High-density symmetry energy above $2n_0$}

Section \ref{sec:experiment_asym} has primarily focused on the physics of nuclear symmetry energy  up to $\approx 2n_0$. This is because of substantial experimental challenges for measuring the symmetry energy using more energetic beams. However, at higher densities, but below the hadron-quark transition density, there are also many interesting issues to be addressed \cite{Li:2014oda,Li:2016eeb}. Therefore, it is worthwhile to explore possible future directions to attack this problem (see also the white paper on \textit{QCD Phase}
\textit{Structure and Interactions at High Baryon Density: Continuation of BES Physics Program with CBM at FAIR} \cite{Almaalol:2022xwv}).

\begin{figure}[!b]
    \centering
    \includegraphics[width=0.99\linewidth]{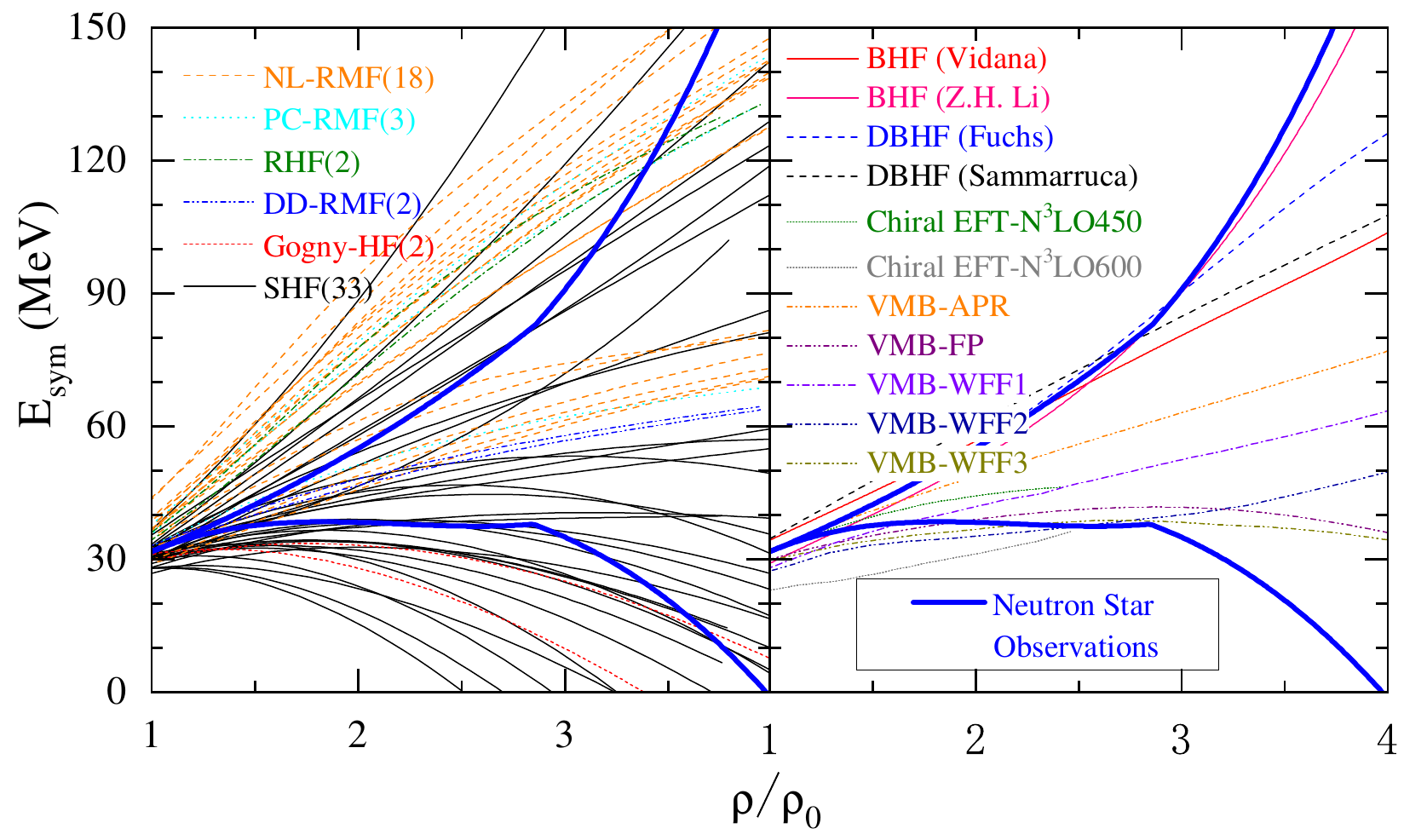}
    \caption{\textit{Left:} Symmetry energy as a function of baryon density as obtained within 60 example models, selected from 6 classes of over 520 phenomenological models and/or energy density functionals. 
    \textit{Right:} Symmetry energy as a function of baryon density as obtained within 11 examples from microscopic and/or {\it ab initio} theories. Thick blue lines are the upper and lower boundaries of symmetry energy from analyses of neutron star observables. Figure from Ref.\ \cite{Zhang:2020zsc}.
    }
    \label{fig:HDEsym}
\end{figure}

Experiments at FRIB400, FAIR, and other high-energy rare isotope beam facilities around the world are expected to provide tremendous resolving power for determining the symmetry energy at densities $\gtrsim 2n_0$. While both the magnitude $E_{\rm sym}(n_0)$ and the slope $L$ of the symmetry energy at $n_0$ have been relatively well determined (with values estimated at $E_{\rm sym}(n_0) \approx 31.7 \pm 3.2\ \txt{MeV}$ and 
$L \approx 58 \pm 19\ \txt{MeV}$~\cite{Li:2013ola, Oertel:2016bki,Li:2017nna,Li:2021thg}, in very good agreement with $\chi$EFT calculations~\cite{Drischler:2020hwi,Drischler:2020yad,Essick:2021ezp,Somasundaram:2020chb}), the curvature $K_{\rm{sym}}$ and skewness $J_{\rm{sym}}$ of the nuclear symmetry energy are still poorly known. In particular, $K_{\rm{sym}}$ is most critical for determining the crust-core transition density and pressure in neutron stars \cite{Xu:2009vi,Providencia:2013dsa,Zhang:2018vbw}. Besides the importance for astrophysics, an experimental determination of the high-density behavior of the nuclear symmetry energy will provide important guidance for developing high-density nuclear many-body theories. Indeed, the density region explored in heavy-ion reactions at BES, HADES, and in the future at FRIB400 and FAIR is mostly beyond the current 
validity range of $\chi$EFT, and it is also where the EOSs predicted by various nuclear many-body theories, especially the symmetry energy contributions, start to diverge broadly (see Fig.~\ref{fig:HDEsym}).

\begin{wrapfigure}{r}{0.55\textwidth}
	\centering
	\vspace{-1mm}
	\includegraphics[width=0.55\textwidth]{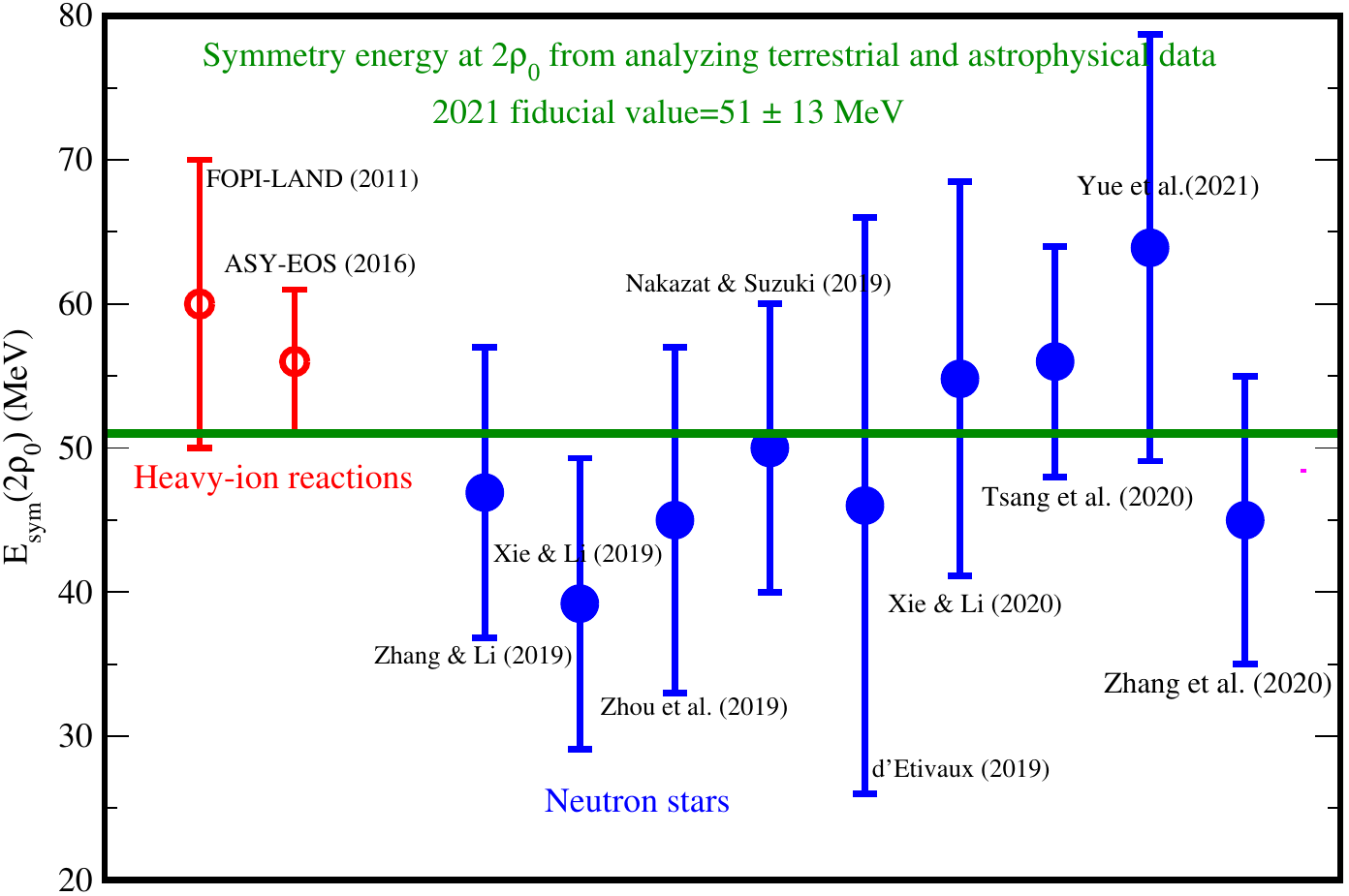}
	\vspace{-9mm}
	\caption{Symmetry energy at twice the saturation density from both heavy-ion reactions and neutron stars. Figure modified from Ref.\ \cite{Li:2021thg}. 
	}
	\vspace{-4mm}
	\label{fig:esym2}
\end{wrapfigure}
Recent neutron star observations have led to some progress in constraining the symmetry energy at suprasaturation densities. Shown in Fig.~\ref{fig:esym2} is a compilation of recent results on the symmetry 
energy at $2n_0$ from two analyses of heavy-ion reactions at GSI and nine independent analyses of neutron star properties by several groups. At 68\% confidence level, these analyses give a mean 
\begin{wrapfigure}{l}{0.45\textwidth}
	\centering
	\vspace{-3.5mm}
	\includegraphics[width=0.44\textwidth]{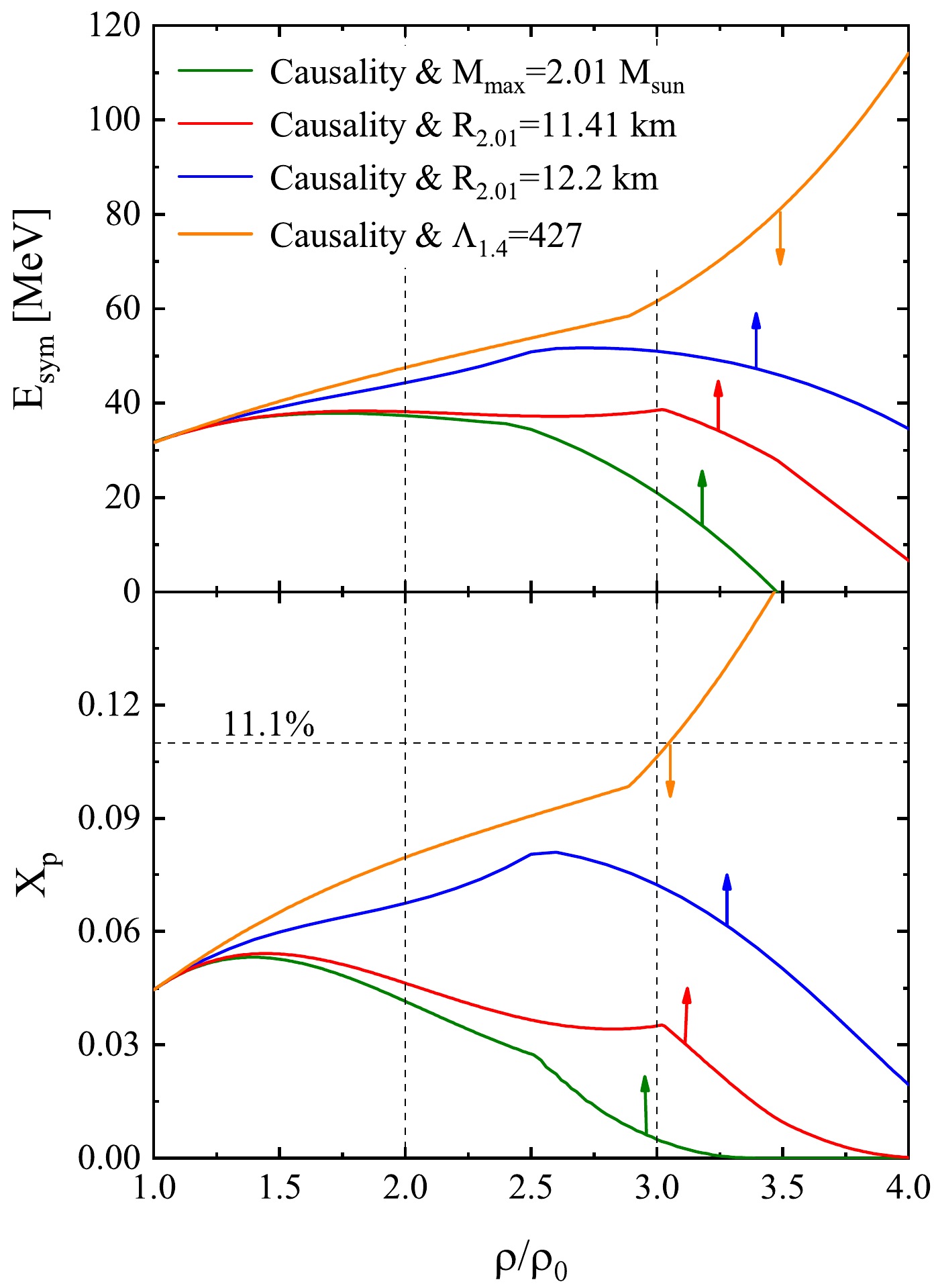}
	\vspace{-1.5mm}
	\caption{Constraints on the high-density symmetry energy and proton fraction in neutron stars from analyzing the tidal polarizability of GW170817 and NICER's observation of PSR J0740+6620. Figure from Ref.\ \cite{Zhang:2021xdt}.
	}
	\vspace{-7mm}
	\label{fig:nicer-esym}
\end{wrapfigure}
value of $E_{\rm{sym}} (2n_0)\approx 51\pm 13$ MeV, as indicated by the green line. Interestingly, $\chi$EFT+MBPT calculations predict a value of $E_{\rm{sym}}(2n_0) \approx 45 \pm 3$ MeV~\cite{Drischler:2021kxf}. Similarly, quantum Monte Carlo calculations using local interactions derived from the $\chi$EFT up to next-to-next-to-leading order predict a value of $E_{\rm{sym}}(2n_0) \approx 46 \pm 4$ MeV~\cite{Lonardoni:2019ypg}. Evidently, the mean value of $E_{\rm{sym}}(2n_0)$ from the analyses mentioned above is consistent with the $\chi$EFT predictions, albeit with large uncertainties. As noted before, $2n_0$ is near the upper validity limit of the current $\chi$EFT theories. Thus, more precise measurements of $E_{\rm{sym}}(2n_0)$ will help to test $\chi$EFT predictions.

Inspecting the results shown in Fig.~\ref{fig:esym2} shows clearly that more work is necessary to reduce the error bars. Most of the neutron star constraints are extracted from radii and tidal deformations of canonical neutron stars with masses around 1.4M$_{\odot}$. These observables are known to be sensitive mostly to the values of pressure around $(1$-$2)n_0$ in neutron stars, and therefore their constraints on $E_{\rm{sym}}(n_B)$ around and above $2n_0$ are not strong.

Observables from more massive neutron stars were expected to place stronger constraints on the high-density symmetry energy. To illustrate how the recent NICER+XMM-Newton's measurements of both the radius and mass of PSR J0740+6620 can influence the constraint on the symmetry energy at densities above $2n_0$, the upper panel of Fig.~\ref{fig:nicer-esym} shows the extracted lower limits of $E_{\rm{sym}}(n_B)$ obtained from directly inverting the TOV equation within a 3-dimensional high-density EOS parameter space \cite{Zhang:2021xdt} for two cases: for the case where only the mass is observed (green line), and for the case where both the mass and radius are observed (red line using the 68\% confidence 
lower radius limit reported by Riley \textit{et al.}~\cite{Riley:2021pdl} and blue line using the radius reported by Miller \textit{et al.}~\cite{Miller:2021qha}). The orange line is the upper limit of the symmetry energy from analyzing the upper limit (68\% confidence) of tidal deformation of GW170817 \cite{Zhang:2018vrx}. The upper limits of the symmetry energy from the upper radius limits reported by both Riley \textit{et al.} and Miller \textit{et al.} are far above the upper limit of symmetry energy from GW170817.

The lower panel in Fig.\ \ref{fig:nicer-esym} shows the corresponding proton fractions in PSR J0740+6620. The influence of knowing both the mass and radius of this most massive neutron star currently known is seen by comparing the green line with the red or blue line, while the difference between the red and blue lines indicates the systematic error from the two independent analyses of the same observational data. Although the estimates of $E_{\rm{sym}}(n_B)$ around $(2$-$3)n_0$ from these analyses are useful compared to the model predictions shown in Fig.~\ref{fig:HDEsym}, much more precise constraints on the $E_{\rm{sym}}(n_B)$ above $2n_0$ are needed.

Pinning down the symmetry energy above $2n_0$ will be very challenging, but achieving this goal will bring a great reward. For example, without a reliable knowledge of the symmetry energy at suprasaturation densities, the density profile of the proton fraction in the core of neutron stars (which has to be higher than about 11\% for the fast cooling to occur) at $\beta-$equilibrium is not determined. Consequently, whether the fast cooling of protoneutron stars occurs through the direct URCA process remains uncertain. Heavy-ion reactions, especially with high-energy radioactive beams, will provide the much-needed data to calibrate nuclear many-body theories and constrain nuclear symmetry energy at densities $\gtrsim 2n_0$. These efforts, in concert with astrophysical research using high-precision X-rays from massive neutron stars (e.g., NICER and STROBE-X \cite{STROBE-XScienceWorkingGroup:2019cyd}), GWs from new LIGO/VIRGO runs and from next-generation detectors such 
\begin{wrapfigure}{r}{0.47\textwidth}
	\centering
	\vspace{-4mm}
	\includegraphics[angle=-90,width=0.47\textwidth]{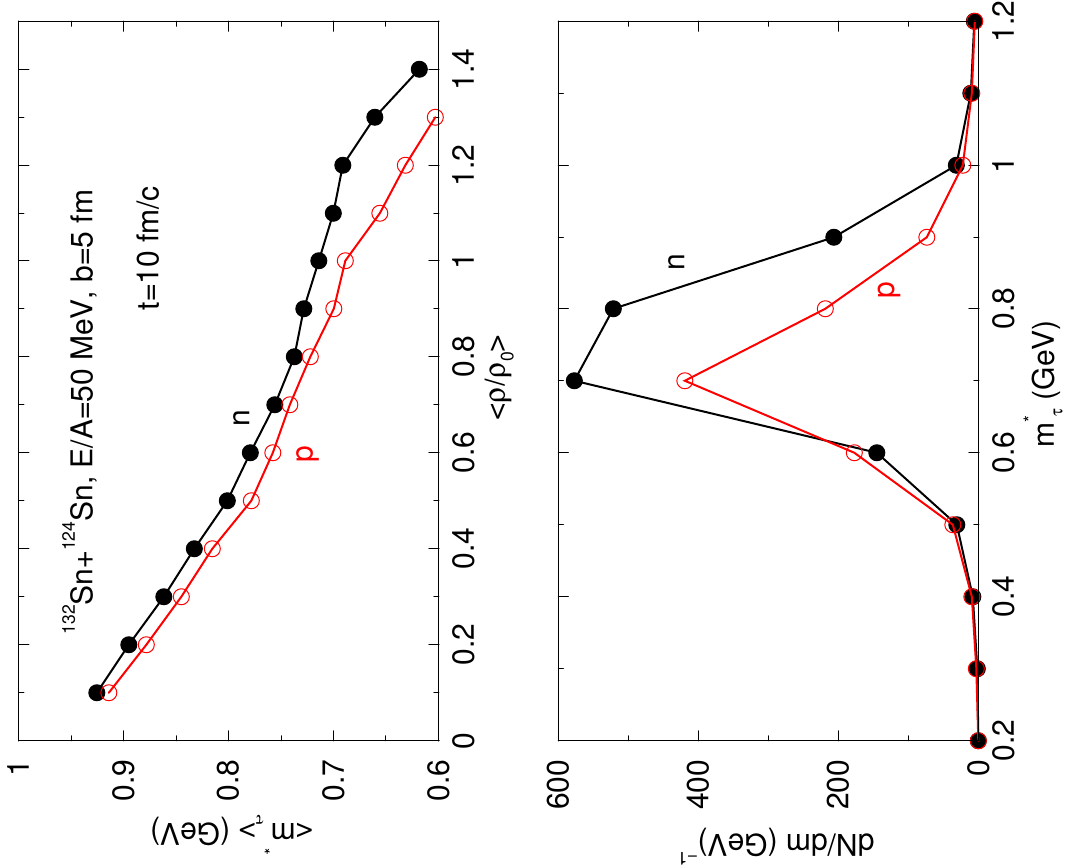}
	\vspace{-3mm}
	\caption{Correlation between the average nucleon effective mass and the average nucleon density (\textit{top}), and the distribution of nucleon effective masses (\textit{bottom}) in the reaction of $^{132}$Sn+$^{124}$Sn at 10 fm/$c$ with a beam energy $\ekin = 50\ A\rm{MeV}$ ($\snn = 1.9\ \txt{GeV}$) and an impact parameter $ b  = 5\ \rm{fm}$, as simulated within the \texttt{IBUU} transport model with an explicitly isospin-dependent single-nucleon potential. Figure from Ref.~\cite{Li:2005jy}.
	}
	\vspace{-15mm}
	\label{ieffect}
\end{wrapfigure}
as the Einstein Telescope and the Cosmic Explorer, and future detection of post-merger high-frequency GWs, will better
constrain $E_{\rm{sym}}(n_B)$ at densities around and above $2n_0$. For these efforts to be fruitful, it is imperative to explore potential observables carrying undistorted information on the symmetry energy above $2n_0$ from neutron stars and their mergers as well as in high-energy heavy-ion reactions.

\subsection{Density-dependence of neutron-proton effective mass splitting in neutron-rich matter}
\label{sec:density-dependence_effective_masses}

The nucleon effective mass is a fundamental quantity characterizing the propagation of a nucleon in a nuclear medium~\cite{Jeukenne:1976uy,Jaminon:1989wj,Sjoberg:1976tq,vanDalen:2005ns}, accounting (to leading order) for effects such as the space-time non-locality of the effective nuclear interactions or Pauli exchange effects. The magnitude and sign of the difference (splitting) between the effective masses of neutrons and protons $\Delta m_{np}^*$ have essential consequences for cosmology, astrophysics, and nuclear physics through influencing, e.g., the equilibrium neutron to proton ratio in the early universe and primordial nucleosynthesis \cite{Steigman:2005uz}, 
properties of mirror nuclei~\cite{Nolen:1969ms}, and the location of drip-lines \cite{Woods:1997cs}. In heavy-ion reactions, $\Delta m_{np}^*$ 
is of importance for isospin-sensitive observables \cite{Li:2003zg, Li:2003ts, Rizzo:2005mk, Giordano:2010pv, Feng:2011xp, Feng:2011mj, Zhang:2014sva, Xie:2014uia}.

The momentum-dependence of the single-nucleon potential is normally characterized by the nucleon effective mass $m^*_{\tau}$ that can be decomposed into an isoscalar and an isovector component \cite{Sartor:1980zza,MAHAUX199253,Li:2018lpy}. Due to our poor knowledge of the momentum dependence of isovector interactions, the isovector nucleon effective mass measured by using the neutron-proton effective mass splitting~$\Delta m^*_{np}$~\cite{Li:2013ola} has not been constrained well \cite{Li:2004zi,Li:2018lpy}. Based on the HVH theorem, $\Delta m^*_{np}$ was 
found approximately proportional to the isospin asymmetry $\delta$ of the medium, with a coefficient depending on the density as well as momentum-dependence of both the isoscalar and isovector nucleon potential \cite{Li:2013ola}. Over the last decade, significant efforts have been made to extract this coefficient at $n_0$. 
A recent survey~\cite{Li:2022gye} of model analyses using data from mostly nucleon-nucleus scattering and giant resonances of heavy nuclei suggests that the $\Delta m^*_{np}$, scaled by the average nucleon mass in free space, ranges from 0 to about~$0.5\delta$~\cite{Whitehead:2020wwb,Xu:2010fh,Charity:2013vha,Li:2014qta,Li:2013ola,Zhang:2015qdp,Kong:2017nil,Xu:2020xib,Xu:2020dhf}.

While experimental efforts to better constrain $\Delta m^*_{np}$ at $n_0$ using heavy-ion reactions with intermediate energy stable 
beams are ongoing (see, e.g., Ref.~\cite{Morfouace:2019jky}), future experiments at FRIB and FRIB400 will enable more sensitive probes of not only $\Delta m^*_{np}$ at $n_0$, but also of its density-dependence (which cannot be probed by the nucleon-nucleus scattering and giant resonances) up to about $2n_0$. As an illustration, shown in Fig.~\ref{ieffect} are the density dependence of the average nucleon 
effective mass (top) and the distribution of the nucleon effective masses (bottom) during a typical FRIB reaction as 
simulated \cite{Li:2005jy} within the \texttt{IBUU} transport model with an explicitly isospin-dependent 
single-nucleon potential \cite{Das:2002fr,Li:2003ts}. 
From the top panel, it is seen that the neutron-proton effective mass splitting is positive and increases with the density 
\begin{wrapfigure}{l}{0.50\textwidth}
	\centering
	\vspace{-5mm}
	\includegraphics[width=1.06\textwidth]{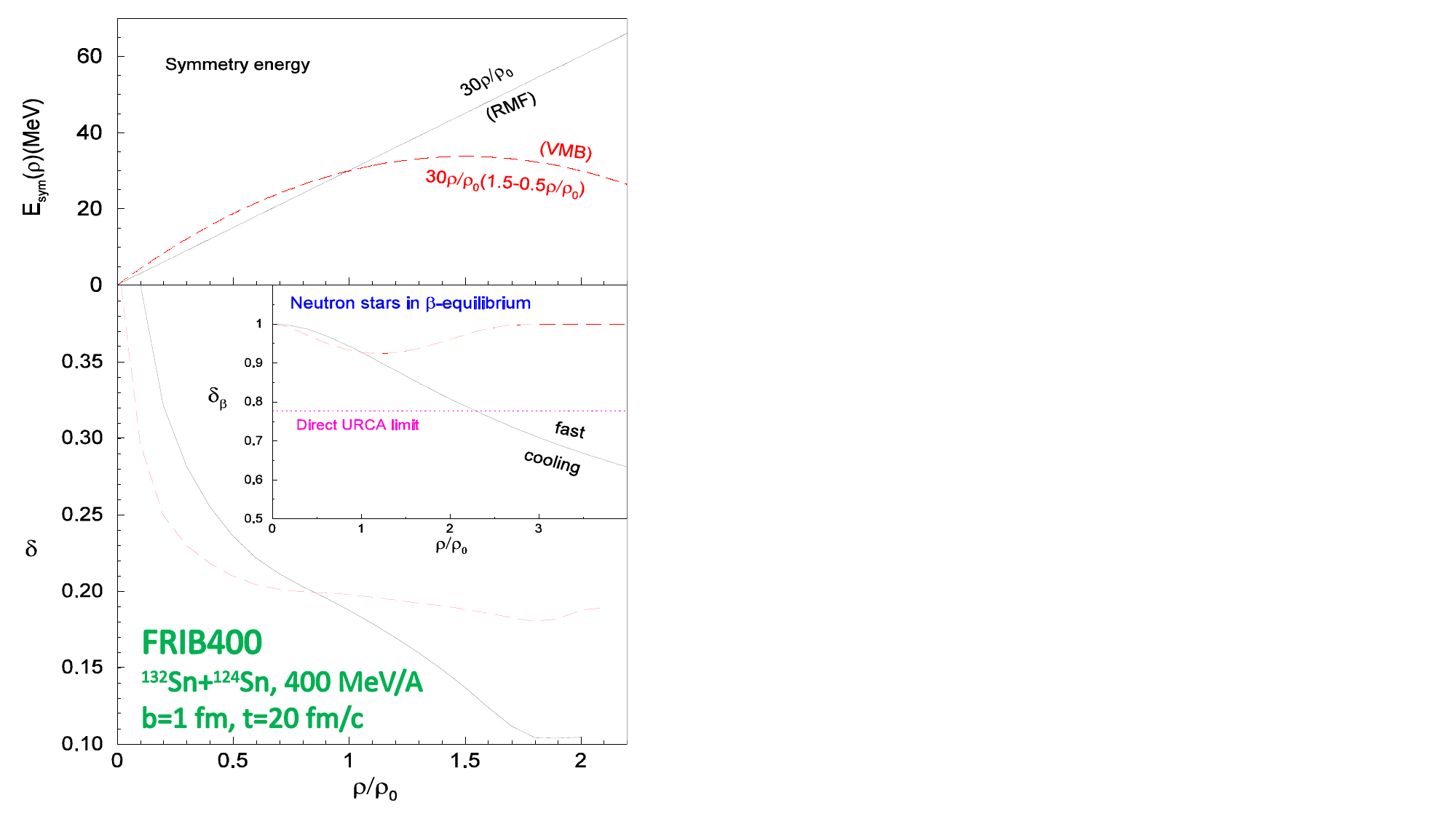}
	\vspace{-8mm}
	\caption{{\it Top:} Density dependence of the nuclear symmetry energy for two typical symmetry energy functionals used in the \texttt{IBUU} simulations. 
		{\it Bottom}: Density dependence of the isospin asymmetry $\delta$ in $^{132}$Sn+$^{124}$Sn collisions at 20 fm/c with a beam energy of 400 MeV/A and an impact parameter of 1 fm, and in the core of neutron stars at $\beta$-equilibrium (\textit{inset}). Figure modified from Ref.\ \cite{Li:2002yda}.
	}
	\vspace{-4mm}
	\label{fig:frib400}
\end{wrapfigure}
up to about $1.3n_0$, consistent with recent $\chi$EFT calculations \cite{Whitehead:2020wwb}. To reach higher densities, more energetic beams are required.

Heavy-ion reactions at FRIB400 will extend the ranges of both density and isospin asymmetry of the medium formed. Shown in the lower panel of Fig.~\ref{fig:frib400} are the isospin asymmetry $\delta$ as a function of density during a typical reaction at FRIB400 (main) and in neutron stars at $\beta$-equilibrium (inset), calculated using the same two typical symmetry energy functionals, shown in the upper panel. 
The $\delta$-$n_B$ relations in both systems show the same isospin fractionation phenomena, e.g., reaching a higher isospin asymmetry when a density functional with a lower symmetry energy is used. One can also see that generally, the low-density regions are more neutron-rich than the high-density regions. These $\delta$-$n_B$ relations are the fundamental origins of all isospin-sensitive observables in both heavy-ion reactions and neutron stars.

A number of observables in heavy-ion reactions have been proposed as promising messengers of the underlying momentum-dependence of the isovector potential and the corresponding neutron-proton effective mass splitting, see, e.g.,~\cite{Baran:2004ih,DiToro:2006zz,Li:2008gp} for reviews. The momentum dependence of the single-nucleon potential affects the reaction dynamics directly through the equations of motion 
and indirectly through the scattering term of nucleons. As the in-medium nucleon-nucleon cross section is proportional to the square of the reduced effective mass of the two colliding nucleons, the nucleon effective mass will affect the nuclear stopping power (which is also described in the literature, especially for nucleon-nucleus scattering, in terms of the nucleon mean free path)~\cite{Negele:1981fkk}.

Consequently, the reaction dynamics and observables of heavy-ion reactions are expected to bear useful information about the density-dependence of the neutron-proton effective mass splitting in neutron-rich matter. The challenge is to find such observables that are both robust and sensitive 
to the variations of the neutron-proton effective mass splitting with density. The nucleon effective mass affects also transport properties of neutron stars, see, e.g., Refs. \cite{Reddy:1997yr,Alford:2022bpp,Hutauruk:2022bii,Sumiyoshi:2021iqq,Shang:2020tvy,Shang:2020kfc}.
Neutron star observables, e.g., neutrino emission and torsional oscillations of neutron stars, may also provide useful information about the density-dependence of neutron-proton effective mass splitting in neutron-rich matter. Explorations of these issues are invaluable.

\newpage
\appendix
\section{Executive Summary}


The nuclear equation of state (EOS) is at the center of numerous theoretical and experimental efforts in nuclear physics, motivated by its crucial role in our understanding of the properties of nuclear matter found on Earth, in neutron stars, and in neutron-star mergers. 
With advances in microscopic theories for nuclear interactions, the availability of experiments probing nuclear matter under conditions not reached before, and the advent of multi-messenger astronomy, the next decade will bring new opportunities for determining the nuclear matter EOS.

\begin{itemize}
    \item \textbf{Profound questions challenging our understanding of strong interactions remain unanswered:}
    It is still unknown whether the transition between a hadronic gas and a quark-gluon plasma, which at zero baryon density is known to be consistent with a crossover transition predicted by Lattice QCD, becomes of first order in the finite-density region of the QCD phase diagram accessible in terrestrial experiments. 
    The isospin-dependence of the EOS, crucial to our understanding of both the structure of neutron-rich nuclei and the properties of neutron stars, is poorly known above nuclear saturation density. 
    Moreover, recent observations of very heavy compact stars indicate that the EOS in neutron-rich matter becomes very stiff at densities of the order of a few times saturation density, leading to values of the speed of sound exceeding $1/\sqrt{3}$ of the speed of light (breaking the conformal limit). 
    Not only is the mechanism behind this striking behavior not known, but it is also unknown whether a similar stiffening occurs in symmetric or nearly-symmetric nuclear matter. \textbf{Resolving these and other questions about the properties of dense nuclear matter is possible by taking advantage of the unique opportunities for studying the nuclear matter EOS in heavy-ion collision experiments. 
    }
    
	\item \textbf{Among controlled terrestrial experiments, collisions of heavy nuclei at intermediate beam kinetic energies (from a few tens of MeV/nucleon to about 25 GeV/nucleon in the fixed-target frame) probe the widest ranges of baryon density and temperature}, enabling studies of nuclear matter from a few tenths to about 5 times the nuclear saturation density and for temperatures from a few to well above a hundred MeV, respectively. 
    In the next decade, numerous efforts worldwide will be devoted to uncovering the dense nuclear matter EOS through heavy-ion collisions, including studies at FRIB where the isospin-dependence of the EOS can be probed in energetic collisions of rare isotopes. 
    \textbf{Modern detectors and refined analysis techniques will yield measurements that will elucidate the dependence of the EOS on density, temperature, and isospin asymmetry.}

    \item \textbf{Hadronic transport simulations are currently the only means of interpreting observables measured in heavy-ion collision experiments at intermediate beam energies}. 
    This means that capitalizing on the enormous scientific effort aimed at uncovering the dense nuclear matter EOS, both at RHIC and at FRIB, depends on the continued development of state-of-the-art hadronic transport simulations. 
	\textbf{Given the imminent results from ongoing and future experimental analyses, there is an urgent need for a theoretical research program that will further inform and accelerate the development of hadronic transport models used to extract the EOS from experimental data. 
	Support for this program is imperative to fully realize the potential of U.S.\ efforts leading the exploration of the dense nuclear matter EOS.}
\end{itemize}

\section*{Acknowledgements}

This White Paper has benefited from talks and discussions at the workshop on \textit{Dense nuclear matter equation of state in heavy-ion collisions} that took place at the Institute for Nuclear Theory (INT), University of Washington (December 5-9, 2022) \cite{INT_workshop}. We thank the INT for its kind hospitality and stimulating research environment.

K.A.\ thanks Hans-Rudolf Schmidt and Arnaud Le F\`{e}vre, and M.S.\ thanks Daniel Cebra for insightful discussions.
P.D.\ and B.T.\ thank Abdou Chbihi, Maria Colonna, Arnaud Le F\`{e}vre, and Giuseppe Verde for discussing complementary international efforts.

A.S.\ thanks J\"{o}rg Aichelin, David Blaschke, Elena Bratkovskaya, Maria Colonna, Dan Cozma, Wick Haxton, Natsumi Ikeno, Gabriele Inghirami, Behruz Kardan, Declan Keane, Arnaud Le~F\`{e}vre, William Llope, Ulrich Mosel, Berndt M\"{u}ller, Witold Nazarewicz, Gra\.{z}yna Odyniec, Panagiota Papakonstantinou, Ralf Rapp, Peter Rau, Bj\"{o}rn Schenke, Srimoyee Sen, Chun Shen, Christian Sturm, Giorgio Torrieri, and Wolfgang Trautmann for insightful discussions and comments on Sections~\ref{sec:introduction}-\ref{sec:combined_constraints}.
K.A.\ and M.S.\ thank Peter Senger and Richard Seto for helpful comments on the draft of Section~\ref{sec:experiment_snm}. 
M.K.\ thanks Navid Abbasi, David Blaschke, Casey Cartwright, Saso Grozdanov, Ulrich Heinz, Gabriele Inghirami, Michal P.~Heller, Jorge Noronha, Jacquelyn Noronha-Hostler, Dirk Rischke, Micha{\l} Spali\'{n}ski, Misha Stephanov, and Giorgio Torrieri for helpful comments on Section~\ref{sec:hydrodynamics}.

This work was supported in part by the INT's U.S.\ Department of Energy grant No.\ DE-FG02-00ER41132. 
K.A.\ acknowledges support from the Bundesministerium f\"{u}r Bildung und Forschung (BMBF, German Federal Ministry of Education and Research) – Project-ID 05P19VTFC1 and Helmholtz Graduate School for Hadron and Ion Research (HGS-HIRe). 
Z.C.\ acknowledges support from the U.S.\ National Science Foundation grant PHYS-2110218. 
P.D.\ acknowledges support by the U.S.\ Department of Energy, Office of Science, under Grant DE-SC0019209.
S.G.\ and I.T.\ acknowledge support from the U.S. Department of Energy, Office of Science, Office of Nuclear Physics, under contract No.~DE-AC52-06NA25396, and by the Office of Advanced Scientific Computing Research, Scientific Discovery through Advanced Computing (SciDAC) NUCLEI program; S.G.\ is also supported by the Department of Energy Early Career Award Program, while the work of I.T.\ is additionally supported by the Laboratory Directed Research and Development program of Los Alamos National Laboratory under project number 20220541ECR.
J.W.H.\ is supported by the U.S. National Science Foundation under grants PHY1652199, PHY2209318, and OAC2103680.
M.K.\ is supported, in part, by the U.S.\ Department of Energy grant DE-SC0012447.
C.-M.K.\ acknowledges support from the U.S. Department of Energy under Award No. DE-SC0015266.
R.K., W.G.L., and M.B.T. are supported by the National Science Foundation under Grant No. PHY-2209145.
B.-A.L.\ is supported in part by the U.S.\ Department of Energy, Office of Science, under Award Number DE-SC0013702, and the CUSTIPEN (China--U.S. Theory Institute for Physics with Exotic Nuclei) under the U.S.\ Department of Energy Grant No. DE-SC0009971.
A.B.M.\ acknowledges support from the U.S.\ Department of Energy grant DE-FG02-93ER40773.
W.G.N.\ is supported by the NASA grant 80NSSC18K1019 and the National Science Foundation grant 2050099. 
S.P.\ and O.S. are supported by the U.S.\ Department of Energy, Office of Science, grant no. DE-FG02-03ER41259. 
M.S.\ is supported by the Alexander von Humboldt Foundation and the U.S.\ Department of Energy grant DE-SC0020651.
R.V.\ acknowledges support from the U.S.\ Department of Energy, Office of Science, Office of Nuclear Physics under Contract DE-AC52-07NA27344.
H.W.\ acknowledges support from the Deutsche Forschungsgemeinschaft (DFG, German Research Foundation) under Germany’s Excellence Strategy EXC-2094-390783311.
H.Z.\ is supported by the National Science Centre, Poland, under grants No.\ 2021/41/B/ST2/02409 and 2020/38/E/ST2/00019, and by the Warsaw University of Technology project grants IDUB-POB-FWEiTE-3 and IDUB-POB POST-DOC PW.

\section*{Author's contributions}

Section \ref{sec:introduction} is primarily written by A.S., with input from P.D., B.-A.L., W.G.L., S.P, and M.B.T. Section~\ref{sec:model_simulations_of_HICs} is primarily written by P.D., A.S., and H.W. Section~\ref{sec:microscopic_calculations_of_the_EOS} is primarily written by C.D., S.G., J.W.H., and I.T. Section~\ref{sec:neutron_star_theory} is primarily written by W.G.N. Section~\ref{sec:experiment_snm} is primarily written by K.A., M.S, and H.Z. Section~\ref{sec:experiment_asym} is primarily written by K.W.B., Z.C., R.K., A.B.M., and M.B.T. Section~\ref{sec:combined_constraints} is primarily written by W.G.N. and M.B.T. Section~\ref{sec:applications_of_hadron_transport} is primarily written by R.V.\ and H.W. Section~\ref{sec:hydrodynamics} is primarily written by M.K. Section~\ref{sec:exploratory_directions} is primarily written by B.-A.L. All primary authors participated in discussions and provided in-depth comments on all versions of the manuscript. A.S.\ is the primary editor of the manuscript.

The list of authors starts with the primary editor, followed by primary authors listed in alphabetical order, which are then followed by endorsing authors likewise listed in alphabetical order. The primary authors provided the content of this White Paper, with individual contributions outlined above. The endorsing authors are members of the nuclear physics community who support the message of the White Paper. Many of the endorsing authors provided extensive comments and valuable suggestions based on the first public version of the manuscript.

\bibliography{main,noninspire}

\end{document}